\documentclass[longbibliography,
 reprint,
superscriptaddress,
 amsmath,amssymb,
 aps,
prb,
floatfix,
bibnotes,
]{revtex4-2}
\usepackage[english]{babel}
\usepackage{comment}
\usepackage{amsmath}
\usepackage{bbm}
\usepackage{natbib}
\usepackage{graphicx}
\usepackage{dcolumn}
\usepackage{bm}
\usepackage{siunitx}
\usepackage{mathtools}
\usepackage{float}
\usepackage{booktabs}
\usepackage[utf8]{inputenc}
\usepackage[T1]{fontenc}
\usepackage{xcolor}
\usepackage{url}
\usepackage[labelstoglobalaux]{bibunits}
\makeatletter
\renewcommand{\@bibunitname}{\jobname-supp}
\makeatother
\usepackage{hyperref}
\hypersetup{colorlinks=true,allcolors=blue}

\begin{document}
\preprint{APS/123-QED}

\title{Supercurrent-Driven N\'eel Torque in Superconductor/Altermagnet Hybrids}
\author{Hamed Vakili}
\thanks{\href{mailto:hvakilitaleghani2@nebraska.edu}{hvakilitaleghani2@nebraska.edu}}
\affiliation{Department of Physics and Astronomy and Nebraska Center for Materials and Nanoscience, University of Nebraska, Lincoln, Nebraska 68588, USA}
\author{Moaz Ali}
\affiliation{Department of Physics and Astronomy and Nebraska Center for Materials and Nanoscience, University of Nebraska, Lincoln, Nebraska 68588, USA}
\author{Igor \v{Z}uti\'{c}}
\affiliation{University at Buffalo, State University of New York, Buffalo, New York 14260-1500, USA}
\author{Alexey A. Kovalev}
\thanks{\href{mailto:alexey.kovalev@unl.edu}{alexey.kovalev@unl.edu}}
\affiliation{Department of Physics and Astronomy and Nebraska Center for Materials and Nanoscience, University of Nebraska, Lincoln, Nebraska 68588, USA}
\date{\today}

\begin{abstract}
We predict a supercurrent-driven N\'eel spin-orbit torque in a superconductor/$d$-wave altermagnet 
heterostructure, associated with the emergence of spin-triplet correlations. The effect can be understood as a consequence of the
supercurrent-induced spin polarization, 
owing to the interplay between spin-orbit coupling and momentum-dependent spin splitting, as found, for example, in altermagnets. 
Remarkably, the supercurrent can be tuned by the N\'eel-vector direction, and 
the supercurrent-induced torque can both propel magnetic domain walls and reverse the N\'eel-vector orientation within a domain wall. These findings establish superconductor/altermagnet heterostructures as a versatile platform for the {supercurrent-driven} control of the N\'eel vector, with potential applications in racetrack memory, dissipationless superconducting electronics, 
and unconventional computing.
\end{abstract}
\pacs{Valid PACS appear here}
\maketitle 

Materials with nonrelativistic spin splitting, including 
altermagnets~\cite{Krempasky2024,Zhu2024:N,Yamada2025:N,Song2025:N,Hayami2019:JPSJ,Smejkal2020SciAdv,Naka2019,Sandratski1981:PSSB,Mazin2021:PNAS,Yuan2021:PRM,PhysRevB.102.014422,Ma2021:NC,Smejkal2022AM,Smejkal2022PRX,Mazin2022,Khodas2026:X,Liu2025:NP,Guo2023MaterTodayPhys},
provide fascinating opportunities to expand the range of spin-dependent phenomena and their possible applications. A motivation to explore these unconventional magnets is often twofold: (i) Their inherent properties arising from the momentum-dependent spin splitting, described by the spin space groups~\cite{BrinkmanElliott1966SpinSpaceGroups,LitvinOpechowski1974SpinGroups,Sandratskii1979SpinSpaceGroups,Liu2022:PRX,Chen2024:PRX}, (ii) their ability to transform a large class of the neighboring materials in heterostructures with proximity effects~\cite{Zutic2019:MT}.

Focusing on altermagnets, there is a growing understanding how to take advantage of their zero net magnetization, ultrafast dynamics, tunability of the spin splitting and multiferroicity, nontrivial topology, transport, thermal, or optical response~\cite{Gonzalez2021,PhysRevLett.128.197202,Bose2022,Karube2022,PhysRevLett.134.176401,Duan2025:PRL,Gu2025:PRL,
Urru2025:PRB,Fernandes2024:PRB,Antonenko2025:PRL,Chen2026:X,
PhysRevB.108.L180401,PhysRevB.110.094427,PhysRevB.110.144421,Wu2025ChinPhysLett,Schwartz2025,Cao2025:PRL}. However, despite a number of studies exploring how altermagnets may transform properties of their heterostructures~\cite{Papaj2023,Ghorashi2024:PRL,Ouassou2023,Giil2024,PhysRevB.110.L140506,PhysRevLett.134.026001}, a more detailed understanding of the underlying altermagnetic proximity effects in the normal and superconducting state is just beginning to emerge~\cite{kqy8-myz1}. 

In this work, we reveal unexplored opportunities in superconductor/altermagnet (S/AM) heterostructures as a versatile platform for emergent phenomena, ranging from unconventional superconductivity and topological states to superconducting spintronics and nonreciprocal response~\cite{Linder2015,Eschrig2015,Amundsen2024,JDEexp}. Some intuition what to expect is derived from S/ferromagnet heterostructures, where the interplay between spin-singlet Cooper pairs and exchange fields can give rise to equal spin Andreev reflection, spin-triplet superconductivity, long-range proximity effects, $\pi$-junctions, Majorana states, and a Josephson effect with a single superconductor~\cite{Zutic1999:PRB,Buzdin2005,Bergeret2005,Ryazanov2001,Khaire2010:PRL,Valls:2022,Shen2024:PRB,Vezin2020:PRB,Banerjee2014:NC,Duckheim2011:PRB,Nadj-Perge2014:S,Fatin2016:PRL,Gonzalez-Ruano2025:NC,Gungordu2022Majorana,PhysRevResearch.5.033109,VakiliJDE,VakiliMajorana}. 

However, because the exchange fields in AMs are momentum-dependent, it is useful to recall related implications of systems with broken inversion symmetry and momentum-dependent spin-orbit coupling (SOC). These include current-induced spin polarization~\cite{AronovLyandaGeller1989,Edelstein1990,Levitov1985:SPJETP,Edelstein1995}, sometimes termed the Edelstein effect, spin-orbit torque (SOT), first measured in dilute magnetic semiconductors~\cite{Chernyshov2009:NP}, and magnetoanisotropy enhanced in the spin-triplet superconducting state~\cite{Amundsen2024,Hogl2015:PRL,Cai2021:NC,Martinez2020:PRA}.

Specifically, there is a growing interest in SOT which provides an efficient means of manipulating magnetic order through current-induced spin accumulation~\cite{Miron2011,Manchon2019,Tsymbal:2019}. In normal-state systems, SOT has enabled deterministic switching of magnets, with potential for advanced memory applications, unconventional computing, and spin communication~\cite{Wadley2016,Yang2022:N,Liang2025:JPDAP,Dainone2024:N}. 
The N\'eel SOT (NSOT) can induce fast motion of domain walls in antiferromagnets~\cite{PhysRevLett.117.017202}. Extending SOT concepts to superconducting systems has so far been proposed only for ferromagnets~\cite{hals}, while its realization in AMs could enable a tantalizing prospect of an ultrafast 
control of tunable magnetic systems with dissipationless spin currents.
This would be a key breakthrough, because despite decades of reports of spin-triplet superconductivity, demonstrating that the resulting spin currents can modify nonsuperconducting regions has remained elusive~\cite{Amundsen2024}. {Here and below, ``dissipationless'' refers to the superconducting drive current; magnetic dynamics can still dissipate through Gilbert damping, residual quasiparticles, vortices, or phase slips.}
 
\begin{figure}
\centering
\includegraphics[width=1\linewidth]{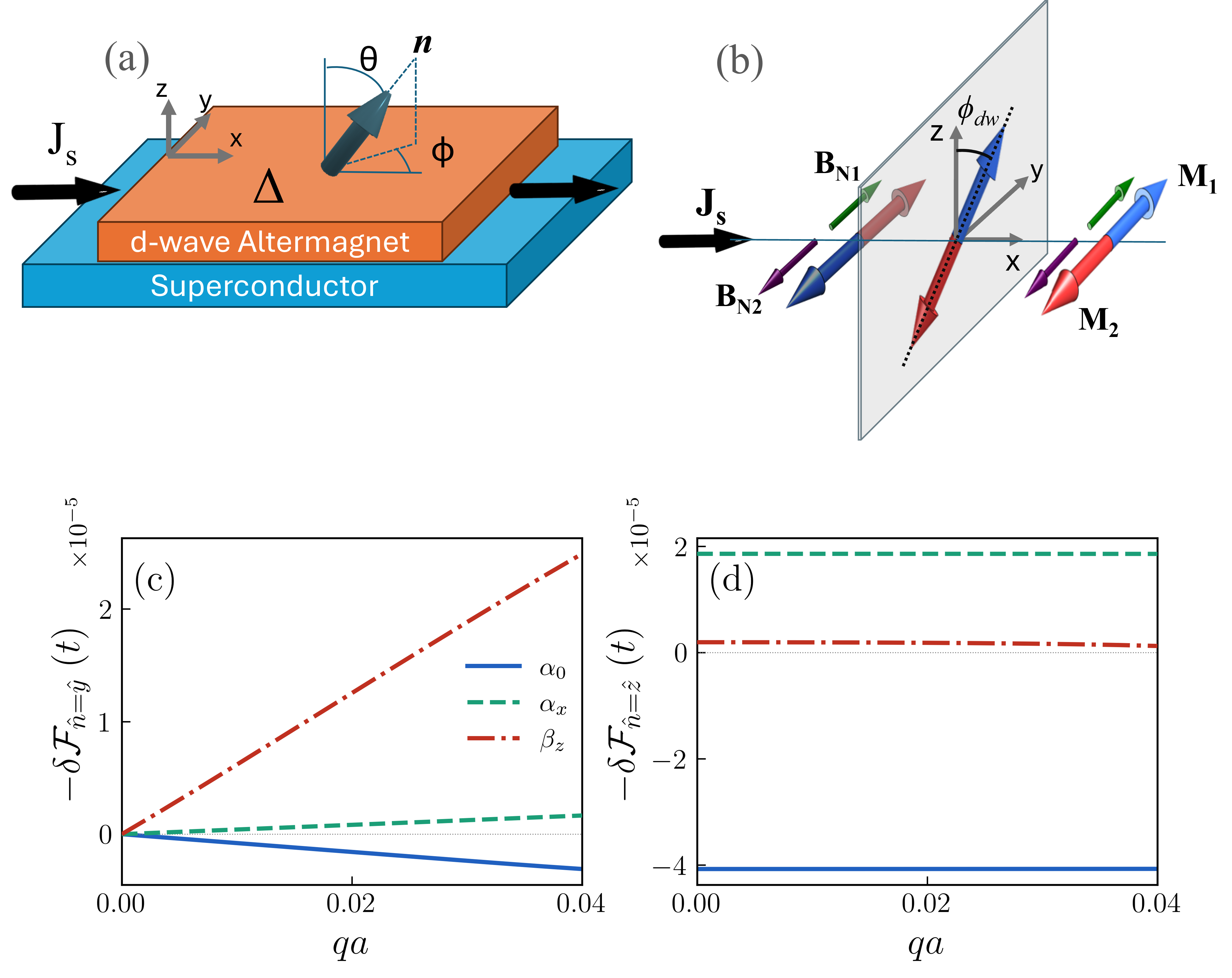}
\caption{(a) S/AM heterostructure with supercurrent density $\mathbf J_s \parallel \hat{x}$, N\'eel vector $\mathbf n$, and proximity-induced pairing gap $\Delta$ in AM. 
(b) $\mathbf J_s$-induced staggered NSOT fields $(\mathbf B_{N1}, \mathbf B_{N2})$ and a $180^\circ$ domain wall with sublattice magnetizations $\mathbf M_1$ and $\mathbf M_2$, where $\mathbf n=(\mathbf M_1-\mathbf M_2)/|\mathbf M_1-\mathbf M_2|$. 
(c) $\mathbf J_s$-induced N\'eel-field contribution, shown as the free-energy density difference $\delta\mathcal{F}_{\mathbf n}=\mathcal{F}_{\mathbf n}-\mathcal{F}_{\hat{x}}$, for $\mathbf n=\hat{y}$ as a function of $qa$, where $a$ is the lattice spacing. 
(d) Same as in (c), but for the uniaxial anisotropy with $\mathbf n=\hat{z}$. 
All energy parameters are given using t. 
The common parameters in Eq.~(\ref{eq:H_main})
are $t_2=0.1$, $t_3=0.6$, $t_4=0.1$, $\tilde{\mu}=-2.4$, $\Delta_0=0.02$, and $J_{\mathrm{ex}}=0.2$. 
For the SOC 
in Eq.~(\ref{eq:Hsoc_main}), 
the corresponding parameters are $\alpha_0=0.15$, $\alpha_x=0.1$, and $\beta_z=0.1$. 
The temperature is $T=0.1T_c$.}
\label{fig:deltamap}
\end{figure}

{Here we uncover NSOT induced by supercurrent flow in an $s$-wave S/$d$-wave AM heterostructure, where the order parameter is spin singlet but interfacial SOC generates induced triplet anomalous correlations,} 
depicted in Fig.~\ref{fig:deltamap}(a).
Using the finite-$\mathbf q$ Bogoliubov--de Gennes (BdG) Hamiltonian, in which $\hbar \mathbf{q}$ denotes the Cooper-pair center-of-mass momentum, together with self-consistent BdG calculations, we demonstrate that SOC in a noncentrosymmetric S region, in combination with altermagnetic interactions, induces a staggered spin density that can in turn generate an NSOT. We show that supercurrent-induced switching of the N\'eel-vector orientation, ${\mathbf n}$ 
is possible within a magnetic domain wall, while the domain wall itself can be propelled for certain ${\mathbf n}$  
alignments. These results establish S/AM heterostructures as a promising platform for superconducting spintronics~\cite{Cai2023:AQT} 
with functionalities beyond those achievable with conventional ferromagnets and antiferromagnets.

\textit{General theory and analytical results.}---
We use the Bogoliubov--de Gennes (BdG) theory to describe an S/AM heterostructure. The superconductivity may be intrinsic or proximity-induced. We employ the wave vector ${\bm k}$-dependent and finite-$\mathbf{q}$ BdG Hamiltonian, 
\begin{equation} 
\mathcal{H}^{{\bf k},{\bf q}}_{\mathrm{BdG}}(\mathbf n) = \begin{pmatrix} H_{{\bf k}+{\bf q}/2} & \hat{\Delta}_{\bf k} \\ \hat{\Delta}^{\dagger}_{\bf k} & -\,H^{*}_{ -{\bf k}+{\bf q}/2} \end{pmatrix}, 
\end{equation}
where $\hat{\Delta}_{\bf k}$ is the $\bf q$-independent gap function, $H_{\bf k}$ is the normal-state Hamiltonian of the AM, 
and the supercurrent is introduced through 
Cooper-pair center-of-mass momentum $\hbar\mathbf{q}$. 
We define  the free energy density
\begin{equation}
 \mathcal{F}({\bf q},\mathbf n)= \mathcal{F}_{\rm sc}-\frac{1}{2\beta V}\sum_{{\bf k},n}
\mathrm{Tr}\ln\!\Big(i\omega_n-\mathcal{H}^{{\bf k},{\bf q}}_{\mathrm{BdG}}  (\mathbf n)\Big),
\label{eq:F}
\end{equation}
with $\omega_n = (2n+1)\pi/\beta$ denoting the Matsubara frequency, $\beta = 1/(k_B T)$, where $T$ is the temperature, and $V$ the surface area or volume of the system. Here, $\mathcal F_{\rm sc}$ 
is the superconducting component of the free-energy density, which is independent of altermagnetic interactions, and can be used to relate the supercurrent density
${\bf J}_s = \nu_s\,{\bf q}$, where $\nu_s=\partial^{2}\mathcal{F}_{\mathrm{sc}}(q)/\partial q^{2}$. 
We focus first on the last term in Eq.~\eqref{eq:F} and neglect possible contributions arising from a self-consistent treatment  
of $\hat{\Delta}_{\mathbf{k}}$, which 
we consider to be $s$-wave,  
$\hat{\Delta}_{\mathbf{k}}=\Delta i \sigma_y$, with a $k$-independent pairing gap, 
while $\sigma_y$ is the Pauli matrix in the spin space.
The gap follows the BCS temperature dependence, $\Delta=\Delta_0\tanh\!\left(1.74\sqrt{T_c/T-1}\right)$, 
where $T_c$ is the critical temperature. {Thus, we assume neither a triplet pairing interaction nor a triplet order parameter. Nevertheless, inversion-breaking SOC and magnetism generally cause the anomalous block of the BdG Green function to acquire induced triplet components. Fermionic antisymmetry requires relative-even-parity triplet correlations, including local $s$-wave triplet correlations, to be odd in Matsubara frequency, whereas relative-odd-parity triplet correlations may be even in frequency. These correlations should therefore be distinguished from an intrinsic triplet pairing gap.}

From Fig.~\ref{fig:deltamap}(b) we can infer the presence of ${\bf J}_s$-induced staggered NSOT field
($\mathbf{B}_{N1}, \mathbf{B}_{N2}$) and define 
the related  effective N\'eel field, as $\mathbf{h} =
|{\bf M}_1-{\bf M}_2|(\mathbf{B}_{N1} - \mathbf{B}_{N2})/2$, along $\hat{y}$. 
More generally, 
$\mathbf h({\bf q},\mathbf n)\equiv-
\partial \mathcal{F}/\partial {\mathbf n}$,  
which can be expressed as a sum over Matsubara frequencies
\begin{equation}
h_i({\bf q},{\mathbf n})
=
-\frac{1}{2\beta V}\sum_{{\bf k}, n}
\mathrm{Tr}\!\left[
\mathcal{G}({\bf k},{\bf q},i\omega_n)\,U_i({\bf k},{\bf q})
\right],
\label{eq:Heff_Matsubara}
\end{equation}
where $\mathcal{G}({\bf k},{\bf q},i\omega_n)=(i\omega_n-\mathcal{H}^{ {\bf k},{\bf q}}_{\mathrm{BdG}})^{-1}$, 
and $U_i({\bf k},{\bf q})\;\equiv\;\partial \mathcal{H}^{{\bf k},{\bf q}}_{\mathrm{BdG}}(\mathbf n)/\partial n_i$. 
The corresponding 
NSOT becomes
\begin{equation}
\boldsymbol\tau_\mathbf n(\mathbf q)={\mathbf n}\times\mathbf h(\mathbf q,\mathbf n), 
\end{equation}
where we  note that this torque is 
staggered between the magnetic sublattices and
has a field-like form~\cite{Tsymbal:2019}, 
{allowing it to be generated by a superconducting drive current rather than a normal Ohmic current. This does not imply lossless magnetic dynamics: Gilbert damping and residual quasiparticle, vortex, or phase-slip channels can still contribute.}
In the limit $J_{ex} \mu/t \ll \Delta$, 
where $\mu$ is the chemical potential, and $t$ is the effective hopping strength
that parametrizes the kinetic energy, 
we can expand the last term in Eq.~\eqref{eq:F} to
the second order in  exchange coupling
$J_{ex}$. We identify a field term at 
the first order
\begin{align}
\mathcal {F}^{(1)}({\bf q},{\mathbf n})
=\frac{1}{2\beta V}\sum_{{\bf k}, n} n_i\,\mathrm{Tr}\!\big[\mathcal G_0\, U_i\big],
\label{eq:F1}
\end{align}
and a magnetic anisotropy at the 
second order
\begin{align}
\mathcal{F}^{(2)}({\bf q},\mathbf n)
=\frac{1}{4\beta V}\sum_{{\bf k}, n} n_i n_j\,
\mathrm{Tr}\!\big[\mathcal G_0\, U_i\,\mathcal G_0\, U_j\big],
\label{eq:F2}
\end{align}
where $\mathcal G_0({\bf k},{\bf q},i\omega_n)$ is evaluated at $J_{ex}=0$. {Outside this weak-exchange regime, higher-order contributions are no longer negligible, and the full free energy must therefore be evaluated numerically using Eq.~(\ref{eq:F}).}

We consider a realistic 2D minimal 4-band model for AM which has shown to match ab-initio bandstructures
~\cite{Minimal_models,lh4w-byk7} 
\begin{equation}
H_{2D} = \epsilon_{0,k} + t_{x,k}\tau_x + t_{z,k}\tau_z  + \tau_z J_{ex} (\mathbf n\cdot\boldsymbol\sigma),
\label{eq:H_main}
\end{equation}
where $\epsilon_{0,k}=-t(\cos k_x+\cos k_y)+t_2\cos k_x\cos k_y-\tilde{\mu}$, $t_{x,k}=t_3\cos\frac{k_x}{2}\cos\frac{k_y}{2}$, and $t_{z,k}=t_4(\cos k_x-\cos k_y)$. Here \(\tau_i\) act in sublattice space, \(\sigma_i\) in spin space.
The altermagnetic interaction is chosen so that, in momentum space, the corresponding spin splitting transforms as
$f(\mathbf{k})\sim (k_x^2-k_y^2)$ near the bottom of the band. 
We also add symmetry-allowed SOC terms assuming the [001] interface~\cite{Deterministic_AM,Note}
\begin{equation}
H_{\rm SOC}(\mathbf k)
=
(\alpha_0+\alpha_x\tau_x)(\mathbf g_-\cdot\boldsymbol\sigma)
+\beta_z\tau_z(\mathbf g_+\cdot\boldsymbol\sigma),
\label{eq:Hsoc_main}
\end{equation}
where $\mathbf g_\pm=(\sin(k_y),\pm\sin(k_x),0)$, and $\alpha_0$, $\alpha_x$, and $\beta_z$ parametrize SOC. {Equation~\eqref{eq:Hsoc_main} is a symmetry-based interfacial SOC model rather than a material-specific first-principles parametrization; intrinsic material SOC terms, including possible $\tau_y$ structures, can renormalize $\gamma$ and the anisotropies without removing the allowed linear coupling. SOC-induced weak ferromagnetism can also be symmetry-allowed~\cite{McClartyRau2024PRL,Jo2025PRL}, but it is expected to be much smaller than the altermagnetic sublattice moment. In simulations, adding such a small moment did not meaningfully change the supercurrent angular dependence or the domain-wall free-energy trends. Note that a simplified 2-band model of AM, would not capture the correct shape of SOC terms as it does not distinguish between spin and sublattice degrees of freedoms. }

{In a uniform-superflow state, Neumann's principle applied to the
surface antisymmetry point group of the $[001]$ interface constrains
the leading contribution to the free-energy density that is bilinear
in the condensate momentum $\mathbf q$ and the N\'eel vector
$\mathbf n$. Here, $\mathbf q$ transforms as an in-plane polar
vector, whereas $\mathbf n$ is a sublattice-odd axial vector. For the
relevant surface antisymmetry point group
${}^{2}4\,{}^{1}m_x\,{}^{2}m_d$~\cite{Deterministic_AM}, the unique symmetry-allowed
bilinear invariant is}
\begin{equation}
\mathcal{F}^{(1)}(\mathbf q,\mathbf n)
=
J_{\mathrm{ex}}\gamma
\left(q_x n_y+q_y n_x\right),
\label{eq:F_expand_main_vec}
\end{equation}
where $\gamma$ is a microscopic response coefficient generated by
the interfacial SOC. It scales with the relevant SOC parameters and
vanishes when all interfacial SOC couplings are set to zero, as shown
by the analysis of Eq.~\eqref{eq:F1} in the Supplemental
Material~\cite{Note}. This term describes an in-plane N\'eel field induced by the supercurrent. Note that a $k_xk_y$ spin splitting will result in the rotated form of Eq.~\eqref{eq:F_expand_main_vec}. The Supplemental Material also gives a $g$-wave altermagnet example~\cite{Note}.
{Additional torque components can become symmetry-allowed in unconventional superconducting phases, but only when the superconducting order lowers the symmetry of the full BdG system~\cite{ChakrabortyBlackSchaffer2025Constraints}. The appearance of triplet components in the anomalous Green function due to SOC and magnetism does not by itself lower the symmetry or imply an intrinsic triplet order parameter; only a symmetry-breaking component of the self-consistent superconducting order can modify the allowed torque tensor~\cite{HeinsdorfFranz2026}.}

Further expansion in $\mathbf n$ has the symmetry dictated form 
\begin{equation}
\mathcal{F}^{(2)}({\bf q},{\mathbf n})
=
-J_{ex}^2A_{z}\,n_z^2,
\label{eq:F0_main_nz}
\end{equation}
where $A_z$ scales with the square of the SOC strength, as follows from Eq.~\eqref{eq:F2}, which has been analyzed in~\cite{Note}.
This dependence on the N\'eel vector corresponds to a  
uniaxial magnetic 
anisotropy~\cite{Tsymbal:2019}.

\begin{figure}
\centering
\includegraphics[width=1\linewidth] 
{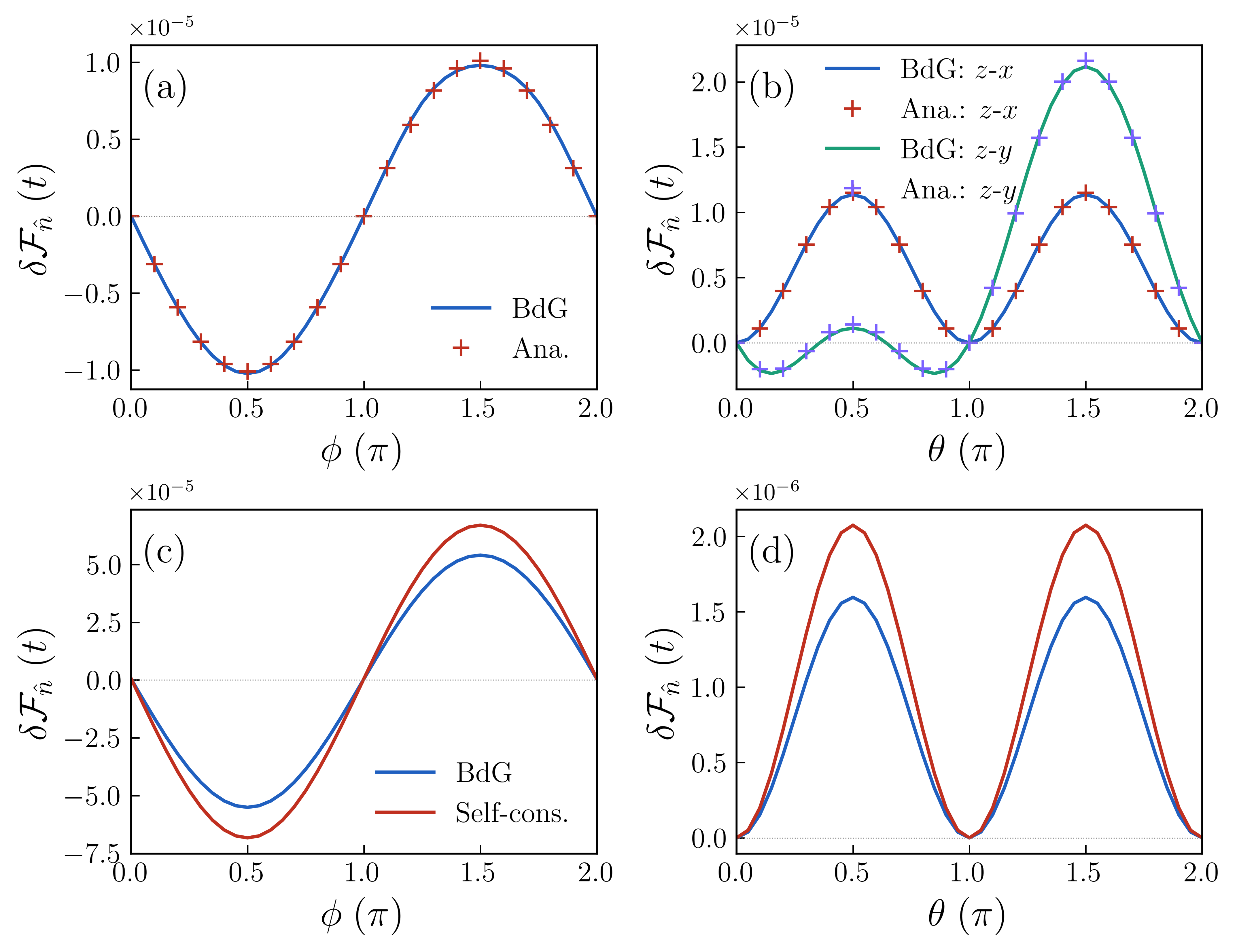}
\caption{ 
(a) Free-energy density from Eq.~\eqref{eq:F} (marked BdG) together with the analytical results (marked Ana.) from Eqs.~\eqref{eq:F1} and \eqref{eq:F2}, for the N\'eel vector $\mathbf n$ rotating in the $\hat{x}$-$\hat{y}$ plane. 
(b) Same as in (a), but for rotation in the $\hat{z}$-$\hat{x}$ and $\hat{z}$-$\hat{y}$ planes. 
BdG and self-consistent free-energy densities for (c) 
$\mathbf n$ rotating in the $\hat{x}$-$\hat{y}$ plane
and (d) in the $\hat{z}$-$\hat{x}$ plane. 
All energy parameters are given using $t$. 
For (a) and (b), $t_2=0$, $t_3=0.3$, $t_4=0.1$, $\tilde{\mu}=-2.1$, $\Delta_0=0.03$, $\beta_z=0.1$, and $J_{\mathrm{ex}}=0.1$. 
For (c) and (d), $t_2=0$, $t_3=0.3$, $t_4=0.1$, $\tilde{\mu}=-2.2$, $V_s=2.5$, $\beta_z=0.1$, and $J_{\mathrm{ex}}=0.1$. 
In all panels, $\mathbf q a=(qa,0,0)$ and $T=0.1T_c$; $qa=0.01$ in (a) and (b), and $qa=0.2$ in (c) and (d). 
The self-consistent calculations in (c) and (d) are 
for a system 
$L_x\times L_y=30a\times25a$.}
    \label{fig:analyticals}
\end{figure}

In Figs.~\ref{fig:deltamap}(c) and \ref{fig:deltamap}(d), we show the quasiparticle free energy $\mathcal{F}({\bf q},\mathbf n)$ obtained from Eq.~\eqref{eq:F} using the BdG spectrum, with the different types of SOC included one at a time.
To identify the field-like and anisotropy-like terms, we plot $\delta \mathcal{F}_{\mathbf n}=\mathcal{F}({\bf q},\mathbf n)-\mathcal{F}({\bf q},\hat x)$ for different directions of $\mathbf n$. In Fig.~\ref{fig:deltamap}(c), 
we plot $\delta \mathcal{F}_{\mathbf n}$ for $\mathbf n=\hat y$ as a function of $q$. We find a 
${\mathbf J}_s$-induced, linear-in-$q$ field-like response: The 
effective field lies in-plane and can be used to align the N\'eel order parameter, see Fig.~\ref{fig:deltamap}(b). 
In Fig.~\ref{fig:deltamap}(d), 
we plot $\delta \mathcal{F}_{\mathbf n}$ for $\mathbf n=(0,0,1)$ as a function of $q$ (parameters listed in the caption). The result shows a uniaxial anisotropy along the $\hat z$ axis that is present even at $q=0$,
and weakly modified by superflow-induced corrections of the form $q^2$. {This $q=0$ anisotropy is the SOC-induced $O(J_{ex}^2)$ contribution in Eq.~\eqref{eq:F0_main_nz}; it is distinct from the linear-in-$q$ N\'eel field in Eq.~\eqref{eq:F_expand_main_vec}.}

In Fig.~\ref{fig:analyticals}, we sweep the N\'eel-vector orientation in two orthogonal rotation planes and compare the analytical results with the numerical BdG evaluation. In panel~(a), ${\mathbf n}$ is rotated within the $\hat{x}$--$\hat{y}$ plane, and in 
(b) 
within the $\hat{z}$--$\hat{x}$ and $\hat{z}$--$\hat{y}$ planes. We use representative superflow strengths of $qa=0.01$ for (a) and (b). 
The results are consistent with Eqs.~\eqref{eq:F_expand_main_vec} and \eqref{eq:F0_main_nz}. {The difference between the $\hat z$--$\hat x$ and $\hat z$--$\hat y$ sweeps at finite $q$ is caused by the imposed superflow direction: for $\mathbf q=q\hat x$, Eq.~\eqref{eq:F_expand_main_vec} gives a linear term $\propto q_x n_y$, so only rotations with nonzero $n_y$ sample this contribution. At $q=0$, the two planes are equivalent up to the common $n_z^2$ anisotropy.}
Figures~\ref{fig:analyticals}(c) and \ref{fig:analyticals}(d) benchmark the $\hat{x}$--$\hat{y}$- and $\hat{z}$--$\hat{x}$-plane angular sweeps, respectively, against a fully self-consistent calculation (details are given in the following section), in which 
$\hat{\Delta}_{\mathbf{k}}$
is determined self-consistently and then inserted into Eq.~\eqref{eq:F} for comparison with the numerical BdG results.  
The total free energy obtained from the self-consistent calculation is divided by the total system size to obtain the free-energy density. The two approaches yield the same characteristic angular harmonics and very similar amplitudes, indicating that self-consistency primarily produces a modest renormalization of the response while leaving the symmetry and functional form of the N\'eel vector dependence intact in the parameter regime considered.

\textit{Self-consistent calculation.}--- 
Here we model S/AM heterostructure
using a tight-binding Hamiltonian on a square lattice with hopping determined by shifts $\mathcal B=\{\hat{x},\hat{y},\eta_+,\eta_-,\nu_+,\nu_-\}$ with $\eta_\pm=\hat{x}\pm\hat{y}$ and $\nu_\pm=(\hat{x}\pm\hat{y})/2$. The total Hamiltonian is given by~\cite{Beenakker2023,Giil2024}
\begin{equation}
H=
\sum_i \frac{1}{2}
\left[
\sum_{\delta\in\mathcal B}
\Psi_{i+\delta}^\dagger T_\delta \Psi_i
+
\Psi_i^\dagger M \Psi_i
+
\Delta_i\,
\Psi_i^\dagger P (\Psi_i^\dagger)^T
\right]
+\text{h.c.}
\end{equation}
where
\begin{align}
T_{\hat{x}}
&=
\left(
-t\tau_0+t_4\tau_z
\right)\sigma_0
+
i\,(-A+B)\sigma_y,
\\[1mm]
T_{\hat{y}}
&=
\left(
-t\tau_0-t_4\tau_z
\right)\sigma_0
+
i\,(A+B)\sigma_x,
\end{align}
and $T_{\eta_\pm}=t_2\,\tau_0\sigma_0/2$, $T_{\nu_\pm}=
t_3\,\tau_x\sigma_0/2$, $M=
-\tilde{\mu}\,\tau_0\sigma_0
+
J_{ex}\,\tau_z(\bm n\cdot\bm\sigma)$, $A=\alpha_0\tau_0+\alpha_x\tau_x$,
$B=\beta_z\tau_z$,
$P=i\tau_0\sigma_y$,
$\Delta_i$ is the superconducting 
pair potential. We define $\Psi_i^\dagger=(a_{i\uparrow}^\dagger,a_{i\downarrow}^\dagger,b_{i\uparrow}^\dagger,b_{i\downarrow}^\dagger)$, where $a_{i\sigma}^\dagger$ and $b_{i\sigma}^\dagger$ are fermionic creation operators for two sublattices. 
The SOC is of the Rashba and Dresselhaus types allowed by symmetry. We apply periodic boundary conditions to suppress edge effects. A uniform superflow is imposed by introducing a phase twist along the $x$ direction using 
the periodic boundary conditions and enforcing a total phase winding of $2\pi$ in the $\hat{x}$ direction. {We have also checked different system sizes and open boundary conditions; they preserve the same current--N\'eel-vector angular harmonics and domain-wall free-energy trends, while open boundaries mainly add edge suppression of $\Delta_i$ and nonuniform edge currents near the boundary.}

We compute $\Delta_i$ self-consistently from the BdG eigenvectors.
Denoting the eigenenergies by $E_n$ and the Nambu eigenvectors at site $i$ by
$(u_{n\uparrow}(i),u_{n\downarrow}(i),v_{n\uparrow}(i),v_{n\downarrow}(i))$,
the superconducting $s$-wave 
pair potential is
\begin{equation}
\Delta_i =
\frac{V_s}{2}\sum_{E_n>0}
\Big[u_{n\uparrow}(i)\,v_{n\downarrow}^\ast(i)
- u_{n\downarrow}(i)\,v_{n\uparrow}^\ast(i)\Big]\,
\tanh\!\left(\frac{\beta E_n}{2}\right),
\label{eq:gap_equation}
\end{equation}
where $V_s$ is the on-site pairing interaction.
The mean-field iteration is stopped when the relative change in the pairing field falls below $10^{-5}$.
For real-space self-consistent calculations, we use the following definition of the total free energy 
\begin{equation}
F =
-\frac{1}{\beta }\sum_{E_n>0}
\ln\!\Bigg[2\cosh\!\Big(\frac{\beta E_{n}}{2}\Big)\Bigg]
+ \sum_i \frac{|\Delta_i|^2}{V_s}.
\end{equation}

In our numerical calculations, we also compute the charge current directly from the bond charge-current operator defined on the $\langle i,i+\delta \rangle$ bond as
\begin{equation} 
\hat{I}_{i\to i+\delta} =\frac{ie}{\hbar}\left(\Psi_i^\dagger\, T_\delta\, \Psi_{i+\delta} - \Psi_{i+\delta}^\dagger\, T_\delta^\dagger\, \Psi_{i}\right). \label{eq:charge_current_op} 
\end{equation}
The charge current across a cut is obtained by summing the expectation value of $\hat{I}_{i\to i+\delta}$
over all bonds $\langle i,i+\delta\rangle$ that cross that cut.

\begin{figure}
\centering
\includegraphics[width=1\linewidth]{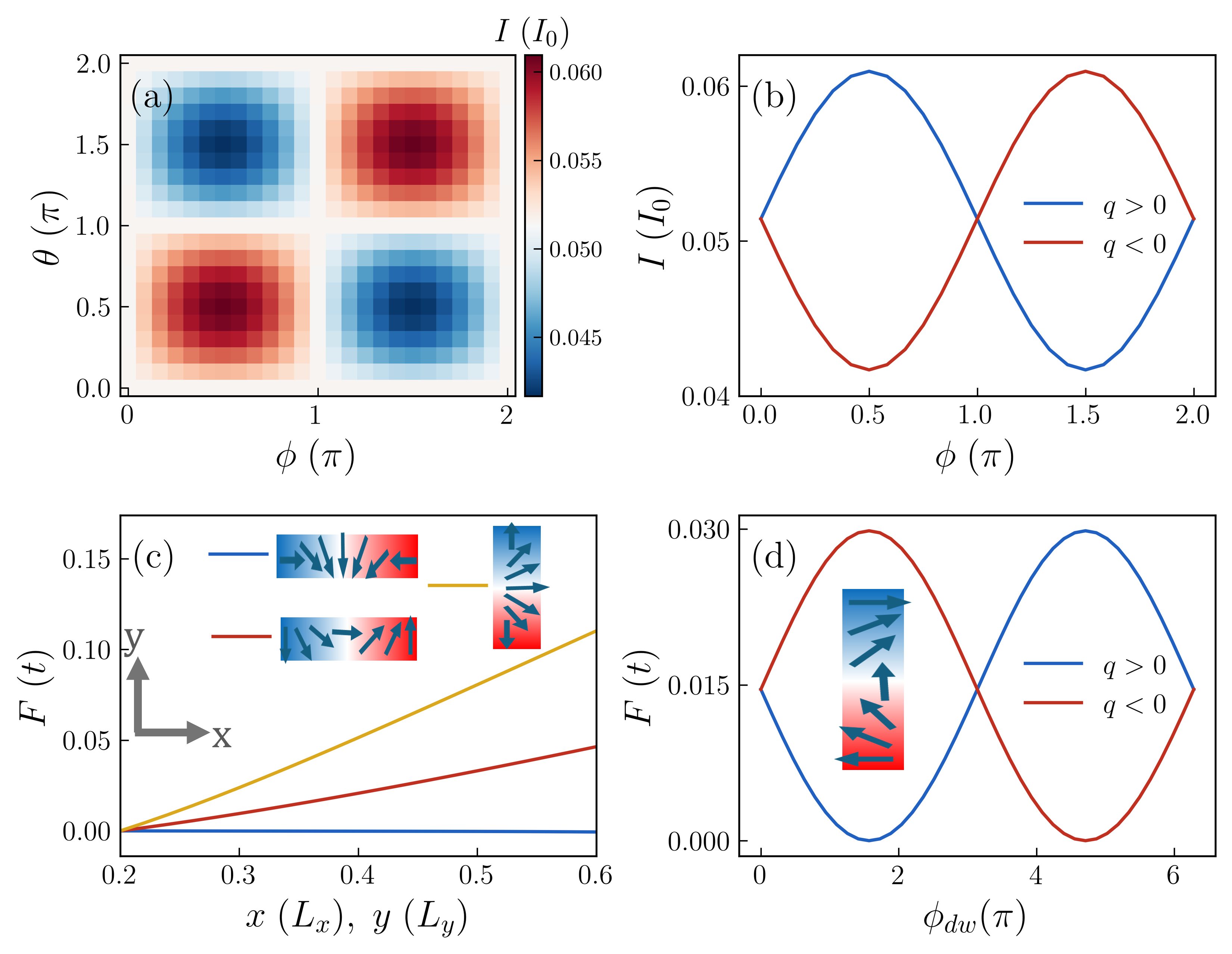}
\caption{ 
(a) The self-consistent  supercurrent, $I$, for a uniform N\'eel vector ${\mathbf n}$ as a function of the spherical angles $\theta$ and $\phi$ and (b) as a function of $\phi$ for opposite signs of $q$ at $\theta=\pi/2$. The same parameters as in Fig.~\ref{fig:analyticals}(c) are used. (c) The self-consistent free energy as a function of the domain-wall position $\mathbf{r}_0$ as the wall moves through S along 
$\hat{x}$ and $\hat{y}$, 
(d) as function of the internal wall angle $\phi_{dw}$ for opposite directions of current. The insets: The domain walls 
in the calculations. We use $\delta_{dw}=4a$. In (c), (d) $L_x=50a$ and $L_y=20a$ are used for a domain wall extending along $\hat{x}$, and $L_x=20a$ and $L_y=50a$ for a domain wall extending along $\hat{y}$. The free energy $F$ is set 
to zero for a domain wall at $x=0.2L_x$ in (c) and at $y=0.2L_y$ in (d). We define $I_0=et/(\hbar a)$. The supercurrent is along $\hat{x}$ in all panels and $|qa|=2\pi/L_x$. }
\label{fig:fen}
\end{figure}

Figure~\ref{fig:fen} connects the microscopic current--N\'eel-vector coupling derived above to two 
functionalities of the S/AM heterostructure: Modulation 
of the supercurrent by the N\'eel-vector alignment and current-controlled N\'eel torques on magnetic textures.
We first demonstrate that the supercurrent is strongly tunable by the N\'eel-vector orientation. As shown in Fig.~\ref{fig:fen}(a), the supercurrent $I({\mathbf n})$ exhibits an angular dependence over the sphere parameterized by $(\theta,\phi)$, reflecting the symmetry-selective coupling between the imposed superflow and the in-plane component of ${\mathbf n}$. This implies that by rotating ${\mathbf n}$ one can modulate the magnitude of the supercurrent.

Figure~\ref{fig:fen}(b) highlights the corresponding current reversal under $\mathbf q\to-\mathbf q$ by plotting $I(\phi)$ for opposite phase twists, $\pm2\pi$. The two curves are related by the expected sign change, confirming that the measured current is controlled by the relative orientation between $\mathbf q$ and the in-plane N\'eel component. Together, panels (a)--(b) establish the possibility of modulating the current using the N\'eel vector orientation of the AM. 

We then show how the same coupling produces a spatially varying free-energy landscape for a domain wall, enabling current-driven texture motion. it is described by the N\'eel vector~\cite{Kosevich1990MagneticSolitons}
\begin{equation}
{\mathbf n} = \big(\sin\theta_{dw}\sin\phi_{dw},\cos\theta_{dw},\,\sin\theta_{dw}\cos\phi_{dw}\big), 
\label{eq:n}
\end{equation}
where $\phi_{dw}$ is the tilt of the domain wall and the angle $\theta_{dw}$ follows a standard 180$^\circ$ wall profile~\cite{Kosevich1990MagneticSolitons}
$\theta_{dw}(x) = 2 \arctan\left( e^{(\mathbf{r} - \mathbf{r}_0)/\delta_{dw}} \right)$,
with $\mathbf{r}_0$ denoting the domain-wall center and $\delta_{dw}$ its 
width. To implement periodic boundary conditions, we place a duplicate of the domain wall at the edge.
In Fig.~\ref{fig:fen}(c), we plot the free energy as the domain-wall center is translated through the superconducting region along the $\hat{x}$ and $\hat{y}$ directions, respectively (see the insets for the wall geometry). For a supercurrent along the $\hat{x}$ axis, the free energy develops a clear linear, position-dependent slope for domain walls separating $\hat{y}$ (up) and $-\hat{y}$ (down) domains, corresponding to a uniform net force on the wall and hence a preferred drift direction. For a domain wall that is globally rotated about the $\hat{z}$ axis by an angle $\pi/2$, the force is negligible. This is because, for this alignment, the N\'eel vector outside the domain wall is parallel to $\mathbf{q}$, and there is no coupling to the supercurrent described by Eq.~\eqref{eq:F_expand_main_vec}. Figure~\ref{fig:fen}(d) shows that complete reversal of the N\'eel vector inside a static domain wall can also be achieved by applying a supercurrent along the $\hat{x}$ axis. {The same panel shows a supercurrent-induced in-plane profile of the wall free energy, because reversing the current selects the opposite internal wall angle. Material-specific anisotropies may favor different wall planes; in-plane or easy-plane anisotropies are known in antiferromagnetic/altermagnetic candidate materials and texture theories~\cite{Kriegner2017PRB,Gomonay2024NPJ,Kravchuk2025PRB}. Altermagnetic domain walls can also carry a small local magnetization-gradient contribution~\cite{Gomonay2024NPJ}; simulations with a spatially wall-localized magnetization profile did not meaningfully change the free-energy slope, preferred wall-motion direction, or internal-angle dependence.}

To assess the experimental relevance of the predicted N\'eel spin-orbit torques, we convert the numerically obtained free-energy scale to physical units. {We explicitly implement the S/AM stack~\cite{Note}}. We take $t = 0.1\;\mathrm{eV}$, $t_4 = 0.15\;\mathrm{eV}$, $\beta_{z} = 10\;\mathrm{meV}$, $J_{ex}=0.18\;\mathrm{eV}$, $\mu = 0.03 \;\mathrm{eV}$, $\Delta_0 = 2\;\mathrm{meV}$, and $a=0.99\;\mathrm{nm}$~\cite{mejkal2022,Zhang2024,arxiv_KRu4O8}. The effective field acting on the N\'eel vector is $B_\text{N}=a^2|\mathcal{F}^{(1)}|/m_s$ \cite{PhysRevLett.117.017202,Tang2025}, where $m_s$ is the sublattice magnetic moment per unit cell. The amplitude of the linear-in-$q$ free-energy density is {numerically} calculated to be {$|\mathcal{F}^{(1)}|\sim 10^{-7}\,\text{eV}/a^2$} at $q=10^{-3}$\AA$^{-1}$ and interfacial coupling of $t_\perp = 10\,\mathrm{meV}$ . Taking $m_s= 0.7\,\mu_B$~\cite{arxiv_KRu4O8}, we obtain $B_\text{N}\approx 2.2~\mathrm{mT}$. According 
to previous studies
~\cite{PhysRevLett.117.017202}, a N\'eel SOT field of this magnitude should be sufficient for the fast motion of a domain wall. Thus, we find that dissipationless currents in proximitized altermagnetic thin films can be used for efficient control of the N\'eel vector, with potential applications in racetrack memory technologies~\cite{Parkin2015MemoryOnTheRacetrack}.

{The phase gradient used in this estimate is also compatible with standard superconducting-current scales~\cite{PhysRevLett.134.026001,Tinkham2004,Henrich2012PRB,ClemBerggren2011PRB}. A related quasiclassical treatment of thermally driven AF domain-wall motion
in superconductor/antiferromagnet bilayers showed that time-dependent
corrections to the superconducting quasiparticle distribution caused by the
moving magnetic texture are negligible for realistic domain-wall velocities
and widths \cite{Bobkov2020}.}

\textit{Conclusion.}---
We have demonstrated that supercurrents in S/AM heterostructures generate a N\'eel spin-orbit torque with symmetry properties dictated by the $d$-wave spin splitting of the altermagnet. Using both a finite-$\mathbf{q}$ BdG framework and fully self-consistent calculations, we established that the interplay of Rashba spin--orbit coupling and altermagnetic interaction produces two distinct contributions to the magnetic free energy: a linear-in-supercurrent N\'eel torque and a smaller uniaxial anisotropy. These effects can arise in both intrinsic and proximity-induced superconductivity, as confirmed by approaches with and without self-consistent determination of the superconducting pairing. {The mechanism is enabled by SOC and singlet--triplet mixing in the anomalous correlations, but it does not require a triplet order parameter or a net ferromagnetic moment. The drive current is superconducting, while material-specific damping and quasiparticle channels determine the residual losses during magnetic dynamics.}

These results establish S/AM heterostructures as a versatile platform for superconducting spintronics, in which the momentum-dependent spin splitting of the altermagnet enables qualitatively new functionalities that are usually associated with ferromagnets. Future directions include exploring finite-temperature and dynamical regimes, as well as experimental realization in candidate altermagnetic materials such as KRu$_4$O$_8$~\cite{Zhang2024}, Mn$_5$Si$_3$~\cite{Reichlova2024Mn5Si3AHE}, VNb$_3$S$_6$~\cite{Zhu2025MagneticGeometry}, MnTe~\cite{Krempasky2024}, and CrSb~\cite{Reimers2024}, where the interface-induced NSOT response might be allowed by the antisymmetry group for specific surface orientations~\cite{Deterministic_AM}. {First-principles calculations of the interfacial SOC and exchange parameters for specific surfaces would provide useful quantitative input for material-specific device estimates.} We further envision 
applications of N\'eel-vector-controlled dissipationless transport and supercurrent-controlled N\'eel-vector dynamics, enabling current-driven domain-wall motion for memory applications~\cite{Parkin2015MemoryOnTheRacetrack,Ryu2013ChiralSpinTorque,PhysRevB.108.174516}, as well as applications in superconducting dissipationless logic and electronic devices~\cite{Edwards2026}.

\textit{Acknowledgments.}--- 
We gratefully acknowledge useful discussions with Kirill Belashchenko and Stuart Parkin. This work was supported by the U.S. Department of Energy, Office of Science, Basic Energy Sciences, under Awards No. DE-SC0021019
and DE-SC0004890 (I.\v{Z}.). 
This work used the Holland Computing Center of the University of Nebraska, which receives support from the UNL Office of Research and Innovation, and the Nebraska Research Initiative.

\bibliography{SAM_revised}

\begin{thebibliography}{5}%
\makeatletter
\providecommand \@ifxundefined [1]{%
 \@ifx{#1\undefined}
}%
\providecommand \@ifnum [1]{%
 \ifnum #1\expandafter \@firstoftwo
 \else \expandafter \@secondoftwo
 \fi
}%
\providecommand \@ifx [1]{%
 \ifx #1\expandafter \@firstoftwo
 \else \expandafter \@secondoftwo
 \fi
}%
\providecommand \natexlab [1]{#1}%
\providecommand \enquote  [1]{``#1''}%
\providecommand \bibnamefont  [1]{#1}%
\providecommand \bibfnamefont [1]{#1}%
\providecommand \citenamefont [1]{#1}%
\providecommand \href@noop [0]{\@secondoftwo}%
\providecommand \href [0]{\begingroup \@sanitize@url \@href}%
\providecommand \@href[1]{\@@startlink{#1}\@@href}%
\providecommand \@@href[1]{\endgroup#1\@@endlink}%
\providecommand \@sanitize@url [0]{\catcode `\\12\catcode `\$12\catcode
  `\&12\catcode `\#12\catcode `\^12\catcode `\_12\catcode `\%12\relax}%
\providecommand \@@startlink[1]{}%
\providecommand \@@endlink[0]{}%
\providecommand \url  [0]{\begingroup\@sanitize@url \@url }%
\providecommand \@url [1]{\endgroup\@href {#1}{\urlprefix }}%
\providecommand \urlprefix  [0]{URL }%
\providecommand \Eprint [0]{\href }%
\providecommand \doibase [0]{https://doi.org/}%
\providecommand \selectlanguage [0]{\@gobble}%
\providecommand \bibinfo  [0]{\@secondoftwo}%
\providecommand \bibfield  [0]{\@secondoftwo}%
\providecommand \translation [1]{[#1]}%
\providecommand \BibitemOpen [0]{}%
\providecommand \bibitemStop [0]{}%
\providecommand \bibitemNoStop [0]{.\EOS\space}%
\providecommand \EOS [0]{\spacefactor3000\relax}%
\providecommand \BibitemShut  [1]{\csname bibitem#1\endcsname}%
\let\auto@bib@innerbib\@empty
\bibitem [{\citenamefont {Roig}\ \emph {et~al.}(2024)\citenamefont {Roig},
  \citenamefont {Kreisel}, \citenamefont {Yu}, \citenamefont {Andersen},\ and\
  \citenamefont {Agterberg}}]{Minimal_models}%
  \BibitemOpen
  \bibfield  {author} {\bibinfo {author} {\bibfnamefont {M.}~\bibnamefont
  {Roig}}, \bibinfo {author} {\bibfnamefont {A.}~\bibnamefont {Kreisel}},
  \bibinfo {author} {\bibfnamefont {Y.}~\bibnamefont {Yu}}, \bibinfo {author}
  {\bibfnamefont {B.~M.}\ \bibnamefont {Andersen}},\ and\ \bibinfo {author}
  {\bibfnamefont {D.~F.}\ \bibnamefont {Agterberg}},\ }\bibfield  {title}
  {\bibinfo {title} {Minimal models for altermagnetism},\ }\href
  {https://doi.org/10.1103/PhysRevB.110.144412} {\bibfield  {journal} {\bibinfo
   {journal} {Phys. Rev. B}\ }\textbf {\bibinfo {volume} {110}},\ \bibinfo
  {pages} {144412} (\bibinfo {year} {2024})}\BibitemShut {NoStop}%
\bibitem [{\citenamefont {Belashchenko}()}]{Deterministic_AM}%
  \BibitemOpen
  \bibfield  {author} {\bibinfo {author} {\bibfnamefont {K.~D.}\ \bibnamefont
  {Belashchenko}},\ }\bibfield  {title} {\bibinfo {title} {Deterministic
  electrical switching in altermagnets via surface antisymmetry groups},\
  }\href {https://arxiv.org/abs/2603.06537} {\bibinfo  {journal}
  {arXiv:2603.06537}\ }\BibitemShut {NoStop}%
\bibitem [{\citenamefont {Belashchenko}(2025)}]{MnTe_strain}%
  \BibitemOpen
\bibfield  {journal} {  }\bibfield  {author} {\bibinfo {author} {\bibfnamefont
  {K.~D.}\ \bibnamefont {Belashchenko}},\ }\bibfield  {title} {\bibinfo {title}
  {Giant strain-induced spin splitting effect in mnte, a $g$-wave altermagnetic
  semiconductor},\ }\href {https://doi.org/10.1103/PhysRevLett.134.086701}
  {\bibfield  {journal} {\bibinfo  {journal} {Phys. Rev. Lett.}\ }\textbf
  {\bibinfo {volume} {134}},\ \bibinfo {pages} {086701} (\bibinfo {year}
  {2025})}\BibitemShut {NoStop}%
\bibitem [{\citenamefont {Heinsdorf}\ and\ \citenamefont
  {Franz}(2026)}]{HeinsdorfFranz2026}%
  \BibitemOpen
  \bibfield  {author} {\bibinfo {author} {\bibfnamefont {N.}~\bibnamefont
  {Heinsdorf}}\ and\ \bibinfo {author} {\bibfnamefont {M.}~\bibnamefont
  {Franz}},\ }\bibfield  {title} {\bibinfo {title} {Proximitizing altermagnets
  with conventional superconductors},\ }\href
  {https://doi.org/10.1103/rhmg-j1fv} {\bibfield  {journal} {\bibinfo
  {journal} {Physical Review B}\ }\textbf {\bibinfo {volume} {113}},\ \bibinfo
  {pages} {L020501} (\bibinfo {year} {2026})}\BibitemShut {NoStop}%
\bibitem [{\citenamefont {Bergeret}\ \emph {et~al.}(2005)\citenamefont
  {Bergeret}, \citenamefont {Volkov},\ and\ \citenamefont
  {Efetov}}]{Bergeret2005}%
  \BibitemOpen
  \bibfield  {author} {\bibinfo {author} {\bibfnamefont {F.~S.}\ \bibnamefont
  {Bergeret}}, \bibinfo {author} {\bibfnamefont {A.~F.}\ \bibnamefont
  {Volkov}},\ and\ \bibinfo {author} {\bibfnamefont {K.~B.}\ \bibnamefont
  {Efetov}},\ }\bibfield  {title} {\bibinfo {title} {Odd triplet
  superconductivity and related phenomena in superconductor-ferromagnet
  structures},\ }\href {https://doi.org/10.1103/RevModPhys.77.1321} {\bibfield
  {journal} {\bibinfo  {journal} {Rev. Mod. Phys.}\ }\textbf {\bibinfo {volume}
  {77}},\ \bibinfo {pages} {1321} (\bibinfo {year} {2005})}\BibitemShut
  {NoStop}%
\end{thebibliography}%


\begin{thebibliography}{115}%
\makeatletter
\providecommand \@ifxundefined [1]{%
 \@ifx{#1\undefined}
}%
\providecommand \@ifnum [1]{%
 \ifnum #1\expandafter \@firstoftwo
 \else \expandafter \@secondoftwo
 \fi
}%
\providecommand \@ifx [1]{%
 \ifx #1\expandafter \@firstoftwo
 \else \expandafter \@secondoftwo
 \fi
}%
\providecommand \natexlab [1]{#1}%
\providecommand \enquote  [1]{``#1''}%
\providecommand \bibnamefont  [1]{#1}%
\providecommand \bibfnamefont [1]{#1}%
\providecommand \citenamefont [1]{#1}%
\providecommand \href@noop [0]{\@secondoftwo}%
\providecommand \href [0]{\begingroup \@sanitize@url \@href}%
\providecommand \@href[1]{\@@startlink{#1}\@@href}%
\providecommand \@@href[1]{\endgroup#1\@@endlink}%
\providecommand \@sanitize@url [0]{\catcode `\\12\catcode `\$12\catcode
  `\&12\catcode `\#12\catcode `\^12\catcode `\_12\catcode `\%12\relax}%
\providecommand \@@startlink[1]{}%
\providecommand \@@endlink[0]{}%
\providecommand \url  [0]{\begingroup\@sanitize@url \@url }%
\providecommand \@url [1]{\endgroup\@href {#1}{\urlprefix }}%
\providecommand \urlprefix  [0]{URL }%
\providecommand \Eprint [0]{\href }%
\providecommand \doibase [0]{https://doi.org/}%
\providecommand \selectlanguage [0]{\@gobble}%
\providecommand \bibinfo  [0]{\@secondoftwo}%
\providecommand \bibfield  [0]{\@secondoftwo}%
\providecommand \translation [1]{[#1]}%
\providecommand \BibitemOpen [0]{}%
\providecommand \bibitemStop [0]{}%
\providecommand \bibitemNoStop [0]{.\EOS\space}%
\providecommand \EOS [0]{\spacefactor3000\relax}%
\providecommand \BibitemShut  [1]{\csname bibitem#1\endcsname}%
\let\auto@bib@innerbib\@empty
\bibitem [{\citenamefont {Krempask\'{y}}\ \emph {et~al.}(2024)\citenamefont
  {Krempask\'{y}}, \citenamefont {\v{S}mejkal}, \citenamefont {D'Souza},
  \citenamefont {Hajlaoui}, \citenamefont {Springholz}, \citenamefont
  {Uhli\v{r}ov\'{a}}, \citenamefont {Alarab}, \citenamefont {Constantinou},
  \citenamefont {Strocov}, \citenamefont {Usanov}, \citenamefont {Pudelko},
  \citenamefont {Gonz\'{a}lez-Hern\'{a}ndez}, \citenamefont {Birk~Hellenes},
  \citenamefont {Jansa}, \citenamefont {Reichlov\'{a}}, \citenamefont
  {\v{S}ob\'{a}\v{n}}, \citenamefont {Gonzalez~Betancourt}, \citenamefont
  {Wadley}, \citenamefont {Sinova}, \citenamefont {Kriegner}, \citenamefont
  {Min\'{a}r}, \citenamefont {Dil},\ and\ \citenamefont
  {Jungwirth}}]{Krempasky2024}%
  \BibitemOpen
  \bibfield  {author} {\bibinfo {author} {\bibfnamefont {J.}~\bibnamefont
  {Krempask\'{y}}}, \bibinfo {author} {\bibfnamefont {L.}~\bibnamefont
  {\v{S}mejkal}}, \bibinfo {author} {\bibfnamefont {S.~W.}\ \bibnamefont
  {D'Souza}}, \bibinfo {author} {\bibfnamefont {M.}~\bibnamefont {Hajlaoui}},
  \bibinfo {author} {\bibfnamefont {G.}~\bibnamefont {Springholz}}, \bibinfo
  {author} {\bibfnamefont {K.}~\bibnamefont {Uhli\v{r}ov\'{a}}}, \bibinfo
  {author} {\bibfnamefont {F.}~\bibnamefont {Alarab}}, \bibinfo {author}
  {\bibfnamefont {P.~C.}\ \bibnamefont {Constantinou}}, \bibinfo {author}
  {\bibfnamefont {V.}~\bibnamefont {Strocov}}, \bibinfo {author} {\bibfnamefont
  {D.}~\bibnamefont {Usanov}}, \bibinfo {author} {\bibfnamefont {W.~R.}\
  \bibnamefont {Pudelko}}, \bibinfo {author} {\bibfnamefont {R.}~\bibnamefont
  {Gonz\'{a}lez-Hern\'{a}ndez}}, \bibinfo {author} {\bibfnamefont
  {A.}~\bibnamefont {Birk~Hellenes}}, \bibinfo {author} {\bibfnamefont
  {Z.}~\bibnamefont {Jansa}}, \bibinfo {author} {\bibfnamefont
  {H.}~\bibnamefont {Reichlov\'{a}}}, \bibinfo {author} {\bibfnamefont
  {Z.}~\bibnamefont {\v{S}ob\'{a}\v{n}}}, \bibinfo {author} {\bibfnamefont
  {R.~D.}\ \bibnamefont {Gonzalez~Betancourt}}, \bibinfo {author}
  {\bibfnamefont {P.}~\bibnamefont {Wadley}}, \bibinfo {author} {\bibfnamefont
  {J.}~\bibnamefont {Sinova}}, \bibinfo {author} {\bibfnamefont
  {D.}~\bibnamefont {Kriegner}}, \bibinfo {author} {\bibfnamefont
  {J.}~\bibnamefont {Min\'{a}r}}, \bibinfo {author} {\bibfnamefont {J.~H.}\
  \bibnamefont {Dil}},\ and\ \bibinfo {author} {\bibfnamefont {T.}~\bibnamefont
  {Jungwirth}},\ }\bibfield  {title} {\bibinfo {title} {Altermagnetic lifting
  of {Kramers} spin degeneracy},\ }\href
  {https://doi.org/10.1038/s41586-023-06907-7} {\bibfield  {journal} {\bibinfo
  {journal} {Nature}\ }\textbf {\bibinfo {volume} {626}},\ \bibinfo {pages}
  {517} (\bibinfo {year} {2024})}\BibitemShut {NoStop}%
\bibitem [{\citenamefont {Zhu}\ \emph {et~al.}(2024)\citenamefont {Zhu},
  \citenamefont {Chen}, \citenamefont {Liu}, \citenamefont {Liu}, \citenamefont
  {Liu}, \citenamefont {Zha}, \citenamefont {Qu}, \citenamefont {Hong},
  \citenamefont {Li}, \citenamefont {Jiang} \emph {et~al.}}]{Zhu2024:N}%
  \BibitemOpen
  \bibfield  {author} {\bibinfo {author} {\bibfnamefont {Y.-P.}\ \bibnamefont
  {Zhu}}, \bibinfo {author} {\bibfnamefont {X.}~\bibnamefont {Chen}}, \bibinfo
  {author} {\bibfnamefont {X.-R.}\ \bibnamefont {Liu}}, \bibinfo {author}
  {\bibfnamefont {Y.}~\bibnamefont {Liu}}, \bibinfo {author} {\bibfnamefont
  {P.}~\bibnamefont {Liu}}, \bibinfo {author} {\bibfnamefont {H.}~\bibnamefont
  {Zha}}, \bibinfo {author} {\bibfnamefont {G.}~\bibnamefont {Qu}}, \bibinfo
  {author} {\bibfnamefont {C.}~\bibnamefont {Hong}}, \bibinfo {author}
  {\bibfnamefont {J.}~\bibnamefont {Li}}, \bibinfo {author} {\bibfnamefont
  {Z.}~\bibnamefont {Jiang}}, \emph {et~al.},\ }\bibfield  {title} {\bibinfo
  {title} {Observation of plaid-like spin splitting in a noncoplanar
  antiferromagnet},\ }\href {https://doi.org/10.1038/s41586-024-07023-w}
  {\bibfield  {journal} {\bibinfo  {journal} {Nature}\ }\textbf {\bibinfo
  {volume} {626}},\ \bibinfo {pages} {523} (\bibinfo {year}
  {2024})}\BibitemShut {NoStop}%
\bibitem [{\citenamefont {Yamada}\ \emph {et~al.}(2025)\citenamefont {Yamada},
  \citenamefont {Birch}, \citenamefont {Baral}, \citenamefont {Okumura},
  \citenamefont {Nakano}, \citenamefont {Gao}, \citenamefont {Ezawa},
  \citenamefont {Nomoto}, \citenamefont {Masell}, \citenamefont {Ishihara},
  \citenamefont {Kolincio}, \citenamefont {Belopolski}, \citenamefont
  {Sagayama}, \citenamefont {Nakao}, \citenamefont {Ohishi}, \citenamefont
  {Ohhara}, \citenamefont {Kiyanagi}, \citenamefont {Nakajima}, \citenamefont
  {Tokura}, \citenamefont {Arima}, \citenamefont {Motome}, \citenamefont
  {Hirschmann},\ and\ \citenamefont {Hirschberger}}]{Yamada2025:N}%
  \BibitemOpen
  \bibfield  {author} {\bibinfo {author} {\bibfnamefont {R.}~\bibnamefont
  {Yamada}}, \bibinfo {author} {\bibfnamefont {M.~T.}\ \bibnamefont {Birch}},
  \bibinfo {author} {\bibfnamefont {P.~R.}\ \bibnamefont {Baral}}, \bibinfo
  {author} {\bibfnamefont {S.}~\bibnamefont {Okumura}}, \bibinfo {author}
  {\bibfnamefont {R.}~\bibnamefont {Nakano}}, \bibinfo {author} {\bibfnamefont
  {S.}~\bibnamefont {Gao}}, \bibinfo {author} {\bibfnamefont {M.}~\bibnamefont
  {Ezawa}}, \bibinfo {author} {\bibfnamefont {T.}~\bibnamefont {Nomoto}},
  \bibinfo {author} {\bibfnamefont {J.}~\bibnamefont {Masell}}, \bibinfo
  {author} {\bibfnamefont {Y.}~\bibnamefont {Ishihara}}, \bibinfo {author}
  {\bibfnamefont {K.~K.}\ \bibnamefont {Kolincio}}, \bibinfo {author}
  {\bibfnamefont {I.}~\bibnamefont {Belopolski}}, \bibinfo {author}
  {\bibfnamefont {H.}~\bibnamefont {Sagayama}}, \bibinfo {author}
  {\bibfnamefont {H.}~\bibnamefont {Nakao}}, \bibinfo {author} {\bibfnamefont
  {K.}~\bibnamefont {Ohishi}}, \bibinfo {author} {\bibfnamefont
  {T.}~\bibnamefont {Ohhara}}, \bibinfo {author} {\bibfnamefont
  {R.}~\bibnamefont {Kiyanagi}}, \bibinfo {author} {\bibfnamefont
  {T.}~\bibnamefont {Nakajima}}, \bibinfo {author} {\bibfnamefont
  {Y.}~\bibnamefont {Tokura}}, \bibinfo {author} {\bibfnamefont {T.-h.}\
  \bibnamefont {Arima}}, \bibinfo {author} {\bibfnamefont {Y.}~\bibnamefont
  {Motome}}, \bibinfo {author} {\bibfnamefont {M.~M.}\ \bibnamefont
  {Hirschmann}},\ and\ \bibinfo {author} {\bibfnamefont {M.}~\bibnamefont
  {Hirschberger}},\ }\bibfield  {title} {\bibinfo {title} {A metallic $p$-wave
  magnet with commensurate spin helix},\ }\href
  {https://doi.org/10.1038/s41586-025-09633-4} {\bibfield  {journal} {\bibinfo
  {journal} {Nature}\ }\textbf {\bibinfo {volume} {646}},\ \bibinfo {pages}
  {837} (\bibinfo {year} {2025})}\BibitemShut {NoStop}%
\bibitem [{\citenamefont {Song}\ \emph {et~al.}(2025)\citenamefont {Song},
  \citenamefont {Stavri{\'c}}, \citenamefont {Barone}, \citenamefont
  {Droghetti}, \citenamefont {Antonenko}, \citenamefont {Venderbos},
  \citenamefont {Occhialini}, \citenamefont {Ilyas}, \citenamefont
  {Erge{\c{c}}en}, \citenamefont {Gedik}, \citenamefont {Cheong}, \citenamefont
  {Fernandes}, \citenamefont {Picozzi},\ and\ \citenamefont
  {Comin}}]{Song2025:N}%
  \BibitemOpen
  \bibfield  {author} {\bibinfo {author} {\bibfnamefont {Q.}~\bibnamefont
  {Song}}, \bibinfo {author} {\bibfnamefont {S.}~\bibnamefont {Stavri{\'c}}},
  \bibinfo {author} {\bibfnamefont {P.}~\bibnamefont {Barone}}, \bibinfo
  {author} {\bibfnamefont {A.}~\bibnamefont {Droghetti}}, \bibinfo {author}
  {\bibfnamefont {D.~S.}\ \bibnamefont {Antonenko}}, \bibinfo {author}
  {\bibfnamefont {J.~W.~F.}\ \bibnamefont {Venderbos}}, \bibinfo {author}
  {\bibfnamefont {C.~A.}\ \bibnamefont {Occhialini}}, \bibinfo {author}
  {\bibfnamefont {B.}~\bibnamefont {Ilyas}}, \bibinfo {author} {\bibfnamefont
  {E.}~\bibnamefont {Erge{\c{c}}en}}, \bibinfo {author} {\bibfnamefont
  {N.}~\bibnamefont {Gedik}}, \bibinfo {author} {\bibfnamefont {S.-W.}\
  \bibnamefont {Cheong}}, \bibinfo {author} {\bibfnamefont {R.~M.}\
  \bibnamefont {Fernandes}}, \bibinfo {author} {\bibfnamefont {S.}~\bibnamefont
  {Picozzi}},\ and\ \bibinfo {author} {\bibfnamefont {R.}~\bibnamefont
  {Comin}},\ }\bibfield  {title} {\bibinfo {title} {{Electrical switching of a
  $p$-wave magnet}},\ }\href {https://doi.org/10.1038/s41586-025-08838-x}
  {\bibfield  {journal} {\bibinfo  {journal} {Nature}\ }\textbf {\bibinfo
  {volume} {642}},\ \bibinfo {pages} {64} (\bibinfo {year} {2025})}\BibitemShut
  {NoStop}%
\bibitem [{\citenamefont {Hayami}\ \emph {et~al.}(2019)\citenamefont {Hayami},
  \citenamefont {Yanagi},\ and\ \citenamefont {Kusunose}}]{Hayami2019:JPSJ}%
  \BibitemOpen
  \bibfield  {author} {\bibinfo {author} {\bibfnamefont {S.}~\bibnamefont
  {Hayami}}, \bibinfo {author} {\bibfnamefont {Y.}~\bibnamefont {Yanagi}},\
  and\ \bibinfo {author} {\bibfnamefont {H.}~\bibnamefont {Kusunose}},\
  }\bibfield  {title} {\bibinfo {title} {Momentum-dependent spin splitting by
  collinear antiferromagnetic ordering},\ }\href
  {https://doi.org/10.7566/JPSJ.88.123702} {\bibfield  {journal} {\bibinfo
  {journal} {J. Phys. Soc. Jpn.}\ }\textbf {\bibinfo {volume} {88}},\ \bibinfo
  {pages} {123702} (\bibinfo {year} {2019})}\BibitemShut {NoStop}%
\bibitem [{\citenamefont {{\v{S}}mejkal}\ \emph {et~al.}(2020)\citenamefont
  {{\v{S}}mejkal}, \citenamefont {Gonzalez-Hernandez}, \citenamefont
  {Jungwirth},\ and\ \citenamefont {Sinova}}]{Smejkal2020SciAdv}%
  \BibitemOpen
  \bibfield  {author} {\bibinfo {author} {\bibfnamefont {L.}~\bibnamefont
  {{\v{S}}mejkal}}, \bibinfo {author} {\bibfnamefont {R.}~\bibnamefont
  {Gonzalez-Hernandez}}, \bibinfo {author} {\bibfnamefont {T.}~\bibnamefont
  {Jungwirth}},\ and\ \bibinfo {author} {\bibfnamefont {J.}~\bibnamefont
  {Sinova}},\ }\bibfield  {title} {\bibinfo {title} {Crystal time-reversal
  symmetry breaking and spontaneous {Hall} effect in collinear
  antiferromagnets},\ }\href {https://doi.org/10.1126/sciadv.aaz8809}
  {\bibfield  {journal} {\bibinfo  {journal} {Sci. Adv.}\ }\textbf {\bibinfo
  {volume} {6}},\ \bibinfo {pages} {eaaz8809} (\bibinfo {year}
  {2020})}\BibitemShut {NoStop}%
\bibitem [{\citenamefont {Naka}\ \emph {et~al.}(2019)\citenamefont {Naka},
  \citenamefont {Hayami}, \citenamefont {Kusunose}, \citenamefont {Yanagi},
  \citenamefont {Motome},\ and\ \citenamefont {Seo}}]{Naka2019}%
  \BibitemOpen
  \bibfield  {author} {\bibinfo {author} {\bibfnamefont {M.}~\bibnamefont
  {Naka}}, \bibinfo {author} {\bibfnamefont {S.}~\bibnamefont {Hayami}},
  \bibinfo {author} {\bibfnamefont {H.}~\bibnamefont {Kusunose}}, \bibinfo
  {author} {\bibfnamefont {Y.}~\bibnamefont {Yanagi}}, \bibinfo {author}
  {\bibfnamefont {Y.}~\bibnamefont {Motome}},\ and\ \bibinfo {author}
  {\bibfnamefont {H.}~\bibnamefont {Seo}},\ }\bibfield  {title} {\bibinfo
  {title} {Spin current generation in organic antiferromagnets},\ }\href
  {https://doi.org/10.1038/s41467-019-12229-y} {\bibfield  {journal} {\bibinfo
  {journal} {Nat. Commun.}\ }\textbf {\bibinfo {volume} {10}},\ \bibinfo
  {pages} {4305} (\bibinfo {year} {2019})}\BibitemShut {NoStop}%
\bibitem [{\citenamefont {Sandratski}\ \emph {et~al.}(1981)\citenamefont
  {Sandratski}, \citenamefont {Egorov},\ and\ \citenamefont
  {Berdyshev}}]{Sandratski1981:PSSB}%
  \BibitemOpen
  \bibfield  {author} {\bibinfo {author} {\bibfnamefont {L.~M.}\ \bibnamefont
  {Sandratski}}, \bibinfo {author} {\bibfnamefont {R.~F.}\ \bibnamefont
  {Egorov}},\ and\ \bibinfo {author} {\bibfnamefont {A.~A.}\ \bibnamefont
  {Berdyshev}},\ }\bibfield  {title} {\bibinfo {title} {{Energy Band Structure
  and Electronic Properties of {NiAs} Type Compounds {II}. Antiferromagnetic
  Manganese Telluride}},\ }\href {https://doi.org/10.1002/pssb.2221030220}
  {\bibfield  {journal} {\bibinfo  {journal} {phys. stat. sol. (b)}\ }\textbf
  {\bibinfo {volume} {103}},\ \bibinfo {pages} {511} (\bibinfo {year}
  {1981})}\BibitemShut {NoStop}%
\bibitem [{\citenamefont {Mazin}\ \emph {et~al.}(2021)\citenamefont {Mazin},
  \citenamefont {Koepernik}, \citenamefont {Johannes}, \citenamefont
  {{Gonz{\'a}lez-Hern{\'a}ndez}},\ and\ \citenamefont
  {{\v{S}}mejkal}}]{Mazin2021:PNAS}%
  \BibitemOpen
  \bibfield  {author} {\bibinfo {author} {\bibfnamefont {I.~I.}\ \bibnamefont
  {Mazin}}, \bibinfo {author} {\bibfnamefont {K.}~\bibnamefont {Koepernik}},
  \bibinfo {author} {\bibfnamefont {M.~D.}\ \bibnamefont {Johannes}}, \bibinfo
  {author} {\bibfnamefont {R.}~\bibnamefont {{Gonz{\'a}lez-Hern{\'a}ndez}}},\
  and\ \bibinfo {author} {\bibfnamefont {L.}~\bibnamefont {{\v{S}}mejkal}},\
  }\bibfield  {title} {\bibinfo {title} {Prediction of unconventional magnetism
  in doped {FeSb$_2$}},\ }\href {https://doi.org/10.1073/pnas.2108924118}
  {\bibfield  {journal} {\bibinfo  {journal} {Proc. Natl. Acad. Sci. U.S.A.}\
  }\textbf {\bibinfo {volume} {118}},\ \bibinfo {pages} {e2108924118} (\bibinfo
  {year} {2021})}\BibitemShut {NoStop}%
\bibitem [{\citenamefont {Yuan}\ \emph {et~al.}(2021)\citenamefont {Yuan},
  \citenamefont {Wang}, \citenamefont {Luo},\ and\ \citenamefont
  {Zunger}}]{Yuan2021:PRM}%
  \BibitemOpen
  \bibfield  {author} {\bibinfo {author} {\bibfnamefont {L.-D.}\ \bibnamefont
  {Yuan}}, \bibinfo {author} {\bibfnamefont {Z.}~\bibnamefont {Wang}}, \bibinfo
  {author} {\bibfnamefont {J.-W.}\ \bibnamefont {Luo}},\ and\ \bibinfo {author}
  {\bibfnamefont {A.}~\bibnamefont {Zunger}},\ }\bibfield  {title} {\bibinfo
  {title} {{Prediction of {low-Z} collinear and noncollinear antiferromagnetic
  compounds having momentum-dependent spin splitting even without spin-orbit
  coupling}},\ }\href {https://doi.org/10.1103/PhysRevMaterials.5.014409}
  {\bibfield  {journal} {\bibinfo  {journal} {Phys. Rev. Mater.}\ }\textbf
  {\bibinfo {volume} {5}},\ \bibinfo {pages} {014409} (\bibinfo {year}
  {2021})}\BibitemShut {NoStop}%
\bibitem [{\citenamefont {Yuan}\ \emph {et~al.}(2020)\citenamefont {Yuan},
  \citenamefont {Wang}, \citenamefont {Luo}, \citenamefont {Rashba},\ and\
  \citenamefont {Zunger}}]{PhysRevB.102.014422}%
  \BibitemOpen
  \bibfield  {author} {\bibinfo {author} {\bibfnamefont {L.-D.}\ \bibnamefont
  {Yuan}}, \bibinfo {author} {\bibfnamefont {Z.}~\bibnamefont {Wang}}, \bibinfo
  {author} {\bibfnamefont {J.-W.}\ \bibnamefont {Luo}}, \bibinfo {author}
  {\bibfnamefont {E.~I.}\ \bibnamefont {Rashba}},\ and\ \bibinfo {author}
  {\bibfnamefont {A.}~\bibnamefont {Zunger}},\ }\bibfield  {title} {\bibinfo
  {title} {Giant momentum-dependent spin splitting in centrosymmetric low-{$Z$}
  antiferromagnets},\ }\href {https://doi.org/10.1103/PhysRevB.102.014422}
  {\bibfield  {journal} {\bibinfo  {journal} {Phys. Rev. B}\ }\textbf {\bibinfo
  {volume} {102}},\ \bibinfo {pages} {014422} (\bibinfo {year}
  {2020})}\BibitemShut {NoStop}%
\bibitem [{\citenamefont {Ma}\ \emph {et~al.}(2021)\citenamefont {Ma},
  \citenamefont {Hu}, \citenamefont {Li}, \citenamefont {Liu}, \citenamefont
  {Yao}, \citenamefont {Jia},\ and\ \citenamefont {Liu}}]{Ma2021:NC}%
  \BibitemOpen
  \bibfield  {author} {\bibinfo {author} {\bibfnamefont {H.-Y.}\ \bibnamefont
  {Ma}}, \bibinfo {author} {\bibfnamefont {M.}~\bibnamefont {Hu}}, \bibinfo
  {author} {\bibfnamefont {N.}~\bibnamefont {Li}}, \bibinfo {author}
  {\bibfnamefont {J.}~\bibnamefont {Liu}}, \bibinfo {author} {\bibfnamefont
  {W.}~\bibnamefont {Yao}}, \bibinfo {author} {\bibfnamefont {J.-F.}\
  \bibnamefont {Jia}},\ and\ \bibinfo {author} {\bibfnamefont {J.}~\bibnamefont
  {Liu}},\ }\bibfield  {title} {\bibinfo {title} {{Multifunctional
  antiferromagnetic materials with giant piezomagnetism and noncollinear spin
  current}},\ }\href {https://doi.org/10.1038/s41467-021-23127-7} {\bibfield
  {journal} {\bibinfo  {journal} {Nat. Commun.}\ }\textbf {\bibinfo {volume}
  {12}},\ \bibinfo {pages} {2846} (\bibinfo {year} {2021})}\BibitemShut
  {NoStop}%
\bibitem [{\citenamefont {{\v{S}}mejkal}\ \emph
  {et~al.}(2022{\natexlab{a}})\citenamefont {{\v{S}}mejkal}, \citenamefont
  {Sinova},\ and\ \citenamefont {Jungwirth}}]{Smejkal2022AM}%
  \BibitemOpen
  \bibfield  {author} {\bibinfo {author} {\bibfnamefont {L.}~\bibnamefont
  {{\v{S}}mejkal}}, \bibinfo {author} {\bibfnamefont {J.}~\bibnamefont
  {Sinova}},\ and\ \bibinfo {author} {\bibfnamefont {T.}~\bibnamefont
  {Jungwirth}},\ }\bibfield  {title} {\bibinfo {title} {Emerging research
  landscape of altermagnetism},\ }\href
  {https://doi.org/10.1103/PhysRevX.12.040501} {\bibfield  {journal} {\bibinfo
  {journal} {Phys. Rev. X}\ }\textbf {\bibinfo {volume} {12}},\ \bibinfo
  {pages} {040501} (\bibinfo {year} {2022}{\natexlab{a}})}\BibitemShut
  {NoStop}%
\bibitem [{\citenamefont {{\v{S}}mejkal}\ \emph
  {et~al.}(2022{\natexlab{b}})\citenamefont {{\v{S}}mejkal}, \citenamefont
  {Sinova},\ and\ \citenamefont {Jungwirth}}]{Smejkal2022PRX}%
  \BibitemOpen
  \bibfield  {author} {\bibinfo {author} {\bibfnamefont {L.}~\bibnamefont
  {{\v{S}}mejkal}}, \bibinfo {author} {\bibfnamefont {J.}~\bibnamefont
  {Sinova}},\ and\ \bibinfo {author} {\bibfnamefont {T.}~\bibnamefont
  {Jungwirth}},\ }\bibfield  {title} {\bibinfo {title} {Beyond conventional
  ferromagnetism and antiferromagnetism: {A} phase with nonrelativistic spin
  and crystal rotation symmetry},\ }\href
  {https://doi.org/10.1103/PhysRevX.12.031042} {\bibfield  {journal} {\bibinfo
  {journal} {Phys. Rev. X}\ }\textbf {\bibinfo {volume} {12}},\ \bibinfo
  {pages} {031042} (\bibinfo {year} {2022}{\natexlab{b}})}\BibitemShut
  {NoStop}%
\bibitem [{\citenamefont {Mazin}(2022)}]{Mazin2022}%
  \BibitemOpen
  \bibfield  {author} {\bibinfo {author} {\bibfnamefont {I.}~\bibnamefont
  {Mazin}} (\bibinfo {collaboration} {The PRX Editors}),\ }\bibfield  {title}
  {\bibinfo {title} {Editorial: Altermagnetism---{A} new punch line of
  fundamental magnetism},\ }\href {https://doi.org/10.1103/PhysRevX.12.040002}
  {\bibfield  {journal} {\bibinfo  {journal} {Phys. Rev. X}\ }\textbf {\bibinfo
  {volume} {12}},\ \bibinfo {pages} {040002} (\bibinfo {year}
  {2022})}\BibitemShut {NoStop}%
\bibitem [{\citenamefont {Khodas}\ \emph {et~al.}()\citenamefont {Khodas},
  \citenamefont {\v{S}mejkal},\ and\ \citenamefont {Mazin}}]{Khodas2026:X}%
  \BibitemOpen
  \bibfield  {author} {\bibinfo {author} {\bibfnamefont {M.}~\bibnamefont
  {Khodas}}, \bibinfo {author} {\bibfnamefont {L.}~\bibnamefont
  {\v{S}mejkal}},\ and\ \bibinfo {author} {\bibfnamefont {I.~I.}\ \bibnamefont
  {Mazin}},\ }\bibfield  {title} {\bibinfo {title} {{Nonrelativistic-Ising
  superconductivity in $p$-wave magnets}},\ }\href
  {https://arxiv.org/abs/2601.19829} {\bibinfo  {journal} {arXiv:2601.19829}\
  }\BibitemShut {NoStop}%
\bibitem [{\citenamefont {Liu}\ \emph {et~al.}(2025)\citenamefont {Liu},
  \citenamefont {Dai},\ and\ \citenamefont {Bl{\"u}gel}}]{Liu2025:NP}%
  \BibitemOpen
\bibfield  {journal} {  }\bibfield  {author} {\bibinfo {author} {\bibfnamefont
  {Q.}~\bibnamefont {Liu}}, \bibinfo {author} {\bibfnamefont {X.}~\bibnamefont
  {Dai}},\ and\ \bibinfo {author} {\bibfnamefont {S.}~\bibnamefont
  {Bl{\"u}gel}},\ }\bibfield  {title} {\bibinfo {title} {Different facets of
  unconventional magnetism},\ }\href
  {https://doi.org/10.1038/s41567-024-02750-3} {\bibfield  {journal} {\bibinfo
  {journal} {Nat. Phys.}\ }\textbf {\bibinfo {volume} {21}},\ \bibinfo {pages}
  {329} (\bibinfo {year} {2025})}\BibitemShut {NoStop}%
\bibitem [{\citenamefont {Guo}\ \emph {et~al.}(2023)\citenamefont {Guo},
  \citenamefont {Liu}, \citenamefont {Janson}, \citenamefont {Fulga},
  \citenamefont {van~den Brink},\ and\ \citenamefont
  {Facio}}]{Guo2023MaterTodayPhys}%
  \BibitemOpen
  \bibfield  {author} {\bibinfo {author} {\bibfnamefont {Y.}~\bibnamefont
  {Guo}}, \bibinfo {author} {\bibfnamefont {H.}~\bibnamefont {Liu}}, \bibinfo
  {author} {\bibfnamefont {O.}~\bibnamefont {Janson}}, \bibinfo {author}
  {\bibfnamefont {I.~C.}\ \bibnamefont {Fulga}}, \bibinfo {author}
  {\bibfnamefont {J.}~\bibnamefont {van~den Brink}},\ and\ \bibinfo {author}
  {\bibfnamefont {J.~I.}\ \bibnamefont {Facio}},\ }\bibfield  {title} {\bibinfo
  {title} {Spin-split collinear antiferromagnets: {A} large-scale ab-initio
  study},\ }\href {https://doi.org/10.1016/j.mtphys.2023.100991} {\bibfield
  {journal} {\bibinfo  {journal} {Mater. Today Phys.}\ }\textbf {\bibinfo
  {volume} {32}},\ \bibinfo {pages} {100991} (\bibinfo {year}
  {2023})}\BibitemShut {NoStop}%
\bibitem [{\citenamefont {Brinkman}\ and\ \citenamefont
  {Elliott}(1966)}]{BrinkmanElliott1966SpinSpaceGroups}%
  \BibitemOpen
  \bibfield  {author} {\bibinfo {author} {\bibfnamefont {W.~F.}\ \bibnamefont
  {Brinkman}}\ and\ \bibinfo {author} {\bibfnamefont {R.~J.}\ \bibnamefont
  {Elliott}},\ }\bibfield  {title} {\bibinfo {title} {Theory of spin-space
  groups},\ }\href {https://doi.org/10.1098/rspa.1966.0211} {\bibfield
  {journal} {\bibinfo  {journal} {Proc. R. Soc. Lond. A}\ }\textbf {\bibinfo
  {volume} {294}},\ \bibinfo {pages} {343} (\bibinfo {year}
  {1966})}\BibitemShut {NoStop}%
\bibitem [{\citenamefont {Litvin}\ and\ \citenamefont
  {Opechowski}(1974)}]{LitvinOpechowski1974SpinGroups}%
  \BibitemOpen
  \bibfield  {author} {\bibinfo {author} {\bibfnamefont {D.~B.}\ \bibnamefont
  {Litvin}}\ and\ \bibinfo {author} {\bibfnamefont {W.}~\bibnamefont
  {Opechowski}},\ }\bibfield  {title} {\bibinfo {title} {Spin groups},\ }\href
  {https://doi.org/10.1016/0031-8914(74)90157-8} {\bibfield  {journal}
  {\bibinfo  {journal} {Physica}\ }\textbf {\bibinfo {volume} {76}},\ \bibinfo
  {pages} {538} (\bibinfo {year} {1974})}\BibitemShut {NoStop}%
\bibitem [{\citenamefont {Sandratskii}(1979)}]{Sandratskii1979SpinSpaceGroups}%
  \BibitemOpen
  \bibfield  {author} {\bibinfo {author} {\bibfnamefont {L.~M.}\ \bibnamefont
  {Sandratskii}},\ }\bibfield  {title} {\bibinfo {title} {Classification of
  single-electron states in a crystal on the basis of spin space groups},\
  }\href {https://doi.org/10.1007/BF00891386} {\bibfield  {journal} {\bibinfo
  {journal} {Sov. Phys. J.}\ }\textbf {\bibinfo {volume} {22}},\ \bibinfo
  {pages} {941} (\bibinfo {year} {1979})}\BibitemShut {NoStop}%
\bibitem [{\citenamefont {Liu}\ \emph {et~al.}(2022)\citenamefont {Liu},
  \citenamefont {Li}, \citenamefont {Han}, \citenamefont {Wan},\ and\
  \citenamefont {Liu}}]{Liu2022:PRX}%
  \BibitemOpen
  \bibfield  {author} {\bibinfo {author} {\bibfnamefont {P.}~\bibnamefont
  {Liu}}, \bibinfo {author} {\bibfnamefont {J.}~\bibnamefont {Li}}, \bibinfo
  {author} {\bibfnamefont {J.}~\bibnamefont {Han}}, \bibinfo {author}
  {\bibfnamefont {X.}~\bibnamefont {Wan}},\ and\ \bibinfo {author}
  {\bibfnamefont {Q.}~\bibnamefont {Liu}},\ }\bibfield  {title} {\bibinfo
  {title} {Spin-group symmetry in magnetic materials with negligible spin-orbit
  coupling},\ }\href {https://doi.org/10.1103/PhysRevX.12.021016} {\bibfield
  {journal} {\bibinfo  {journal} {Phys. Rev. X}\ }\textbf {\bibinfo {volume}
  {12}},\ \bibinfo {pages} {021016} (\bibinfo {year} {2022})}\BibitemShut
  {NoStop}%
\bibitem [{\citenamefont {Chen}\ \emph {et~al.}(2024)\citenamefont {Chen},
  \citenamefont {Ren}, \citenamefont {Zhu}, \citenamefont {Yu}, \citenamefont
  {Zhang}, \citenamefont {Liu}, \citenamefont {Li}, \citenamefont {Liu},
  \citenamefont {Li},\ and\ \citenamefont {Liu}}]{Chen2024:PRX}%
  \BibitemOpen
  \bibfield  {author} {\bibinfo {author} {\bibfnamefont {X.}~\bibnamefont
  {Chen}}, \bibinfo {author} {\bibfnamefont {J.}~\bibnamefont {Ren}}, \bibinfo
  {author} {\bibfnamefont {Y.}~\bibnamefont {Zhu}}, \bibinfo {author}
  {\bibfnamefont {Y.}~\bibnamefont {Yu}}, \bibinfo {author} {\bibfnamefont
  {A.}~\bibnamefont {Zhang}}, \bibinfo {author} {\bibfnamefont
  {P.}~\bibnamefont {Liu}}, \bibinfo {author} {\bibfnamefont {J.}~\bibnamefont
  {Li}}, \bibinfo {author} {\bibfnamefont {Y.}~\bibnamefont {Liu}}, \bibinfo
  {author} {\bibfnamefont {C.}~\bibnamefont {Li}},\ and\ \bibinfo {author}
  {\bibfnamefont {Q.}~\bibnamefont {Liu}},\ }\bibfield  {title} {\bibinfo
  {title} {Enumeration and representation theory of spin space groups},\ }\href
  {https://doi.org/10.1103/PhysRevX.14.031038} {\bibfield  {journal} {\bibinfo
  {journal} {Phys. Rev. X}\ }\textbf {\bibinfo {volume} {14}},\ \bibinfo
  {pages} {031038} (\bibinfo {year} {2024})}\BibitemShut {NoStop}%
\bibitem [{\citenamefont {{\v{Z}}uti{\'c}}\ \emph {et~al.}(2019)\citenamefont
  {{\v{Z}}uti{\'c}}, \citenamefont {Matos-Abiague}, \citenamefont {Scharf},
  \citenamefont {Dery},\ and\ \citenamefont {Belashchenko}}]{Zutic2019:MT}%
  \BibitemOpen
  \bibfield  {author} {\bibinfo {author} {\bibfnamefont {I.}~\bibnamefont
  {{\v{Z}}uti{\'c}}}, \bibinfo {author} {\bibfnamefont {A.}~\bibnamefont
  {Matos-Abiague}}, \bibinfo {author} {\bibfnamefont {B.}~\bibnamefont
  {Scharf}}, \bibinfo {author} {\bibfnamefont {H.}~\bibnamefont {Dery}},\ and\
  \bibinfo {author} {\bibfnamefont {K.}~\bibnamefont {Belashchenko}},\
  }\bibfield  {title} {\bibinfo {title} {Proximitized materials},\ }\href
  {https://doi.org/10.1016/j.mattod.2018.05.003} {\bibfield  {journal}
  {\bibinfo  {journal} {Mater. Today}\ }\textbf {\bibinfo {volume} {22}},\
  \bibinfo {pages} {85} (\bibinfo {year} {2019})}\BibitemShut {NoStop}%
\bibitem [{\citenamefont {Gonz\'{a}lez-Hern\'{a}ndez}\ \emph
  {et~al.}(2021)\citenamefont {Gonz\'{a}lez-Hern\'{a}ndez}, \citenamefont
  {\v{S}mejkal}, \citenamefont {V\'{y}born\'{y}}, \citenamefont {Yahagi},
  \citenamefont {Sinova}, \citenamefont {Jungwirth},\ and\ \citenamefont
  {\v{Z}elezn\'{y}}}]{Gonzalez2021}%
  \BibitemOpen
  \bibfield  {author} {\bibinfo {author} {\bibfnamefont {R.}~\bibnamefont
  {Gonz\'{a}lez-Hern\'{a}ndez}}, \bibinfo {author} {\bibfnamefont
  {L.}~\bibnamefont {\v{S}mejkal}}, \bibinfo {author} {\bibfnamefont
  {K.}~\bibnamefont {V\'{y}born\'{y}}}, \bibinfo {author} {\bibfnamefont
  {Y.}~\bibnamefont {Yahagi}}, \bibinfo {author} {\bibfnamefont
  {J.}~\bibnamefont {Sinova}}, \bibinfo {author} {\bibfnamefont
  {T.}~\bibnamefont {Jungwirth}},\ and\ \bibinfo {author} {\bibfnamefont
  {J.}~\bibnamefont {\v{Z}elezn\'{y}}},\ }\bibfield  {title} {\bibinfo {title}
  {Efficient electrical spin splitter based on nonrelativistic collinear
  antiferromagnetism},\ }\href {https://doi.org/10.1103/PhysRevLett.126.127701}
  {\bibfield  {journal} {\bibinfo  {journal} {Phys. Rev. Lett.}\ }\textbf
  {\bibinfo {volume} {126}},\ \bibinfo {pages} {127701} (\bibinfo {year}
  {2021})}\BibitemShut {NoStop}%
\bibitem [{\citenamefont {Bai}\ \emph {et~al.}(2022)\citenamefont {Bai},
  \citenamefont {Han}, \citenamefont {Feng}, \citenamefont {Zhou},
  \citenamefont {Su}, \citenamefont {Wang}, \citenamefont {Liao}, \citenamefont
  {Zhu}, \citenamefont {Chen}, \citenamefont {Pan}, \citenamefont {Fan},\ and\
  \citenamefont {Song}}]{PhysRevLett.128.197202}%
  \BibitemOpen
  \bibfield  {author} {\bibinfo {author} {\bibfnamefont {H.}~\bibnamefont
  {Bai}}, \bibinfo {author} {\bibfnamefont {L.}~\bibnamefont {Han}}, \bibinfo
  {author} {\bibfnamefont {X.~Y.}\ \bibnamefont {Feng}}, \bibinfo {author}
  {\bibfnamefont {Y.~J.}\ \bibnamefont {Zhou}}, \bibinfo {author}
  {\bibfnamefont {R.~X.}\ \bibnamefont {Su}}, \bibinfo {author} {\bibfnamefont
  {Q.}~\bibnamefont {Wang}}, \bibinfo {author} {\bibfnamefont {L.~Y.}\
  \bibnamefont {Liao}}, \bibinfo {author} {\bibfnamefont {W.~X.}\ \bibnamefont
  {Zhu}}, \bibinfo {author} {\bibfnamefont {X.~Z.}\ \bibnamefont {Chen}},
  \bibinfo {author} {\bibfnamefont {F.}~\bibnamefont {Pan}}, \bibinfo {author}
  {\bibfnamefont {X.~L.}\ \bibnamefont {Fan}},\ and\ \bibinfo {author}
  {\bibfnamefont {C.}~\bibnamefont {Song}},\ }\bibfield  {title} {\bibinfo
  {title} {Observation of spin splitting torque in a collinear antiferromagnet
  {${\mathrm{RuO}}_{2}$}},\ }\href
  {https://doi.org/10.1103/PhysRevLett.128.197202} {\bibfield  {journal}
  {\bibinfo  {journal} {Phys. Rev. Lett.}\ }\textbf {\bibinfo {volume} {128}},\
  \bibinfo {pages} {197202} (\bibinfo {year} {2022})}\BibitemShut {NoStop}%
\bibitem [{\citenamefont {Bose}\ \emph {et~al.}(2022)\citenamefont {Bose},
  \citenamefont {Schreiber}, \citenamefont {Jain}, \citenamefont {Shao},
  \citenamefont {Nair}, \citenamefont {Sun}, \citenamefont {Zhang},
  \citenamefont {Muller}, \citenamefont {Tsymbal}, \citenamefont {Schlom},\
  and\ \citenamefont {Ralph}}]{Bose2022}%
  \BibitemOpen
  \bibfield  {author} {\bibinfo {author} {\bibfnamefont {A.}~\bibnamefont
  {Bose}}, \bibinfo {author} {\bibfnamefont {N.~J.}\ \bibnamefont {Schreiber}},
  \bibinfo {author} {\bibfnamefont {R.}~\bibnamefont {Jain}}, \bibinfo {author}
  {\bibfnamefont {D.-F.}\ \bibnamefont {Shao}}, \bibinfo {author}
  {\bibfnamefont {H.~P.}\ \bibnamefont {Nair}}, \bibinfo {author}
  {\bibfnamefont {J.}~\bibnamefont {Sun}}, \bibinfo {author} {\bibfnamefont
  {X.~S.}\ \bibnamefont {Zhang}}, \bibinfo {author} {\bibfnamefont {D.~A.}\
  \bibnamefont {Muller}}, \bibinfo {author} {\bibfnamefont {E.~Y.}\
  \bibnamefont {Tsymbal}}, \bibinfo {author} {\bibfnamefont {D.~G.}\
  \bibnamefont {Schlom}},\ and\ \bibinfo {author} {\bibfnamefont {D.~C.}\
  \bibnamefont {Ralph}},\ }\bibfield  {title} {\bibinfo {title} {Tilted spin
  current generated by the collinear antiferromagnet ruthenium dioxide},\
  }\href {https://doi.org/10.1038/s41928-022-00744-8} {\bibfield  {journal}
  {\bibinfo  {journal} {Nat. Electron.}\ }\textbf {\bibinfo {volume} {5}},\
  \bibinfo {pages} {267} (\bibinfo {year} {2022})}\BibitemShut {NoStop}%
\bibitem [{\citenamefont {Karube}\ \emph {et~al.}(2022)\citenamefont {Karube},
  \citenamefont {Tanaka}, \citenamefont {Sugawara}, \citenamefont {Kadoguchi},
  \citenamefont {Kohda},\ and\ \citenamefont {Nitta}}]{Karube2022}%
  \BibitemOpen
  \bibfield  {author} {\bibinfo {author} {\bibfnamefont {S.}~\bibnamefont
  {Karube}}, \bibinfo {author} {\bibfnamefont {T.}~\bibnamefont {Tanaka}},
  \bibinfo {author} {\bibfnamefont {D.}~\bibnamefont {Sugawara}}, \bibinfo
  {author} {\bibfnamefont {N.}~\bibnamefont {Kadoguchi}}, \bibinfo {author}
  {\bibfnamefont {M.}~\bibnamefont {Kohda}},\ and\ \bibinfo {author}
  {\bibfnamefont {J.}~\bibnamefont {Nitta}},\ }\bibfield  {title} {\bibinfo
  {title} {Observation of spin-splitter torque in collinear antiferromagnetic
  {${\mathrm{RuO}}_{2}$}},\ }\href
  {https://doi.org/10.1103/PhysRevLett.129.137201} {\bibfield  {journal}
  {\bibinfo  {journal} {Phys. Rev. Lett.}\ }\textbf {\bibinfo {volume} {129}},\
  \bibinfo {pages} {137201} (\bibinfo {year} {2022})}\BibitemShut {NoStop}%
\bibitem [{\citenamefont {Vakili}\ \emph {et~al.}(2025)\citenamefont {Vakili},
  \citenamefont {Schwartz},\ and\ \citenamefont
  {Kovalev}}]{PhysRevLett.134.176401}%
  \BibitemOpen
  \bibfield  {author} {\bibinfo {author} {\bibfnamefont {H.}~\bibnamefont
  {Vakili}}, \bibinfo {author} {\bibfnamefont {E.}~\bibnamefont {Schwartz}},\
  and\ \bibinfo {author} {\bibfnamefont {A.~A.}\ \bibnamefont {Kovalev}},\
  }\bibfield  {title} {\bibinfo {title} {Spin-transfer torque in altermagnets
  with magnetic textures},\ }\href
  {https://doi.org/10.1103/PhysRevLett.134.176401} {\bibfield  {journal}
  {\bibinfo  {journal} {Phys. Rev. Lett.}\ }\textbf {\bibinfo {volume} {134}},\
  \bibinfo {pages} {176401} (\bibinfo {year} {2025})}\BibitemShut {NoStop}%
\bibitem [{\citenamefont {Duan}\ \emph {et~al.}(2025)\citenamefont {Duan},
  \citenamefont {Zhang}, \citenamefont {Zhu}, \citenamefont {Liu},
  \citenamefont {Zhang}, \citenamefont {\ifmmode \check{Z}\else
  \v{Z}\fi{}uti\ifmmode~\acute{c}\else \'{c}\fi{}},\ and\ \citenamefont
  {Zhou}}]{Duan2025:PRL}%
  \BibitemOpen
  \bibfield  {author} {\bibinfo {author} {\bibfnamefont {X.}~\bibnamefont
  {Duan}}, \bibinfo {author} {\bibfnamefont {J.}~\bibnamefont {Zhang}},
  \bibinfo {author} {\bibfnamefont {Z.}~\bibnamefont {Zhu}}, \bibinfo {author}
  {\bibfnamefont {Y.}~\bibnamefont {Liu}}, \bibinfo {author} {\bibfnamefont
  {Z.}~\bibnamefont {Zhang}}, \bibinfo {author} {\bibfnamefont
  {I.}~\bibnamefont {\ifmmode \check{Z}\else
  \v{Z}\fi{}uti\ifmmode~\acute{c}\else \'{c}\fi{}}},\ and\ \bibinfo {author}
  {\bibfnamefont {T.}~\bibnamefont {Zhou}},\ }\bibfield  {title} {\bibinfo
  {title} {Antiferroelectric altermagnets: Antiferroelectricity alters
  magnets},\ }\href {https://doi.org/10.1103/PhysRevLett.134.106801} {\bibfield
   {journal} {\bibinfo  {journal} {Phys. Rev. Lett.}\ }\textbf {\bibinfo
  {volume} {134}},\ \bibinfo {pages} {106801} (\bibinfo {year}
  {2025})}\BibitemShut {NoStop}%
\bibitem [{\citenamefont {Gu}\ \emph {et~al.}(2025)\citenamefont {Gu},
  \citenamefont {Liu}, \citenamefont {Zhu}, \citenamefont {Yananose},
  \citenamefont {Chen}, \citenamefont {Hu}, \citenamefont {Stroppa},\ and\
  \citenamefont {Liu}}]{Gu2025:PRL}%
  \BibitemOpen
  \bibfield  {author} {\bibinfo {author} {\bibfnamefont {M.}~\bibnamefont
  {Gu}}, \bibinfo {author} {\bibfnamefont {Y.}~\bibnamefont {Liu}}, \bibinfo
  {author} {\bibfnamefont {H.}~\bibnamefont {Zhu}}, \bibinfo {author}
  {\bibfnamefont {K.}~\bibnamefont {Yananose}}, \bibinfo {author}
  {\bibfnamefont {X.}~\bibnamefont {Chen}}, \bibinfo {author} {\bibfnamefont
  {Y.}~\bibnamefont {Hu}}, \bibinfo {author} {\bibfnamefont {A.}~\bibnamefont
  {Stroppa}},\ and\ \bibinfo {author} {\bibfnamefont {Q.}~\bibnamefont {Liu}},\
  }\bibfield  {title} {\bibinfo {title} {Ferroelectric switchable
  altermagnetism},\ }\href {https://doi.org/10.1103/PhysRevLett.134.106802}
  {\bibfield  {journal} {\bibinfo  {journal} {Phys. Rev. Lett.}\ }\textbf
  {\bibinfo {volume} {134}},\ \bibinfo {pages} {106802} (\bibinfo {year}
  {2025})}\BibitemShut {NoStop}%
\bibitem [{\citenamefont {Urru}\ \emph {et~al.}(2025)\citenamefont {Urru},
  \citenamefont {Seleznev}, \citenamefont {Teng}, \citenamefont {Park},
  \citenamefont {Reyes-Lillo},\ and\ \citenamefont {Rabe}}]{Urru2025:PRB}%
  \BibitemOpen
  \bibfield  {author} {\bibinfo {author} {\bibfnamefont {A.}~\bibnamefont
  {Urru}}, \bibinfo {author} {\bibfnamefont {D.}~\bibnamefont {Seleznev}},
  \bibinfo {author} {\bibfnamefont {Y.}~\bibnamefont {Teng}}, \bibinfo {author}
  {\bibfnamefont {S.~Y.}\ \bibnamefont {Park}}, \bibinfo {author}
  {\bibfnamefont {S.~E.}\ \bibnamefont {Reyes-Lillo}},\ and\ \bibinfo {author}
  {\bibfnamefont {K.~M.}\ \bibnamefont {Rabe}},\ }\bibfield  {title} {\bibinfo
  {title} {{G-type antiferromagnetic BiFeO$_3$ is a multiferroic $g$-wave
  altermagnet}},\ }\href@noop {} {\bibfield  {journal} {\bibinfo  {journal}
  {Phys. Rev. B}\ }\textbf {\bibinfo {volume} {112}},\ \bibinfo {pages}
  {104411} (\bibinfo {year} {2025})}\BibitemShut {NoStop}%
\bibitem [{\citenamefont {Fernandes}\ \emph {et~al.}(2024)\citenamefont
  {Fernandes}, \citenamefont {de~Carvalho}, \citenamefont {Birol},\ and\
  \citenamefont {Pereira}}]{Fernandes2024:PRB}%
  \BibitemOpen
  \bibfield  {author} {\bibinfo {author} {\bibfnamefont {R.~M.}\ \bibnamefont
  {Fernandes}}, \bibinfo {author} {\bibfnamefont {V.~S.}\ \bibnamefont
  {de~Carvalho}}, \bibinfo {author} {\bibfnamefont {T.}~\bibnamefont {Birol}},\
  and\ \bibinfo {author} {\bibfnamefont {R.~G.}\ \bibnamefont {Pereira}},\
  }\bibfield  {title} {\bibinfo {title} {{Topological transition from nodal to
  nodeless Zeeman splitting in altermagnets}},\ }\href
  {https://doi.org/10.1103/PhysRevB.109.024404} {\bibfield  {journal} {\bibinfo
   {journal} {Phys. Rev. B}\ }\textbf {\bibinfo {volume} {109}},\ \bibinfo
  {pages} {024404} (\bibinfo {year} {2024})}\BibitemShut {NoStop}%
\bibitem [{\citenamefont {Antonenko}\ \emph {et~al.}(2025)\citenamefont
  {Antonenko}, \citenamefont {Fernandes},\ and\ \citenamefont
  {Venderbos}}]{Antonenko2025:PRL}%
  \BibitemOpen
  \bibfield  {author} {\bibinfo {author} {\bibfnamefont {D.~S.}\ \bibnamefont
  {Antonenko}}, \bibinfo {author} {\bibfnamefont {R.~M.}\ \bibnamefont
  {Fernandes}},\ and\ \bibinfo {author} {\bibfnamefont {J.~W.~F.}\ \bibnamefont
  {Venderbos}},\ }\bibfield  {title} {\bibinfo {title} {Mirror {Chern} bands
  and {Weyl} nodal loops in altermagnets},\ }\href
  {https://doi.org/10.1103/PhysRevLett.134.096703} {\bibfield  {journal}
  {\bibinfo  {journal} {Phys. Rev. Lett.}\ }\textbf {\bibinfo {volume} {134}},\
  \bibinfo {pages} {096703} (\bibinfo {year} {2025})}\BibitemShut {NoStop}%
\bibitem [{\citenamefont {Chen}\ \emph {et~al.}()\citenamefont {Chen},
  \citenamefont {Zhang}, \citenamefont {Hao}, \citenamefont {Qian},
  \citenamefont {Zhu}, \citenamefont {\v{Z}uti\'c}, \citenamefont {Zhang},\
  and\ \citenamefont {Zhou}}]{Chen2026:X}%
  \BibitemOpen
  \bibfield  {author} {\bibinfo {author} {\bibfnamefont {X.}~\bibnamefont
  {Chen}}, \bibinfo {author} {\bibfnamefont {J.}~\bibnamefont {Zhang}},
  \bibinfo {author} {\bibfnamefont {B.}~\bibnamefont {Hao}}, \bibinfo {author}
  {\bibfnamefont {J.}~\bibnamefont {Qian}}, \bibinfo {author} {\bibfnamefont
  {Z.}~\bibnamefont {Zhu}}, \bibinfo {author} {\bibfnamefont {I.}~\bibnamefont
  {\v{Z}uti\'c}}, \bibinfo {author} {\bibfnamefont {Z.}~\bibnamefont {Zhang}},\
  and\ \bibinfo {author} {\bibfnamefont {T.}~\bibnamefont {Zhou}},\ }\bibfield
  {title} {\bibinfo {title} {Altermagnets enable gate-switchable helical and
  chiral topological transport with spin-valley-momentum-locked dual
  protection},\ }\href {https://arxiv.org/abs/2603.06487} {\bibinfo  {journal}
  {arXiv:2603.06487}\ }\BibitemShut {NoStop}%
\bibitem [{\citenamefont {Cui}\ \emph {et~al.}(2023)\citenamefont {Cui},
  \citenamefont {Zeng}, \citenamefont {Cui}, \citenamefont {Yu},\ and\
  \citenamefont {Yang}}]{PhysRevB.108.L180401}%
  \BibitemOpen
\bibfield  {journal} {  }\bibfield  {author} {\bibinfo {author} {\bibfnamefont
  {Q.}~\bibnamefont {Cui}}, \bibinfo {author} {\bibfnamefont {B.}~\bibnamefont
  {Zeng}}, \bibinfo {author} {\bibfnamefont {P.}~\bibnamefont {Cui}}, \bibinfo
  {author} {\bibfnamefont {T.}~\bibnamefont {Yu}},\ and\ \bibinfo {author}
  {\bibfnamefont {H.}~\bibnamefont {Yang}},\ }\bibfield  {title} {\bibinfo
  {title} {Efficient spin {Seebeck} and spin {Nernst} effects of magnons in
  altermagnets},\ }\href {https://doi.org/10.1103/PhysRevB.108.L180401}
  {\bibfield  {journal} {\bibinfo  {journal} {Phys. Rev. B}\ }\textbf {\bibinfo
  {volume} {108}},\ \bibinfo {pages} {L180401} (\bibinfo {year}
  {2023})}\BibitemShut {NoStop}%
\bibitem [{\citenamefont {Wei\ss{}enhofer}\ and\ \citenamefont
  {Marmodoro}(2024)}]{PhysRevB.110.094427}%
  \BibitemOpen
  \bibfield  {author} {\bibinfo {author} {\bibfnamefont {M.}~\bibnamefont
  {Wei\ss{}enhofer}}\ and\ \bibinfo {author} {\bibfnamefont {A.}~\bibnamefont
  {Marmodoro}},\ }\bibfield  {title} {\bibinfo {title} {Atomistic spin dynamics
  simulations of magnonic spin {Seebeck} and spin {Nernst} effects in
  altermagnets},\ }\href {https://doi.org/10.1103/PhysRevB.110.094427}
  {\bibfield  {journal} {\bibinfo  {journal} {Phys. Rev. B}\ }\textbf {\bibinfo
  {volume} {110}},\ \bibinfo {pages} {094427} (\bibinfo {year}
  {2024})}\BibitemShut {NoStop}%
\bibitem [{\citenamefont {Yershov}\ \emph {et~al.}(2024)\citenamefont
  {Yershov}, \citenamefont {Kravchuk}, \citenamefont {Daghofer},\ and\
  \citenamefont {van~den Brink}}]{PhysRevB.110.144421}%
  \BibitemOpen
  \bibfield  {author} {\bibinfo {author} {\bibfnamefont {K.~V.}\ \bibnamefont
  {Yershov}}, \bibinfo {author} {\bibfnamefont {V.~P.}\ \bibnamefont
  {Kravchuk}}, \bibinfo {author} {\bibfnamefont {M.}~\bibnamefont {Daghofer}},\
  and\ \bibinfo {author} {\bibfnamefont {J.}~\bibnamefont {van~den Brink}},\
  }\bibfield  {title} {\bibinfo {title} {Fluctuation-induced piezomagnetism in
  local moment altermagnets},\ }\href
  {https://doi.org/10.1103/PhysRevB.110.144421} {\bibfield  {journal} {\bibinfo
   {journal} {Phys. Rev. B}\ }\textbf {\bibinfo {volume} {110}},\ \bibinfo
  {pages} {144421} (\bibinfo {year} {2024})}\BibitemShut {NoStop}%
\bibitem [{\citenamefont {Wu}\ \emph {et~al.}(2025)\citenamefont {Wu},
  \citenamefont {Dong}, \citenamefont {Zhu}, \citenamefont {Zheng},\ and\
  \citenamefont {Zhang}}]{Wu2025ChinPhysLett}%
  \BibitemOpen
  \bibfield  {author} {\bibinfo {author} {\bibfnamefont {K.}~\bibnamefont
  {Wu}}, \bibinfo {author} {\bibfnamefont {J.}~\bibnamefont {Dong}}, \bibinfo
  {author} {\bibfnamefont {M.}~\bibnamefont {Zhu}}, \bibinfo {author}
  {\bibfnamefont {F.}~\bibnamefont {Zheng}},\ and\ \bibinfo {author}
  {\bibfnamefont {J.}~\bibnamefont {Zhang}},\ }\bibfield  {title} {\bibinfo
  {title} {Magnon splitting and magnon spin transport in altermagnets},\ }\href
  {https://doi.org/10.1088/0256-307X/42/7/070702} {\bibfield  {journal}
  {\bibinfo  {journal} {Chin. Phys. Lett.}\ }\textbf {\bibinfo {volume} {42}},\
  \bibinfo {pages} {070702} (\bibinfo {year} {2025})}\BibitemShut {NoStop}%
\bibitem [{\citenamefont {Schwartz}\ \emph {et~al.}()\citenamefont {Schwartz},
  \citenamefont {Vakili},\ and\ \citenamefont {Kovalev}}]{Schwartz2025}%
  \BibitemOpen
  \bibfield  {author} {\bibinfo {author} {\bibfnamefont {E.}~\bibnamefont
  {Schwartz}}, \bibinfo {author} {\bibfnamefont {H.}~\bibnamefont {Vakili}},\
  and\ \bibinfo {author} {\bibfnamefont {A.~A.}\ \bibnamefont {Kovalev}},\
  }\bibfield  {title} {\bibinfo {title} {Theory of thermomagnonic torques in
  altermagnets},\ }\href {https://arxiv.org/abs/2512.14660} {\bibinfo
  {journal} {arXiv:2512.14660}\ }\BibitemShut {NoStop}%
\bibitem [{\citenamefont {Cao}\ \emph {et~al.}(2025)\citenamefont {Cao},
  \citenamefont {Denisov}, \citenamefont {Liu},\ and\ \citenamefont
  {\v{Z}uti\'c}}]{Cao2025:PRL}%
  \BibitemOpen
\bibfield  {journal} {  }\bibfield  {author} {\bibinfo {author} {\bibfnamefont
  {J.~D.}\ \bibnamefont {Cao}}, \bibinfo {author} {\bibfnamefont {K.~S.}\
  \bibnamefont {Denisov}}, \bibinfo {author} {\bibfnamefont {Y.}~\bibnamefont
  {Liu}},\ and\ \bibinfo {author} {\bibfnamefont {I.}~\bibnamefont
  {\v{Z}uti\'c}},\ }\bibfield  {title} {\bibinfo {title} {{Symmetry
  Classification for Alternating Excitons in Two-Dimensional Altermagnets}},\
  }\href {https://doi.org/10.1103/PhysRevLett.135.266703} {\bibfield  {journal}
  {\bibinfo  {journal} {Phys. Rev. Lett.}\ }\textbf {\bibinfo {volume} {135}},\
  \bibinfo {pages} {266703} (\bibinfo {year} {2025})}\BibitemShut {NoStop}%
\bibitem [{\citenamefont {Papaj}(2023)}]{Papaj2023}%
  \BibitemOpen
  \bibfield  {author} {\bibinfo {author} {\bibfnamefont {M.}~\bibnamefont
  {Papaj}},\ }\bibfield  {title} {\bibinfo {title} {{Andreev} reflection at the
  altermagnetic metal-superconductor interface},\ }\href
  {https://doi.org/10.1103/PhysRevB.108.L060508} {\bibfield  {journal}
  {\bibinfo  {journal} {Phys. Rev. B}\ }\textbf {\bibinfo {volume} {108}},\
  \bibinfo {pages} {L060508} (\bibinfo {year} {2023})}\BibitemShut {NoStop}%
\bibitem [{\citenamefont {Ghorashi}\ \emph {et~al.}(2024)\citenamefont
  {Ghorashi}, \citenamefont {Hughes},\ and\ \citenamefont
  {Cano}}]{Ghorashi2024:PRL}%
  \BibitemOpen
  \bibfield  {author} {\bibinfo {author} {\bibfnamefont {S.~A.~A.}\
  \bibnamefont {Ghorashi}}, \bibinfo {author} {\bibfnamefont {T.~L.}\
  \bibnamefont {Hughes}},\ and\ \bibinfo {author} {\bibfnamefont
  {J.}~\bibnamefont {Cano}},\ }\bibfield  {title} {\bibinfo {title}
  {Altermagnetic routes to {Majorana} modes in zero net magnetization},\ }\href
  {https://doi.org/10.1103/PhysRevLett.133.106601} {\bibfield  {journal}
  {\bibinfo  {journal} {Phys. Rev. Lett.}\ }\textbf {\bibinfo {volume} {133}},\
  \bibinfo {pages} {106601} (\bibinfo {year} {2024})}\BibitemShut {NoStop}%
\bibitem [{\citenamefont {Ouassou}\ \emph {et~al.}(2023)\citenamefont
  {Ouassou}, \citenamefont {Brataas},\ and\ \citenamefont
  {Linder}}]{Ouassou2023}%
  \BibitemOpen
  \bibfield  {author} {\bibinfo {author} {\bibfnamefont {J.~A.}\ \bibnamefont
  {Ouassou}}, \bibinfo {author} {\bibfnamefont {A.}~\bibnamefont {Brataas}},\
  and\ \bibinfo {author} {\bibfnamefont {J.}~\bibnamefont {Linder}},\
  }\bibfield  {title} {\bibinfo {title} {dc {Josephson} effect in
  altermagnets},\ }\href {https://doi.org/10.1103/PhysRevLett.131.076003}
  {\bibfield  {journal} {\bibinfo  {journal} {Phys. Rev. Lett.}\ }\textbf
  {\bibinfo {volume} {131}},\ \bibinfo {pages} {076003} (\bibinfo {year}
  {2023})}\BibitemShut {NoStop}%
\bibitem [{\citenamefont {Giil}\ and\ \citenamefont {Linder}(2024)}]{Giil2024}%
  \BibitemOpen
  \bibfield  {author} {\bibinfo {author} {\bibfnamefont {H.~G.}\ \bibnamefont
  {Giil}}\ and\ \bibinfo {author} {\bibfnamefont {J.}~\bibnamefont {Linder}},\
  }\bibfield  {title} {\bibinfo {title} {Superconductor-altermagnet memory
  functionality without stray fields},\ }\href
  {https://doi.org/10.1103/PhysRevB.109.134511} {\bibfield  {journal} {\bibinfo
   {journal} {Phys. Rev. B}\ }\textbf {\bibinfo {volume} {109}},\ \bibinfo
  {pages} {134511} (\bibinfo {year} {2024})}\BibitemShut {NoStop}%
\bibitem [{\citenamefont {Giil}\ \emph {et~al.}(2024)\citenamefont {Giil},
  \citenamefont {Brekke}, \citenamefont {Linder},\ and\ \citenamefont
  {Brataas}}]{PhysRevB.110.L140506}%
  \BibitemOpen
  \bibfield  {author} {\bibinfo {author} {\bibfnamefont {H.~G.}\ \bibnamefont
  {Giil}}, \bibinfo {author} {\bibfnamefont {B.}~\bibnamefont {Brekke}},
  \bibinfo {author} {\bibfnamefont {J.}~\bibnamefont {Linder}},\ and\ \bibinfo
  {author} {\bibfnamefont {A.}~\bibnamefont {Brataas}},\ }\bibfield  {title}
  {\bibinfo {title} {Quasiclassical theory of superconducting spin-splitter
  effects and spin-filtering via altermagnets},\ }\href
  {https://doi.org/10.1103/PhysRevB.110.L140506} {\bibfield  {journal}
  {\bibinfo  {journal} {Phys. Rev. B}\ }\textbf {\bibinfo {volume} {110}},\
  \bibinfo {pages} {L140506} (\bibinfo {year} {2024})}\BibitemShut {NoStop}%
\bibitem [{\citenamefont {Hu}\ \emph {et~al.}(2025)\citenamefont {Hu},
  \citenamefont {Matsyshyn},\ and\ \citenamefont
  {Song}}]{PhysRevLett.134.026001}%
  \BibitemOpen
  \bibfield  {author} {\bibinfo {author} {\bibfnamefont {J.-X.}\ \bibnamefont
  {Hu}}, \bibinfo {author} {\bibfnamefont {O.}~\bibnamefont {Matsyshyn}},\ and\
  \bibinfo {author} {\bibfnamefont {J.~C.~W.}\ \bibnamefont {Song}},\
  }\bibfield  {title} {\bibinfo {title} {Nonlinear superconducting
  magnetoelectric effect},\ }\href
  {https://doi.org/10.1103/PhysRevLett.134.026001} {\bibfield  {journal}
  {\bibinfo  {journal} {Phys. Rev. Lett.}\ }\textbf {\bibinfo {volume} {134}},\
  \bibinfo {pages} {026001} (\bibinfo {year} {2025})}\BibitemShut {NoStop}%
\bibitem [{\citenamefont {Zhu}\ \emph {et~al.}(2026)\citenamefont {Zhu},
  \citenamefont {Huang}, \citenamefont {Chen}, \citenamefont {Cui},
  \citenamefont {Duan}, \citenamefont {Zhang}, \citenamefont {\ifmmode
  \check{Z}\else \v{Z}\fi{}uti\ifmmode~\acute{c}\else \'{c}\fi{}},\ and\
  \citenamefont {Zhou}}]{kqy8-myz1}%
  \BibitemOpen
  \bibfield  {author} {\bibinfo {author} {\bibfnamefont {Z.}~\bibnamefont
  {Zhu}}, \bibinfo {author} {\bibfnamefont {R.}~\bibnamefont {Huang}}, \bibinfo
  {author} {\bibfnamefont {X.}~\bibnamefont {Chen}}, \bibinfo {author}
  {\bibfnamefont {Z.}~\bibnamefont {Cui}}, \bibinfo {author} {\bibfnamefont
  {X.}~\bibnamefont {Duan}}, \bibinfo {author} {\bibfnamefont {J.}~\bibnamefont
  {Zhang}}, \bibinfo {author} {\bibfnamefont {I.}~\bibnamefont {\ifmmode
  \check{Z}\else \v{Z}\fi{}uti\ifmmode~\acute{c}\else \'{c}\fi{}}},\ and\
  \bibinfo {author} {\bibfnamefont {T.}~\bibnamefont {Zhou}},\ }\bibfield
  {title} {\bibinfo {title} {Altermagnetic proximity effect},\ }\href
  {https://doi.org/10.1103/kqy8-myz1} {\bibfield  {journal} {\bibinfo
  {journal} {Phys. Rev. Lett.}\ }\textbf {\bibinfo {volume} {136}},\ \bibinfo
  {pages} {186702} (\bibinfo {year} {2026})}\BibitemShut {NoStop}%
\bibitem [{\citenamefont {Linder}\ and\ \citenamefont
  {Robinson}(2015)}]{Linder2015}%
  \BibitemOpen
  \bibfield  {author} {\bibinfo {author} {\bibfnamefont {J.}~\bibnamefont
  {Linder}}\ and\ \bibinfo {author} {\bibfnamefont {J.~W.~A.}\ \bibnamefont
  {Robinson}},\ }\bibfield  {title} {\bibinfo {title} {Superconducting
  spintronics},\ }\href {https://doi.org/10.1038/nphys3242} {\bibfield
  {journal} {\bibinfo  {journal} {Nat. Phys.}\ }\textbf {\bibinfo {volume}
  {11}},\ \bibinfo {pages} {307} (\bibinfo {year} {2015})}\BibitemShut
  {NoStop}%
\bibitem [{\citenamefont {Eschrig}(2015)}]{Eschrig2015}%
  \BibitemOpen
  \bibfield  {author} {\bibinfo {author} {\bibfnamefont {M.}~\bibnamefont
  {Eschrig}},\ }\bibfield  {title} {\bibinfo {title} {Spin-polarized
  supercurrents for spintronics: a review of current progress},\ }\href
  {https://doi.org/10.1088/0034-4885/78/10/104501} {\bibfield  {journal}
  {\bibinfo  {journal} {Rep. Prog. Phys.}\ }\textbf {\bibinfo {volume} {78}},\
  \bibinfo {pages} {104501} (\bibinfo {year} {2015})}\BibitemShut {NoStop}%
\bibitem [{\citenamefont {Amundsen}\ \emph {et~al.}(2024)\citenamefont
  {Amundsen}, \citenamefont {Linder}, \citenamefont {Robinson}, \citenamefont
  {\v{Z}uti\'{c}},\ and\ \citenamefont {Banerjee}}]{Amundsen2024}%
  \BibitemOpen
  \bibfield  {author} {\bibinfo {author} {\bibfnamefont {M.}~\bibnamefont
  {Amundsen}}, \bibinfo {author} {\bibfnamefont {J.}~\bibnamefont {Linder}},
  \bibinfo {author} {\bibfnamefont {J.~W.~A.}\ \bibnamefont {Robinson}},
  \bibinfo {author} {\bibfnamefont {I.}~\bibnamefont {\v{Z}uti\'{c}}},\ and\
  \bibinfo {author} {\bibfnamefont {N.}~\bibnamefont {Banerjee}},\ }\bibfield
  {title} {\bibinfo {title} {Colloquium: Spin-orbit effects in superconducting
  hybrid structures},\ }\href {https://doi.org/10.1103/RevModPhys.96.021003}
  {\bibfield  {journal} {\bibinfo  {journal} {Rev. Mod. Phys.}\ }\textbf
  {\bibinfo {volume} {96}},\ \bibinfo {pages} {021003} (\bibinfo {year}
  {2024})}\BibitemShut {NoStop}%
\bibitem [{\citenamefont {Ando}\ \emph {et~al.}(2020)\citenamefont {Ando} \emph
  {et~al.}}]{JDEexp}%
  \BibitemOpen
  \bibfield  {author} {\bibinfo {author} {\bibfnamefont {F.}~\bibnamefont
  {Ando}} \emph {et~al.},\ }\bibfield  {title} {\bibinfo {title} {Observation
  of superconducting diode effect},\ }\href
  {https://doi.org/10.1038/s41586-020-2590-4} {\bibfield  {journal} {\bibinfo
  {journal} {Nature}\ }\textbf {\bibinfo {volume} {584}},\ \bibinfo {pages}
  {373} (\bibinfo {year} {2020})}\BibitemShut {NoStop}%
\bibitem [{\citenamefont {{\v{Z}}uti\'c}\ and\ \citenamefont {{Das
  Sarma}}(1999)}]{Zutic1999:PRB}%
  \BibitemOpen
  \bibfield  {author} {\bibinfo {author} {\bibfnamefont {I.}~\bibnamefont
  {{\v{Z}}uti\'c}}\ and\ \bibinfo {author} {\bibfnamefont {S.}~\bibnamefont
  {{Das Sarma}}},\ }\bibfield  {title} {\bibinfo {title} {Spin-polarized
  transport and {Andreev} reflection in semiconductor/superconductor hybrid
  structures},\ }\href {https://doi.org/10.1103/PhysRevB.60.R16322} {\bibfield
  {journal} {\bibinfo  {journal} {Phys. Rev. B}\ }\textbf {\bibinfo {volume}
  {60}},\ \bibinfo {pages} {R16322} (\bibinfo {year} {1999})}\BibitemShut
  {NoStop}%
\bibitem [{\citenamefont {Buzdin}(2005)}]{Buzdin2005}%
  \BibitemOpen
  \bibfield  {author} {\bibinfo {author} {\bibfnamefont {A.~I.}\ \bibnamefont
  {Buzdin}},\ }\bibfield  {title} {\bibinfo {title} {Proximity effects in
  superconductor-ferromagnet heterostructures},\ }\href
  {https://doi.org/10.1103/RevModPhys.77.935} {\bibfield  {journal} {\bibinfo
  {journal} {Rev. Mod. Phys.}\ }\textbf {\bibinfo {volume} {77}},\ \bibinfo
  {pages} {935} (\bibinfo {year} {2005})}\BibitemShut {NoStop}%
\bibitem [{\citenamefont {Bergeret}\ \emph {et~al.}(2005)\citenamefont
  {Bergeret}, \citenamefont {Volkov},\ and\ \citenamefont
  {Efetov}}]{Bergeret2005}%
  \BibitemOpen
  \bibfield  {author} {\bibinfo {author} {\bibfnamefont {F.~S.}\ \bibnamefont
  {Bergeret}}, \bibinfo {author} {\bibfnamefont {A.~F.}\ \bibnamefont
  {Volkov}},\ and\ \bibinfo {author} {\bibfnamefont {K.~B.}\ \bibnamefont
  {Efetov}},\ }\bibfield  {title} {\bibinfo {title} {Odd triplet
  superconductivity and related phenomena in superconductor-ferromagnet
  structures},\ }\href {https://doi.org/10.1103/RevModPhys.77.1321} {\bibfield
  {journal} {\bibinfo  {journal} {Rev. Mod. Phys.}\ }\textbf {\bibinfo {volume}
  {77}},\ \bibinfo {pages} {1321} (\bibinfo {year} {2005})}\BibitemShut
  {NoStop}%
\bibitem [{\citenamefont {Ryazanov}\ \emph {et~al.}(2001)\citenamefont
  {Ryazanov}, \citenamefont {Oboznov}, \citenamefont {Rusanov}, \citenamefont
  {Veretennikov}, \citenamefont {Golubov},\ and\ \citenamefont
  {Aarts}}]{Ryazanov2001}%
  \BibitemOpen
  \bibfield  {author} {\bibinfo {author} {\bibfnamefont {V.~V.}\ \bibnamefont
  {Ryazanov}}, \bibinfo {author} {\bibfnamefont {V.~A.}\ \bibnamefont
  {Oboznov}}, \bibinfo {author} {\bibfnamefont {A.~Y.}\ \bibnamefont
  {Rusanov}}, \bibinfo {author} {\bibfnamefont {A.~V.}\ \bibnamefont
  {Veretennikov}}, \bibinfo {author} {\bibfnamefont {A.~A.}\ \bibnamefont
  {Golubov}},\ and\ \bibinfo {author} {\bibfnamefont {J.}~\bibnamefont
  {Aarts}},\ }\bibfield  {title} {\bibinfo {title} {Coupling of two
  superconductors through a ferromagnet: Evidence for a $\pi$ junction},\
  }\href {https://doi.org/10.1103/PhysRevLett.86.2427} {\bibfield  {journal}
  {\bibinfo  {journal} {Phys. Rev. Lett.}\ }\textbf {\bibinfo {volume} {86}},\
  \bibinfo {pages} {2427} (\bibinfo {year} {2001})}\BibitemShut {NoStop}%
\bibitem [{\citenamefont {Khaire}\ \emph {et~al.}(2010)\citenamefont {Khaire},
  \citenamefont {Khasawneh}, \citenamefont {{Pratt, Jr.}},\ and\ \citenamefont
  {Birge}}]{Khaire2010:PRL}%
  \BibitemOpen
  \bibfield  {author} {\bibinfo {author} {\bibfnamefont {T.~S.}\ \bibnamefont
  {Khaire}}, \bibinfo {author} {\bibfnamefont {M.~A.}\ \bibnamefont
  {Khasawneh}}, \bibinfo {author} {\bibfnamefont {W.~P.}\ \bibnamefont {{Pratt,
  Jr.}}},\ and\ \bibinfo {author} {\bibfnamefont {N.~O.}\ \bibnamefont
  {Birge}},\ }\bibfield  {title} {\bibinfo {title} {Observation of spin-triplet
  superconductivity in {Co}-based {Josephson} junctions},\ }\href
  {https://doi.org/10.1103/PhysRevLett.104.137002} {\bibfield  {journal}
  {\bibinfo  {journal} {Phys. Rev. Lett.}\ }\textbf {\bibinfo {volume} {104}},\
  \bibinfo {pages} {137002} (\bibinfo {year} {2010})}\BibitemShut {NoStop}%
\bibitem [{\citenamefont {Valls}(2022)}]{Valls:2022}%
  \BibitemOpen
  \bibfield  {author} {\bibinfo {author} {\bibfnamefont {O.~T.}\ \bibnamefont
  {Valls}},\ }\href {https://doi.org/10.1142/12725} {\emph {\bibinfo {title}
  {Superconductor/Ferromagnet Nanostructures}}}\ (\bibinfo  {publisher} {World
  Scientific, Hackensack, NJ},\ \bibinfo {year} {2022})\BibitemShut {NoStop}%
\bibitem [{\citenamefont {Shen}\ \emph {et~al.}(2024)\citenamefont {Shen},
  \citenamefont {Han}, \citenamefont {Vezin}, \citenamefont {Alidoust},\ and\
  \citenamefont {{\v{Z}}uti\'c}}]{Shen2024:PRB}%
  \BibitemOpen
  \bibfield  {author} {\bibinfo {author} {\bibfnamefont {C.}~\bibnamefont
  {Shen}}, \bibinfo {author} {\bibfnamefont {J.~E.}\ \bibnamefont {Han}},
  \bibinfo {author} {\bibfnamefont {T.}~\bibnamefont {Vezin}}, \bibinfo
  {author} {\bibfnamefont {M.}~\bibnamefont {Alidoust}},\ and\ \bibinfo
  {author} {\bibfnamefont {I.}~\bibnamefont {{\v{Z}}uti\'c}},\ }\bibfield
  {title} {\bibinfo {title} {{Signatures of enhanced spin-triplet
  superconductivity induced by interfacial properties}},\ }\href
  {https://doi.org/10.1103/PhysRevB.110.104514} {\bibfield  {journal} {\bibinfo
   {journal} {Phys. Rev. B}\ }\textbf {\bibinfo {volume} {110}},\ \bibinfo
  {pages} {104514} (\bibinfo {year} {2024})}\BibitemShut {NoStop}%
\bibitem [{\citenamefont {Vezin}\ \emph {et~al.}(2020)\citenamefont {Vezin},
  \citenamefont {Shen}, \citenamefont {Han},\ and\ \citenamefont
  {{\v{Z}}uti\'c}}]{Vezin2020:PRB}%
  \BibitemOpen
  \bibfield  {author} {\bibinfo {author} {\bibfnamefont {T.}~\bibnamefont
  {Vezin}}, \bibinfo {author} {\bibfnamefont {C.}~\bibnamefont {Shen}},
  \bibinfo {author} {\bibfnamefont {J.~E.}\ \bibnamefont {Han}},\ and\ \bibinfo
  {author} {\bibfnamefont {I.}~\bibnamefont {{\v{Z}}uti\'c}},\ }\bibfield
  {title} {\bibinfo {title} {Enhanced spin-triplet pairing in magnetic
  junctions with $s$-wave superconductors},\ }\href
  {https://doi.org/10.1103/PhysRevB.101.014515} {\bibfield  {journal} {\bibinfo
   {journal} {Phys. Rev. B}\ }\textbf {\bibinfo {volume} {101}},\ \bibinfo
  {pages} {014515} (\bibinfo {year} {2020})}\BibitemShut {NoStop}%
\bibitem [{\citenamefont {Banerjee}\ \emph {et~al.}(2014)\citenamefont
  {Banerjee}, \citenamefont {Robinson},\ and\ \citenamefont
  {Blamire}}]{Banerjee2014:NC}%
  \BibitemOpen
  \bibfield  {author} {\bibinfo {author} {\bibfnamefont {N.}~\bibnamefont
  {Banerjee}}, \bibinfo {author} {\bibfnamefont {J.~W.~A.}\ \bibnamefont
  {Robinson}},\ and\ \bibinfo {author} {\bibfnamefont {M.~G.}\ \bibnamefont
  {Blamire}},\ }\bibfield  {title} {\bibinfo {title} {Reversible control of
  spin-polarized supercurrents in ferromagnetic {Josephson} junctions},\ }\href
  {https://doi.org/10.1038/ncomms5771} {\bibfield  {journal} {\bibinfo
  {journal} {Nat. Commun.}\ }\textbf {\bibinfo {volume} {5}},\ \bibinfo {pages}
  {4771} (\bibinfo {year} {2014})}\BibitemShut {NoStop}%
\bibitem [{\citenamefont {Duckheim}\ and\ \citenamefont
  {Brouwer}(2011)}]{Duckheim2011:PRB}%
  \BibitemOpen
  \bibfield  {author} {\bibinfo {author} {\bibfnamefont {M.}~\bibnamefont
  {Duckheim}}\ and\ \bibinfo {author} {\bibfnamefont {P.~W.}\ \bibnamefont
  {Brouwer}},\ }\bibfield  {title} {\bibinfo {title} {Andreev reflection from
  noncentrosymmetric superconductors and {Majorana} bound-state generation in
  half-metallic ferromagnets},\ }\href
  {https://doi.org/10.1103/PhysRevB.83.054513} {\bibfield  {journal} {\bibinfo
  {journal} {Phys. Rev. B}\ }\textbf {\bibinfo {volume} {83}},\ \bibinfo
  {pages} {054513} (\bibinfo {year} {2011})}\BibitemShut {NoStop}%
\bibitem [{\citenamefont {Nadj-Perge}\ \emph {et~al.}(2014)\citenamefont
  {Nadj-Perge}, \citenamefont {Drozdov}, \citenamefont {Li}, \citenamefont
  {Chen}, \citenamefont {Jeon}, \citenamefont {Seo}, \citenamefont {MacDonald},
  \citenamefont {Bernevig},\ and\ \citenamefont {Yazdani}}]{Nadj-Perge2014:S}%
  \BibitemOpen
  \bibfield  {author} {\bibinfo {author} {\bibfnamefont {S.}~\bibnamefont
  {Nadj-Perge}}, \bibinfo {author} {\bibfnamefont {I.~K.}\ \bibnamefont
  {Drozdov}}, \bibinfo {author} {\bibfnamefont {J.}~\bibnamefont {Li}},
  \bibinfo {author} {\bibfnamefont {H.}~\bibnamefont {Chen}}, \bibinfo {author}
  {\bibfnamefont {S.}~\bibnamefont {Jeon}}, \bibinfo {author} {\bibfnamefont
  {J.}~\bibnamefont {Seo}}, \bibinfo {author} {\bibfnamefont {A.~H.}\
  \bibnamefont {MacDonald}}, \bibinfo {author} {\bibfnamefont {B.~A.}\
  \bibnamefont {Bernevig}},\ and\ \bibinfo {author} {\bibfnamefont
  {A.}~\bibnamefont {Yazdani}},\ }\bibfield  {title} {\bibinfo {title}
  {Observation of {Majorana} fermions in ferromagnetic atomic chains on a
  superconductor},\ }\href {https://doi.org/10.1126/science.1259327} {\bibfield
   {journal} {\bibinfo  {journal} {Science}\ }\textbf {\bibinfo {volume}
  {346}},\ \bibinfo {pages} {602} (\bibinfo {year} {2014})}\BibitemShut
  {NoStop}%
\bibitem [{\citenamefont {Fatin}\ \emph {et~al.}(2016)\citenamefont {Fatin},
  \citenamefont {Matos-Abiague}, \citenamefont {Scharf},\ and\ \citenamefont
  {\ifmmode \check{Z}\else \v{Z}\fi{}uti\ifmmode~\acute{c}\else
  \'{c}\fi{}}}]{Fatin2016:PRL}%
  \BibitemOpen
  \bibfield  {author} {\bibinfo {author} {\bibfnamefont {G.~L.}\ \bibnamefont
  {Fatin}}, \bibinfo {author} {\bibfnamefont {A.}~\bibnamefont
  {Matos-Abiague}}, \bibinfo {author} {\bibfnamefont {B.}~\bibnamefont
  {Scharf}},\ and\ \bibinfo {author} {\bibfnamefont {I.}~\bibnamefont {\ifmmode
  \check{Z}\else \v{Z}\fi{}uti\ifmmode~\acute{c}\else \'{c}\fi{}}},\ }\bibfield
   {title} {\bibinfo {title} {Wireless {Majorana} bound states: From magnetic
  tunability to braiding},\ }\href
  {https://doi.org/10.1103/PhysRevLett.117.077002} {\bibfield  {journal}
  {\bibinfo  {journal} {Phys. Rev. Lett.}\ }\textbf {\bibinfo {volume} {117}},\
  \bibinfo {pages} {077002} (\bibinfo {year} {2016})}\BibitemShut {NoStop}%
\bibitem [{\citenamefont {Gonz\'{a}lez-Ruano}\ \emph
  {et~al.}(2025)\citenamefont {Gonz\'{a}lez-Ruano}, \citenamefont {Shen},
  \citenamefont {Tuero}, \citenamefont {Tiusan}, \citenamefont {Lu},
  \citenamefont {Han}, \citenamefont {{\v{Z}}uti\'c},\ and\ \citenamefont
  {Aliev}}]{Gonzalez-Ruano2025:NC}%
  \BibitemOpen
  \bibfield  {author} {\bibinfo {author} {\bibfnamefont {C.}~\bibnamefont
  {Gonz\'{a}lez-Ruano}}, \bibinfo {author} {\bibfnamefont {C.}~\bibnamefont
  {Shen}}, \bibinfo {author} {\bibfnamefont {P.}~\bibnamefont {Tuero}},
  \bibinfo {author} {\bibfnamefont {C.}~\bibnamefont {Tiusan}}, \bibinfo
  {author} {\bibfnamefont {Y.}~\bibnamefont {Lu}}, \bibinfo {author}
  {\bibfnamefont {J.~E.}\ \bibnamefont {Han}}, \bibinfo {author} {\bibfnamefont
  {I.}~\bibnamefont {{\v{Z}}uti\'c}},\ and\ \bibinfo {author} {\bibfnamefont
  {F.~G.}\ \bibnamefont {Aliev}},\ }\bibfield  {title} {\bibinfo {title}
  {{Giant shot noise in superconductor/ferromagnet junctions with
  orbital-symmetry-controlled spin-orbit coupling}},\ }\href
  {https://doi.org/10.1038/s41467-025-56456-2} {\bibfield  {journal} {\bibinfo
  {journal} {Nat. Commun.}\ }\textbf {\bibinfo {volume} {16}},\ \bibinfo
  {pages} {9524} (\bibinfo {year} {2025})}\BibitemShut {NoStop}%
\bibitem [{\citenamefont {G{\"u}ng{\"o}rd{\"u}}\ and\ \citenamefont
  {Kovalev}(2022)}]{Gungordu2022Majorana}%
  \BibitemOpen
  \bibfield  {author} {\bibinfo {author} {\bibfnamefont {U.}~\bibnamefont
  {G{\"u}ng{\"o}rd{\"u}}}\ and\ \bibinfo {author} {\bibfnamefont {A.~A.}\
  \bibnamefont {Kovalev}},\ }\bibfield  {title} {\bibinfo {title} {Majorana
  bound states with chiral magnetic textures},\ }\href
  {https://doi.org/10.1063/5.0097008} {\bibfield  {journal} {\bibinfo
  {journal} {J. Appl. Phys.}\ }\textbf {\bibinfo {volume} {132}},\ \bibinfo
  {pages} {041101} (\bibinfo {year} {2022})}\BibitemShut {NoStop}%
\bibitem [{\citenamefont {Konakanchi}\ \emph {et~al.}(2023)\citenamefont
  {Konakanchi}, \citenamefont {V\"ayrynen}, \citenamefont {Chen}, \citenamefont
  {Upadhyaya},\ and\ \citenamefont {Rokhinson}}]{PhysRevResearch.5.033109}%
  \BibitemOpen
  \bibfield  {author} {\bibinfo {author} {\bibfnamefont {S.~T.}\ \bibnamefont
  {Konakanchi}}, \bibinfo {author} {\bibfnamefont {J.~I.}\ \bibnamefont
  {V\"ayrynen}}, \bibinfo {author} {\bibfnamefont {Y.~P.}\ \bibnamefont
  {Chen}}, \bibinfo {author} {\bibfnamefont {P.}~\bibnamefont {Upadhyaya}},\
  and\ \bibinfo {author} {\bibfnamefont {L.~P.}\ \bibnamefont {Rokhinson}},\
  }\bibfield  {title} {\bibinfo {title} {Platform for braiding majorana modes
  with magnetic skyrmions},\ }\href
  {https://doi.org/10.1103/PhysRevResearch.5.033109} {\bibfield  {journal}
  {\bibinfo  {journal} {Phys. Rev. Res.}\ }\textbf {\bibinfo {volume} {5}},\
  \bibinfo {pages} {033109} (\bibinfo {year} {2023})}\BibitemShut {NoStop}%
\bibitem [{\citenamefont {Vakili}\ \emph {et~al.}(2024)\citenamefont {Vakili},
  \citenamefont {Ali},\ and\ \citenamefont {Kovalev}}]{VakiliJDE}%
  \BibitemOpen
  \bibfield  {author} {\bibinfo {author} {\bibfnamefont {H.}~\bibnamefont
  {Vakili}}, \bibinfo {author} {\bibfnamefont {M.}~\bibnamefont {Ali}},\ and\
  \bibinfo {author} {\bibfnamefont {A.~A.}\ \bibnamefont {Kovalev}},\
  }\bibfield  {title} {\bibinfo {title} {Field-free {Josephson} diode effect in
  a $d$-wave superconductor heterostructure},\ }\href
  {https://doi.org/10.1103/PhysRevB.110.104518} {\bibfield  {journal} {\bibinfo
   {journal} {Phys. Rev. B}\ }\textbf {\bibinfo {volume} {110}},\ \bibinfo
  {pages} {104518} (\bibinfo {year} {2024})}\BibitemShut {NoStop}%
\bibitem [{\citenamefont {Vakili}\ \emph {et~al.}(2023)\citenamefont {Vakili},
  \citenamefont {Ali}, \citenamefont {Elekhtiar},\ and\ \citenamefont
  {Kovalev}}]{VakiliMajorana}%
  \BibitemOpen
  \bibfield  {author} {\bibinfo {author} {\bibfnamefont {H.}~\bibnamefont
  {Vakili}}, \bibinfo {author} {\bibfnamefont {M.}~\bibnamefont {Ali}},
  \bibinfo {author} {\bibfnamefont {M.}~\bibnamefont {Elekhtiar}},\ and\
  \bibinfo {author} {\bibfnamefont {A.~A.}\ \bibnamefont {Kovalev}},\
  }\bibfield  {title} {\bibinfo {title} {Majorana bound states in a $d$-wave
  superconductor planar {Josephson} junction},\ }\href
  {https://doi.org/10.1103/PhysRevB.108.L140506} {\bibfield  {journal}
  {\bibinfo  {journal} {Phys. Rev. B}\ }\textbf {\bibinfo {volume} {108}},\
  \bibinfo {pages} {L140506} (\bibinfo {year} {2023})}\BibitemShut {NoStop}%
\bibitem [{\citenamefont {Aronov}\ and\ \citenamefont
  {Lyanda-Geller}(1989)}]{AronovLyandaGeller1989}%
  \BibitemOpen
  \bibfield  {author} {\bibinfo {author} {\bibfnamefont {A.~G.}\ \bibnamefont
  {Aronov}}\ and\ \bibinfo {author} {\bibfnamefont {Y.~B.}\ \bibnamefont
  {Lyanda-Geller}},\ }\bibfield  {title} {\bibinfo {title} {Nuclear electric
  resonance and orientation of carrier spins by an electric field},\ }\href
  {https://doi.org/10.1134/1.13586} {\bibfield  {journal} {\bibinfo  {journal}
  {JETP Lett.}\ }\textbf {\bibinfo {volume} {50}},\ \bibinfo {pages} {431}
  (\bibinfo {year} {1989})},\ \bibinfo {note} {[Pis'ma Zh. Eksp. Teor. Fiz. 50,
  398--400 (1989)]}\BibitemShut {NoStop}%
\bibitem [{\citenamefont {Edelstein}(1990)}]{Edelstein1990}%
  \BibitemOpen
  \bibfield  {author} {\bibinfo {author} {\bibfnamefont {V.~M.}\ \bibnamefont
  {Edelstein}},\ }\bibfield  {title} {\bibinfo {title} {Spin polarization of
  conduction electrons induced by electric current in two-dimensional
  asymmetric electron systems},\ }\href
  {https://doi.org/10.1016/0038-1098(90)90963-C} {\bibfield  {journal}
  {\bibinfo  {journal} {Solid State Commun.}\ }\textbf {\bibinfo {volume}
  {73}},\ \bibinfo {pages} {233} (\bibinfo {year} {1990})}\BibitemShut
  {NoStop}%
\bibitem [{\citenamefont {Levitov}\ \emph {et~al.}(1985)\citenamefont
  {Levitov}, \citenamefont {Nazarov},\ and\ \citenamefont
  {Eliashberg}}]{Levitov1985:SPJETP}%
  \BibitemOpen
  \bibfield  {author} {\bibinfo {author} {\bibfnamefont {L.~S.}\ \bibnamefont
  {Levitov}}, \bibinfo {author} {\bibfnamefont {Y.~V.}\ \bibnamefont
  {Nazarov}},\ and\ \bibinfo {author} {\bibfnamefont {G.~M.}\ \bibnamefont
  {Eliashberg}},\ }\bibfield  {title} {\bibinfo {title} {{Magnetoelectric
  effects in conductors with mirror isomer symmetry}},\ }\href
  {http://www.jetp.ras.ru/cgi-bin/dn/e_061_01_0133.pdf} {\bibfield  {journal}
  {\bibinfo  {journal} {Sov. Phys. JETP}\ }\textbf {\bibinfo {volume} {61}},\
  \bibinfo {pages} {133} (\bibinfo {year} {1985})}\BibitemShut {NoStop}%
\bibitem [{\citenamefont {Edelstein}(1995)}]{Edelstein1995}%
  \BibitemOpen
  \bibfield  {author} {\bibinfo {author} {\bibfnamefont {V.~M.}\ \bibnamefont
  {Edelstein}},\ }\bibfield  {title} {\bibinfo {title} {Magnetoelectric effect
  in polar superconductors},\ }\href
  {https://doi.org/10.1103/PhysRevLett.75.2004} {\bibfield  {journal} {\bibinfo
   {journal} {Phys. Rev. Lett.}\ }\textbf {\bibinfo {volume} {75}},\ \bibinfo
  {pages} {2004} (\bibinfo {year} {1995})}\BibitemShut {NoStop}%
\bibitem [{\citenamefont {Chernyshov}\ \emph {et~al.}(2009)\citenamefont
  {Chernyshov}, \citenamefont {Overby}, \citenamefont {Liu}, \citenamefont
  {Furdyna}, \citenamefont {Lyanda-Geller},\ and\ \citenamefont
  {Rokhinson}}]{Chernyshov2009:NP}%
  \BibitemOpen
  \bibfield  {author} {\bibinfo {author} {\bibfnamefont {A.}~\bibnamefont
  {Chernyshov}}, \bibinfo {author} {\bibfnamefont {M.}~\bibnamefont {Overby}},
  \bibinfo {author} {\bibfnamefont {X.}~\bibnamefont {Liu}}, \bibinfo {author}
  {\bibfnamefont {J.~K.}\ \bibnamefont {Furdyna}}, \bibinfo {author}
  {\bibfnamefont {Y.}~\bibnamefont {Lyanda-Geller}},\ and\ \bibinfo {author}
  {\bibfnamefont {L.~P.}\ \bibnamefont {Rokhinson}},\ }\bibfield  {title}
  {\bibinfo {title} {{Evidence for reversible control of magnetization in a
  ferromagnetic material by means of spin--orbit magnetic field}},\ }\href
  {https://doi.org/10.1038/nphys1362} {\bibfield  {journal} {\bibinfo
  {journal} {Nat. Phys.}\ }\textbf {\bibinfo {volume} {5}},\ \bibinfo {pages}
  {656} (\bibinfo {year} {2009})}\BibitemShut {NoStop}%
\bibitem [{\citenamefont {H\"{o}gl}\ \emph {et~al.}(2015)\citenamefont
  {H\"{o}gl}, \citenamefont {Matos-Abiague}, \citenamefont {{\v{Z}}uti\'c},\
  and\ \citenamefont {Fabian}}]{Hogl2015:PRL}%
  \BibitemOpen
  \bibfield  {author} {\bibinfo {author} {\bibfnamefont {P.}~\bibnamefont
  {H\"{o}gl}}, \bibinfo {author} {\bibfnamefont {A.}~\bibnamefont
  {Matos-Abiague}}, \bibinfo {author} {\bibfnamefont {I.}~\bibnamefont
  {{\v{Z}}uti\'c}},\ and\ \bibinfo {author} {\bibfnamefont {J.}~\bibnamefont
  {Fabian}},\ }\bibfield  {title} {\bibinfo {title} {Magnetoanisotropic
  {Andreev} reflection in ferromagnet/superconductor junctions},\ }\href
  {https://doi.org/10.1103/PhysRevLett.115.116601} {\bibfield  {journal}
  {\bibinfo  {journal} {Phys. Rev. Lett.}\ }\textbf {\bibinfo {volume} {115}},\
  \bibinfo {pages} {116601} (\bibinfo {year} {2015})}\BibitemShut {NoStop}%
\bibitem [{\citenamefont {Cai}\ \emph {et~al.}(2021)\citenamefont {Cai},
  \citenamefont {Yao}, \citenamefont {Lv}, \citenamefont {Ma}, \citenamefont
  {Xing}, \citenamefont {Li}, \citenamefont {Ji}, \citenamefont {Zhou},
  \citenamefont {Shen}, \citenamefont {Jia}, \citenamefont {Xie}, \citenamefont
  {{\v{Z}}uti{\'c}}, \citenamefont {Sun},\ and\ \citenamefont
  {Han}}]{Cai2021:NC}%
  \BibitemOpen
  \bibfield  {author} {\bibinfo {author} {\bibfnamefont {R.}~\bibnamefont
  {Cai}}, \bibinfo {author} {\bibfnamefont {Y.}~\bibnamefont {Yao}}, \bibinfo
  {author} {\bibfnamefont {P.}~\bibnamefont {Lv}}, \bibinfo {author}
  {\bibfnamefont {Y.}~\bibnamefont {Ma}}, \bibinfo {author} {\bibfnamefont
  {W.}~\bibnamefont {Xing}}, \bibinfo {author} {\bibfnamefont {B.}~\bibnamefont
  {Li}}, \bibinfo {author} {\bibfnamefont {Y.}~\bibnamefont {Ji}}, \bibinfo
  {author} {\bibfnamefont {H.}~\bibnamefont {Zhou}}, \bibinfo {author}
  {\bibfnamefont {C.}~\bibnamefont {Shen}}, \bibinfo {author} {\bibfnamefont
  {S.}~\bibnamefont {Jia}}, \bibinfo {author} {\bibfnamefont {X.~C.}\
  \bibnamefont {Xie}}, \bibinfo {author} {\bibfnamefont {I.}~\bibnamefont
  {{\v{Z}}uti{\'c}}}, \bibinfo {author} {\bibfnamefont {Q.-F.}\ \bibnamefont
  {Sun}},\ and\ \bibinfo {author} {\bibfnamefont {W.}~\bibnamefont {Han}},\
  }\bibfield  {title} {\bibinfo {title} {Evidence for anisotropic spin-triplet
  {Andreev} reflection at the {2D} {van der Waals} ferromagnet/superconductor
  interface},\ }\href {https://doi.org/10.1038/s41467-021-27041-w} {\bibfield
  {journal} {\bibinfo  {journal} {Nat. Commun.}\ }\textbf {\bibinfo {volume}
  {12}},\ \bibinfo {pages} {6725} (\bibinfo {year} {2021})}\BibitemShut
  {NoStop}%
\bibitem [{\citenamefont {{Mart{i}nez}}\ \emph {et~al.}(2020)\citenamefont
  {{Mart{i}nez}}, \citenamefont {H\"ogl}, \citenamefont {Gonz\'alez-Ruano},
  \citenamefont {Cascales}, \citenamefont {Tiusan}, \citenamefont {Lu},
  \citenamefont {Hehn}, \citenamefont {Matos-Abiague}, \citenamefont {Fabian},
  \citenamefont {\ifmmode \check{Z}\else \v{Z}\fi{}uti\ifmmode~\acute{c}\else
  \'{c}\fi{}},\ and\ \citenamefont {Aliev}}]{Martinez2020:PRA}%
  \BibitemOpen
  \bibfield  {author} {\bibinfo {author} {\bibfnamefont {I.}~\bibnamefont
  {{Mart{i}nez}}}, \bibinfo {author} {\bibfnamefont {P.}~\bibnamefont
  {H\"ogl}}, \bibinfo {author} {\bibfnamefont {C.}~\bibnamefont
  {Gonz\'alez-Ruano}}, \bibinfo {author} {\bibfnamefont {J.~P.}\ \bibnamefont
  {Cascales}}, \bibinfo {author} {\bibfnamefont {C.}~\bibnamefont {Tiusan}},
  \bibinfo {author} {\bibfnamefont {Y.}~\bibnamefont {Lu}}, \bibinfo {author}
  {\bibfnamefont {M.}~\bibnamefont {Hehn}}, \bibinfo {author} {\bibfnamefont
  {A.}~\bibnamefont {Matos-Abiague}}, \bibinfo {author} {\bibfnamefont
  {J.}~\bibnamefont {Fabian}}, \bibinfo {author} {\bibfnamefont
  {I.}~\bibnamefont {\ifmmode \check{Z}\else
  \v{Z}\fi{}uti\ifmmode~\acute{c}\else \'{c}\fi{}}},\ and\ \bibinfo {author}
  {\bibfnamefont {F.~G.}\ \bibnamefont {Aliev}},\ }\bibfield  {title} {\bibinfo
  {title} {Interfacial spin-orbit coupling: {A} platform for superconducting
  spintronics},\ }\href {https://doi.org/10.1103/PhysRevApplied.13.014030}
  {\bibfield  {journal} {\bibinfo  {journal} {Phys. Rev. Applied}\ }\textbf
  {\bibinfo {volume} {13}},\ \bibinfo {pages} {014030} (\bibinfo {year}
  {2020})}\BibitemShut {NoStop}%
\bibitem [{\citenamefont {Miron}\ \emph {et~al.}(2011)\citenamefont {Miron},
  \citenamefont {Garello}, \citenamefont {Gaudin}, \citenamefont {Zermatten},
  \citenamefont {Costache}, \citenamefont {Auffret}, \citenamefont {Bandiera},
  \citenamefont {Rodmacq}, \citenamefont {Schuhl},\ and\ \citenamefont
  {Gambardella}}]{Miron2011}%
  \BibitemOpen
  \bibfield  {author} {\bibinfo {author} {\bibfnamefont {I.~M.}\ \bibnamefont
  {Miron}}, \bibinfo {author} {\bibfnamefont {K.}~\bibnamefont {Garello}},
  \bibinfo {author} {\bibfnamefont {G.}~\bibnamefont {Gaudin}}, \bibinfo
  {author} {\bibfnamefont {P.-J.}\ \bibnamefont {Zermatten}}, \bibinfo {author}
  {\bibfnamefont {M.~V.}\ \bibnamefont {Costache}}, \bibinfo {author}
  {\bibfnamefont {S.}~\bibnamefont {Auffret}}, \bibinfo {author} {\bibfnamefont
  {S.}~\bibnamefont {Bandiera}}, \bibinfo {author} {\bibfnamefont
  {B.}~\bibnamefont {Rodmacq}}, \bibinfo {author} {\bibfnamefont
  {A.}~\bibnamefont {Schuhl}},\ and\ \bibinfo {author} {\bibfnamefont
  {P.}~\bibnamefont {Gambardella}},\ }\bibfield  {title} {\bibinfo {title}
  {Perpendicular switching of a single ferromagnetic layer induced by in-plane
  current injection},\ }\href {https://doi.org/10.1038/nature10309} {\bibfield
  {journal} {\bibinfo  {journal} {Nature}\ }\textbf {\bibinfo {volume} {476}},\
  \bibinfo {pages} {189} (\bibinfo {year} {2011})}\BibitemShut {NoStop}%
\bibitem [{\citenamefont {Manchon}\ \emph {et~al.}(2019)\citenamefont
  {Manchon}, \citenamefont {{\v{Z}}elezn\'{y}}, \citenamefont {Miron},
  \citenamefont {Jungwirth}, \citenamefont {Sinova}, \citenamefont {Thiaville},
  \citenamefont {Garello},\ and\ \citenamefont {Gambardella}}]{Manchon2019}%
  \BibitemOpen
  \bibfield  {author} {\bibinfo {author} {\bibfnamefont {A.}~\bibnamefont
  {Manchon}}, \bibinfo {author} {\bibfnamefont {J.}~\bibnamefont
  {{\v{Z}}elezn\'{y}}}, \bibinfo {author} {\bibfnamefont {I.~M.}\ \bibnamefont
  {Miron}}, \bibinfo {author} {\bibfnamefont {T.}~\bibnamefont {Jungwirth}},
  \bibinfo {author} {\bibfnamefont {J.}~\bibnamefont {Sinova}}, \bibinfo
  {author} {\bibfnamefont {A.}~\bibnamefont {Thiaville}}, \bibinfo {author}
  {\bibfnamefont {K.}~\bibnamefont {Garello}},\ and\ \bibinfo {author}
  {\bibfnamefont {P.}~\bibnamefont {Gambardella}},\ }\bibfield  {title}
  {\bibinfo {title} {Current-induced spin-orbit torques in ferromagnetic and
  antiferromagnetic systems},\ }\href
  {https://doi.org/10.1103/RevModPhys.91.035004} {\bibfield  {journal}
  {\bibinfo  {journal} {Rev. Mod. Phys.}\ }\textbf {\bibinfo {volume} {91}},\
  \bibinfo {pages} {035004} (\bibinfo {year} {2019})}\BibitemShut {NoStop}%
\bibitem [{\citenamefont {Tsymbal}\ and\ \citenamefont
  {{\v{Z}}uti\'c}(2019)}]{Tsymbal:2019}%
  \BibitemOpen
  \bibinfo {editor} {\bibfnamefont {E.~Y.}\ \bibnamefont {Tsymbal}}\ and\
  \bibinfo {editor} {\bibfnamefont {I.}~\bibnamefont {{\v{Z}}uti\'c}},\ eds.,\
  \href {https://doi.org/10.1201/9780429423079} {\emph {\bibinfo {title}
  {{Spintronics Handbook Spin Transport and Magnetism, 2nd Ed.}}}}\ (\bibinfo
  {publisher} {CRC Press, Taylor \& Francis, Boca Raton, FL},\ \bibinfo {year}
  {2019})\BibitemShut {NoStop}%
\bibitem [{\citenamefont {Wadley}\ \emph {et~al.}(2016)\citenamefont {Wadley},
  \citenamefont {Howells}, \citenamefont {{\v{Z}}elezn\'{y}}, \citenamefont
  {Andrews}, \citenamefont {Hills}, \citenamefont {Campion}, \citenamefont
  {Nov\'{a}k}, \citenamefont {Olejn\'{i}k}, \citenamefont {Maccherozzi},
  \citenamefont {Dhesi}, \citenamefont {Martin}, \citenamefont {Wagner},
  \citenamefont {Wunderlich}, \citenamefont {Freimuth}, \citenamefont
  {Mokrousov}, \citenamefont {Kune{\v{s}}}, \citenamefont {Chauhan},
  \citenamefont {Grzybowski}, \citenamefont {Rushforth}, \citenamefont
  {Edmonds}, \citenamefont {Gallagher},\ and\ \citenamefont
  {Jungwirth}}]{Wadley2016}%
  \BibitemOpen
  \bibfield  {author} {\bibinfo {author} {\bibfnamefont {P.}~\bibnamefont
  {Wadley}}, \bibinfo {author} {\bibfnamefont {B.}~\bibnamefont {Howells}},
  \bibinfo {author} {\bibfnamefont {J.}~\bibnamefont {{\v{Z}}elezn\'{y}}},
  \bibinfo {author} {\bibfnamefont {C.}~\bibnamefont {Andrews}}, \bibinfo
  {author} {\bibfnamefont {V.}~\bibnamefont {Hills}}, \bibinfo {author}
  {\bibfnamefont {R.~P.}\ \bibnamefont {Campion}}, \bibinfo {author}
  {\bibfnamefont {V.}~\bibnamefont {Nov\'{a}k}}, \bibinfo {author}
  {\bibfnamefont {K.}~\bibnamefont {Olejn\'{i}k}}, \bibinfo {author}
  {\bibfnamefont {F.}~\bibnamefont {Maccherozzi}}, \bibinfo {author}
  {\bibfnamefont {S.~S.}\ \bibnamefont {Dhesi}}, \bibinfo {author}
  {\bibfnamefont {S.~Y.}\ \bibnamefont {Martin}}, \bibinfo {author}
  {\bibfnamefont {T.}~\bibnamefont {Wagner}}, \bibinfo {author} {\bibfnamefont
  {J.}~\bibnamefont {Wunderlich}}, \bibinfo {author} {\bibfnamefont
  {F.}~\bibnamefont {Freimuth}}, \bibinfo {author} {\bibfnamefont
  {Y.}~\bibnamefont {Mokrousov}}, \bibinfo {author} {\bibfnamefont
  {J.}~\bibnamefont {Kune{\v{s}}}}, \bibinfo {author} {\bibfnamefont {J.~S.}\
  \bibnamefont {Chauhan}}, \bibinfo {author} {\bibfnamefont {M.~J.}\
  \bibnamefont {Grzybowski}}, \bibinfo {author} {\bibfnamefont {A.~W.}\
  \bibnamefont {Rushforth}}, \bibinfo {author} {\bibfnamefont {K.~W.}\
  \bibnamefont {Edmonds}}, \bibinfo {author} {\bibfnamefont {B.~L.}\
  \bibnamefont {Gallagher}},\ and\ \bibinfo {author} {\bibfnamefont
  {T.}~\bibnamefont {Jungwirth}},\ }\bibfield  {title} {\bibinfo {title}
  {Electrical switching of an antiferromagnet},\ }\href
  {https://doi.org/10.1126/science.aab1031} {\bibfield  {journal} {\bibinfo
  {journal} {Science}\ }\textbf {\bibinfo {volume} {351}},\ \bibinfo {pages}
  {587} (\bibinfo {year} {2016})}\BibitemShut {NoStop}%
\bibitem [{\citenamefont {Yang}\ \emph {et~al.}(2022)\citenamefont {Yang},
  \citenamefont {Valenzuela}, \citenamefont {Chshiev}, \citenamefont {Couet},
  \citenamefont {Dieny}, \citenamefont {Dlubak}, \citenamefont {Fert},
  \citenamefont {Garello}, \citenamefont {Jamet}, \citenamefont {Jeong},
  \citenamefont {Lee}, \citenamefont {Lee}, \citenamefont {Martin},
  \citenamefont {Kar}, \citenamefont {S{\'e}n{\'e}or}, \citenamefont {Shin},\
  and\ \citenamefont {Roche}}]{Yang2022:N}%
  \BibitemOpen
  \bibfield  {author} {\bibinfo {author} {\bibfnamefont {H.}~\bibnamefont
  {Yang}}, \bibinfo {author} {\bibfnamefont {S.~O.}\ \bibnamefont
  {Valenzuela}}, \bibinfo {author} {\bibfnamefont {M.}~\bibnamefont {Chshiev}},
  \bibinfo {author} {\bibfnamefont {S.}~\bibnamefont {Couet}}, \bibinfo
  {author} {\bibfnamefont {B.}~\bibnamefont {Dieny}}, \bibinfo {author}
  {\bibfnamefont {B.}~\bibnamefont {Dlubak}}, \bibinfo {author} {\bibfnamefont
  {A.}~\bibnamefont {Fert}}, \bibinfo {author} {\bibfnamefont {K.}~\bibnamefont
  {Garello}}, \bibinfo {author} {\bibfnamefont {M.}~\bibnamefont {Jamet}},
  \bibinfo {author} {\bibfnamefont {D.-E.}\ \bibnamefont {Jeong}}, \bibinfo
  {author} {\bibfnamefont {K.}~\bibnamefont {Lee}}, \bibinfo {author}
  {\bibfnamefont {T.}~\bibnamefont {Lee}}, \bibinfo {author} {\bibfnamefont
  {M.-B.}\ \bibnamefont {Martin}}, \bibinfo {author} {\bibfnamefont {G.~S.}\
  \bibnamefont {Kar}}, \bibinfo {author} {\bibfnamefont {P.}~\bibnamefont
  {S{\'e}n{\'e}or}}, \bibinfo {author} {\bibfnamefont {H.-J.}\ \bibnamefont
  {Shin}},\ and\ \bibinfo {author} {\bibfnamefont {S.}~\bibnamefont {Roche}},\
  }\bibfield  {title} {\bibinfo {title} {Two-dimensional materials prospects
  for non-volatile spintronic memories},\ }\href
  {https://doi.org/10.1038/s41586-022-04768-0} {\bibfield  {journal} {\bibinfo
  {journal} {Nature}\ }\textbf {\bibinfo {volume} {606}},\ \bibinfo {pages}
  {663} (\bibinfo {year} {2022})}\BibitemShut {NoStop}%
\bibitem [{\citenamefont {Liang}\ \emph {et~al.}(2025)\citenamefont {Liang},
  \citenamefont {Huang}, \citenamefont {Tan}, \citenamefont {Tang},\ and\
  \citenamefont {Xie}}]{Liang2025:JPDAP}%
  \BibitemOpen
  \bibfield  {author} {\bibinfo {author} {\bibfnamefont {Q.}~\bibnamefont
  {Liang}}, \bibinfo {author} {\bibfnamefont {Y.}~\bibnamefont {Huang}},
  \bibinfo {author} {\bibfnamefont {Y.}~\bibnamefont {Tan}}, \bibinfo {author}
  {\bibfnamefont {Y.}~\bibnamefont {Tang}},\ and\ \bibinfo {author}
  {\bibfnamefont {X.}~\bibnamefont {Xie}},\ }\bibfield  {title} {\bibinfo
  {title} {{Recent progress in neuromorphic computing based on spin--orbit
  torque device}},\ }\href {https://doi.org/10.1088/1361-6463/adaa3a}
  {\bibfield  {journal} {\bibinfo  {journal} {J. Phys. D: Appl. Phys.}\
  }\textbf {\bibinfo {volume} {58}},\ \bibinfo {pages} {443001} (\bibinfo
  {year} {2025})}\BibitemShut {NoStop}%
\bibitem [{\citenamefont {Dainone}\ \emph {et~al.}(2024)\citenamefont
  {Dainone}, \citenamefont {Prestes}, \citenamefont {Renucci}, \citenamefont
  {Bouch{\'e}}, \citenamefont {Morassi}, \citenamefont {Devaux}, \citenamefont
  {Lindemann}, \citenamefont {George}, \citenamefont {Jaffr{\`e}s},
  \citenamefont {Lemaitre}, \citenamefont {Xu}, \citenamefont {Stoffel},
  \citenamefont {Chen}, \citenamefont {Lombez}, \citenamefont {Lagarde},
  \citenamefont {Cong}, \citenamefont {Ma}, \citenamefont {Pigeat},
  \citenamefont {Vergnat}, \citenamefont {Rinnert}, \citenamefont {Marie},
  \citenamefont {Han}, \citenamefont {Mangin}, \citenamefont
  {Rojas-S{\'a}nchez}, \citenamefont {Wang}, \citenamefont {Beard},
  \citenamefont {{\v{Z}}uti{\'c}},\ and\ \citenamefont {Lu}}]{Dainone2024:N}%
  \BibitemOpen
  \bibfield  {author} {\bibinfo {author} {\bibfnamefont {P.~A.}\ \bibnamefont
  {Dainone}}, \bibinfo {author} {\bibfnamefont {N.~F.}\ \bibnamefont
  {Prestes}}, \bibinfo {author} {\bibfnamefont {P.}~\bibnamefont {Renucci}},
  \bibinfo {author} {\bibfnamefont {A.}~\bibnamefont {Bouch{\'e}}}, \bibinfo
  {author} {\bibfnamefont {M.}~\bibnamefont {Morassi}}, \bibinfo {author}
  {\bibfnamefont {X.}~\bibnamefont {Devaux}}, \bibinfo {author} {\bibfnamefont
  {M.}~\bibnamefont {Lindemann}}, \bibinfo {author} {\bibfnamefont {J.-M.}\
  \bibnamefont {George}}, \bibinfo {author} {\bibfnamefont {H.}~\bibnamefont
  {Jaffr{\`e}s}}, \bibinfo {author} {\bibfnamefont {A.}~\bibnamefont
  {Lemaitre}}, \bibinfo {author} {\bibfnamefont {B.}~\bibnamefont {Xu}},
  \bibinfo {author} {\bibfnamefont {M.}~\bibnamefont {Stoffel}}, \bibinfo
  {author} {\bibfnamefont {T.}~\bibnamefont {Chen}}, \bibinfo {author}
  {\bibfnamefont {L.}~\bibnamefont {Lombez}}, \bibinfo {author} {\bibfnamefont
  {D.}~\bibnamefont {Lagarde}}, \bibinfo {author} {\bibfnamefont
  {G.}~\bibnamefont {Cong}}, \bibinfo {author} {\bibfnamefont {T.}~\bibnamefont
  {Ma}}, \bibinfo {author} {\bibfnamefont {P.}~\bibnamefont {Pigeat}}, \bibinfo
  {author} {\bibfnamefont {M.}~\bibnamefont {Vergnat}}, \bibinfo {author}
  {\bibfnamefont {H.}~\bibnamefont {Rinnert}}, \bibinfo {author} {\bibfnamefont
  {X.}~\bibnamefont {Marie}}, \bibinfo {author} {\bibfnamefont
  {X.}~\bibnamefont {Han}}, \bibinfo {author} {\bibfnamefont {S.}~\bibnamefont
  {Mangin}}, \bibinfo {author} {\bibfnamefont {J.-C.}\ \bibnamefont
  {Rojas-S{\'a}nchez}}, \bibinfo {author} {\bibfnamefont {J.-P.}\ \bibnamefont
  {Wang}}, \bibinfo {author} {\bibfnamefont {N.~C.}\ \bibnamefont {Beard}},
  \bibinfo {author} {\bibfnamefont {I.}~\bibnamefont {{\v{Z}}uti{\'c}}},\ and\
  \bibinfo {author} {\bibfnamefont {Y.}~\bibnamefont {Lu}},\ }\bibfield
  {title} {\bibinfo {title} {Controlling the helicity of light by electrical
  magnetization switching},\ }\href
  {https://doi.org/10.1038/s41586-024-07125-5} {\bibfield  {journal} {\bibinfo
  {journal} {Nature}\ }\textbf {\bibinfo {volume} {627}},\ \bibinfo {pages}
  {783} (\bibinfo {year} {2024})}\BibitemShut {NoStop}%
\bibitem [{\citenamefont {Gomonay}\ \emph {et~al.}(2016)\citenamefont
  {Gomonay}, \citenamefont {Jungwirth},\ and\ \citenamefont
  {Sinova}}]{PhysRevLett.117.017202}%
  \BibitemOpen
  \bibfield  {author} {\bibinfo {author} {\bibfnamefont {O.}~\bibnamefont
  {Gomonay}}, \bibinfo {author} {\bibfnamefont {T.}~\bibnamefont {Jungwirth}},\
  and\ \bibinfo {author} {\bibfnamefont {J.}~\bibnamefont {Sinova}},\
  }\bibfield  {title} {\bibinfo {title} {High antiferromagnetic domain wall
  velocity induced by {N\'eel} spin-orbit torques},\ }\href
  {https://doi.org/10.1103/PhysRevLett.117.017202} {\bibfield  {journal}
  {\bibinfo  {journal} {Phys. Rev. Lett.}\ }\textbf {\bibinfo {volume} {117}},\
  \bibinfo {pages} {017202} (\bibinfo {year} {2016})}\BibitemShut {NoStop}%
\bibitem [{\citenamefont {Hals}(2016)}]{hals}%
  \BibitemOpen
  \bibfield  {author} {\bibinfo {author} {\bibfnamefont {K.~M.~D.}\
  \bibnamefont {Hals}},\ }\bibfield  {title} {\bibinfo {title}
  {Supercurrent-induced spin-orbit torques},\ }\href
  {https://doi.org/10.1103/PhysRevB.93.115431} {\bibfield  {journal} {\bibinfo
  {journal} {Phys. Rev. B}\ }\textbf {\bibinfo {volume} {93}},\ \bibinfo
  {pages} {115431} (\bibinfo {year} {2016})}\BibitemShut {NoStop}%
\bibitem [{\citenamefont {Cai}\ \emph {et~al.}(2023)\citenamefont {Cai},
  \citenamefont {{\v{Z}}uti\'c},\ and\ \citenamefont {Han}}]{Cai2023:AQT}%
  \BibitemOpen
  \bibfield  {author} {\bibinfo {author} {\bibfnamefont {R.}~\bibnamefont
  {Cai}}, \bibinfo {author} {\bibfnamefont {I.}~\bibnamefont {{\v{Z}}uti\'c}},\
  and\ \bibinfo {author} {\bibfnamefont {W.}~\bibnamefont {Han}},\ }\bibfield
  {title} {\bibinfo {title} {Superconductor/ferromagnet heterostructures: {A}
  platform for superconducting spintronics and quantum computation},\ }\href
  {https://doi.org/10.1002/qute.202200080} {\bibfield  {journal} {\bibinfo
  {journal} {Adv. Quantum Technol.}\ }\textbf {\bibinfo {volume} {6}},\
  \bibinfo {pages} {2200080} (\bibinfo {year} {2023})}\BibitemShut {NoStop}%
\bibitem [{\citenamefont {Roig}\ \emph {et~al.}(2024)\citenamefont {Roig},
  \citenamefont {Kreisel}, \citenamefont {Yu}, \citenamefont {Andersen},\ and\
  \citenamefont {Agterberg}}]{Minimal_models}%
  \BibitemOpen
  \bibfield  {author} {\bibinfo {author} {\bibfnamefont {M.}~\bibnamefont
  {Roig}}, \bibinfo {author} {\bibfnamefont {A.}~\bibnamefont {Kreisel}},
  \bibinfo {author} {\bibfnamefont {Y.}~\bibnamefont {Yu}}, \bibinfo {author}
  {\bibfnamefont {B.~M.}\ \bibnamefont {Andersen}},\ and\ \bibinfo {author}
  {\bibfnamefont {D.~F.}\ \bibnamefont {Agterberg}},\ }\bibfield  {title}
  {\bibinfo {title} {Minimal models for altermagnetism},\ }\href
  {https://doi.org/10.1103/PhysRevB.110.144412} {\bibfield  {journal} {\bibinfo
   {journal} {Phys. Rev. B}\ }\textbf {\bibinfo {volume} {110}},\ \bibinfo
  {pages} {144412} (\bibinfo {year} {2024})}\BibitemShut {NoStop}%
\bibitem [{\citenamefont {Tani}\ and\ \citenamefont
  {Z\"ulicke}(2025)}]{lh4w-byk7}%
  \BibitemOpen
  \bibfield  {author} {\bibinfo {author} {\bibfnamefont {T.}~\bibnamefont
  {Tani}}\ and\ \bibinfo {author} {\bibfnamefont {U.}~\bibnamefont
  {Z\"ulicke}},\ }\bibfield  {title} {\bibinfo {title} {Multipole order in
  two-dimensional altermagnets},\ }\href {https://doi.org/10.1103/lh4w-byk7}
  {\bibfield  {journal} {\bibinfo  {journal} {Phys. Rev. Res.}\ }\textbf
  {\bibinfo {volume} {7}},\ \bibinfo {pages} {043149} (\bibinfo {year}
  {2025})}\BibitemShut {NoStop}%
\bibitem [{\citenamefont {Belashchenko}()}]{Deterministic_AM}%
  \BibitemOpen
  \bibfield  {author} {\bibinfo {author} {\bibfnamefont {K.~D.}\ \bibnamefont
  {Belashchenko}},\ }\bibfield  {title} {\bibinfo {title} {Deterministic
  electrical switching in altermagnets via surface antisymmetry groups},\
  }\href {https://arxiv.org/abs/2603.06537} {\bibinfo  {journal}
  {arXiv:2603.06537}\ }\BibitemShut {NoStop}%
\bibitem [{Not()}]{Note}%
  \BibitemOpen
\bibfield  {journal} {  }\href@noop {} {}\bibinfo {note} {See Supplemental
  Material, which includes Refs.~[90,91,97], for the analytical derivations of
  $\gamma$ and $A_z$ in the four-band model, the symmetry analysis of the
  allowed interfacial SOC terms and its extension to $g$-wave altermagnets, and
  the multilayer S/AM model used for the magnitude estimate.}\BibitemShut
  {Stop}%
\bibitem [{\citenamefont {McClarty}\ and\ \citenamefont
  {Rau}(2024)}]{McClartyRau2024PRL}%
  \BibitemOpen
  \bibfield  {author} {\bibinfo {author} {\bibfnamefont {P.~A.}\ \bibnamefont
  {McClarty}}\ and\ \bibinfo {author} {\bibfnamefont {J.~G.}\ \bibnamefont
  {Rau}},\ }\bibfield  {title} {\bibinfo {title} {Landau theory of
  altermagnetism},\ }\href {https://doi.org/10.1103/PhysRevLett.132.176702}
  {\bibfield  {journal} {\bibinfo  {journal} {Physical Review Letters}\
  }\textbf {\bibinfo {volume} {132}},\ \bibinfo {pages} {176702} (\bibinfo
  {year} {2024})}\BibitemShut {NoStop}%
\bibitem [{\citenamefont {Jo}\ \emph {et~al.}(2025)\citenamefont {Jo},
  \citenamefont {Go}, \citenamefont {Mokrousov}, \citenamefont {Oppeneer},
  \citenamefont {Cheong},\ and\ \citenamefont {Lee}}]{Jo2025PRL}%
  \BibitemOpen
  \bibfield  {author} {\bibinfo {author} {\bibfnamefont {D.}~\bibnamefont
  {Jo}}, \bibinfo {author} {\bibfnamefont {D.}~\bibnamefont {Go}}, \bibinfo
  {author} {\bibfnamefont {Y.}~\bibnamefont {Mokrousov}}, \bibinfo {author}
  {\bibfnamefont {P.~M.}\ \bibnamefont {Oppeneer}}, \bibinfo {author}
  {\bibfnamefont {S.-W.}\ \bibnamefont {Cheong}},\ and\ \bibinfo {author}
  {\bibfnamefont {H.-W.}\ \bibnamefont {Lee}},\ }\bibfield  {title} {\bibinfo
  {title} {Weak ferromagnetism in altermagnets from alternating {$g$}-tensor
  anisotropy},\ }\href {https://doi.org/10.1103/PhysRevLett.134.196703}
  {\bibfield  {journal} {\bibinfo  {journal} {Physical Review Letters}\
  }\textbf {\bibinfo {volume} {134}},\ \bibinfo {pages} {196703} (\bibinfo
  {year} {2025})}\BibitemShut {NoStop}%
\bibitem [{\citenamefont {Chakraborty}\ and\ \citenamefont
  {Black-Schaffer}(2025)}]{ChakrabortyBlackSchaffer2025Constraints}%
  \BibitemOpen
  \bibfield  {author} {\bibinfo {author} {\bibfnamefont {D.}~\bibnamefont
  {Chakraborty}}\ and\ \bibinfo {author} {\bibfnamefont {A.~M.}\ \bibnamefont
  {Black-Schaffer}},\ }\bibfield  {title} {\bibinfo {title} {Constraints on
  superconducting pairing in altermagnets},\ }\href
  {https://doi.org/10.1103/zylh-rqxl} {\bibfield  {journal} {\bibinfo
  {journal} {Physical Review B}\ }\textbf {\bibinfo {volume} {112}},\ \bibinfo
  {pages} {014516} (\bibinfo {year} {2025})}\BibitemShut {NoStop}%
\bibitem [{\citenamefont {Heinsdorf}\ and\ \citenamefont
  {Franz}(2026)}]{HeinsdorfFranz2026}%
  \BibitemOpen
  \bibfield  {author} {\bibinfo {author} {\bibfnamefont {N.}~\bibnamefont
  {Heinsdorf}}\ and\ \bibinfo {author} {\bibfnamefont {M.}~\bibnamefont
  {Franz}},\ }\bibfield  {title} {\bibinfo {title} {Proximitizing altermagnets
  with conventional superconductors},\ }\href
  {https://doi.org/10.1103/rhmg-j1fv} {\bibfield  {journal} {\bibinfo
  {journal} {Physical Review B}\ }\textbf {\bibinfo {volume} {113}},\ \bibinfo
  {pages} {L020501} (\bibinfo {year} {2026})}\BibitemShut {NoStop}%
\bibitem [{\citenamefont {Beenakker}\ and\ \citenamefont
  {Vakhtel}(2023)}]{Beenakker2023}%
  \BibitemOpen
  \bibfield  {author} {\bibinfo {author} {\bibfnamefont {C.~W.~J.}\
  \bibnamefont {Beenakker}}\ and\ \bibinfo {author} {\bibfnamefont
  {T.}~\bibnamefont {Vakhtel}},\ }\bibfield  {title} {\bibinfo {title}
  {Phase-shifted {Andreev} levels in an altermagnet {Josephson} junction},\
  }\href {https://doi.org/10.1103/PhysRevB.108.075425} {\bibfield  {journal}
  {\bibinfo  {journal} {Phys. Rev. B}\ }\textbf {\bibinfo {volume} {108}},\
  \bibinfo {pages} {075425} (\bibinfo {year} {2023})}\BibitemShut {NoStop}%
\bibitem [{\citenamefont {Kosevich}\ \emph {et~al.}(1990)\citenamefont
  {Kosevich}, \citenamefont {Ivanov},\ and\ \citenamefont
  {Kovalev}}]{Kosevich1990MagneticSolitons}%
  \BibitemOpen
  \bibfield  {author} {\bibinfo {author} {\bibfnamefont {A.~M.}\ \bibnamefont
  {Kosevich}}, \bibinfo {author} {\bibfnamefont {B.~A.}\ \bibnamefont
  {Ivanov}},\ and\ \bibinfo {author} {\bibfnamefont {A.~S.}\ \bibnamefont
  {Kovalev}},\ }\bibfield  {title} {\bibinfo {title} {Magnetic solitons},\
  }\href {https://doi.org/10.1016/0370-1573(90)90130-T} {\bibfield  {journal}
  {\bibinfo  {journal} {Phys. Rep.}\ }\textbf {\bibinfo {volume} {194}},\
  \bibinfo {pages} {117} (\bibinfo {year} {1990})}\BibitemShut {NoStop}%
\bibitem [{\citenamefont {Kriegner}\ \emph {et~al.}(2017)\citenamefont
  {Kriegner}, \citenamefont {Reichlova}, \citenamefont {Grenzer}, \citenamefont
  {Schmidt}, \citenamefont {Ressouche}, \citenamefont {Godinho}, \citenamefont
  {Wagner}, \citenamefont {Martin}, \citenamefont {Shick}, \citenamefont
  {Volobuev}, \citenamefont {Springholz}, \citenamefont {Holy}, \citenamefont
  {Wunderlich}, \citenamefont {Jungwirth},\ and\ \citenamefont
  {Vyborny}}]{Kriegner2017PRB}%
  \BibitemOpen
  \bibfield  {author} {\bibinfo {author} {\bibfnamefont {D.}~\bibnamefont
  {Kriegner}}, \bibinfo {author} {\bibfnamefont {H.}~\bibnamefont {Reichlova}},
  \bibinfo {author} {\bibfnamefont {J.}~\bibnamefont {Grenzer}}, \bibinfo
  {author} {\bibfnamefont {W.}~\bibnamefont {Schmidt}}, \bibinfo {author}
  {\bibfnamefont {E.}~\bibnamefont {Ressouche}}, \bibinfo {author}
  {\bibfnamefont {J.}~\bibnamefont {Godinho}}, \bibinfo {author} {\bibfnamefont
  {T.}~\bibnamefont {Wagner}}, \bibinfo {author} {\bibfnamefont {S.~Y.}\
  \bibnamefont {Martin}}, \bibinfo {author} {\bibfnamefont {A.~B.}\
  \bibnamefont {Shick}}, \bibinfo {author} {\bibfnamefont {V.~V.}\ \bibnamefont
  {Volobuev}}, \bibinfo {author} {\bibfnamefont {G.}~\bibnamefont
  {Springholz}}, \bibinfo {author} {\bibfnamefont {V.}~\bibnamefont {Holy}},
  \bibinfo {author} {\bibfnamefont {J.}~\bibnamefont {Wunderlich}}, \bibinfo
  {author} {\bibfnamefont {T.}~\bibnamefont {Jungwirth}},\ and\ \bibinfo
  {author} {\bibfnamefont {K.}~\bibnamefont {Vyborny}},\ }\bibfield  {title}
  {\bibinfo {title} {Magnetic anisotropy in antiferromagnetic hexagonal
  {MnTe}},\ }\href {https://doi.org/10.1103/PhysRevB.96.214418} {\bibfield
  {journal} {\bibinfo  {journal} {Physical Review B}\ }\textbf {\bibinfo
  {volume} {96}},\ \bibinfo {pages} {214418} (\bibinfo {year}
  {2017})}\BibitemShut {NoStop}%
\bibitem [{\citenamefont {Gomonay}\ \emph {et~al.}(2024)\citenamefont
  {Gomonay}, \citenamefont {Kravchuk}, \citenamefont {Jaeschke-Ubiergo},
  \citenamefont {Yershov}, \citenamefont {Jungwirth}, \citenamefont {Smejkal},
  \citenamefont {van~den Brink},\ and\ \citenamefont
  {Sinova}}]{Gomonay2024NPJ}%
  \BibitemOpen
  \bibfield  {author} {\bibinfo {author} {\bibfnamefont {O.}~\bibnamefont
  {Gomonay}}, \bibinfo {author} {\bibfnamefont {V.~P.}\ \bibnamefont
  {Kravchuk}}, \bibinfo {author} {\bibfnamefont {R.}~\bibnamefont
  {Jaeschke-Ubiergo}}, \bibinfo {author} {\bibfnamefont {K.~V.}\ \bibnamefont
  {Yershov}}, \bibinfo {author} {\bibfnamefont {T.}~\bibnamefont {Jungwirth}},
  \bibinfo {author} {\bibfnamefont {L.}~\bibnamefont {Smejkal}}, \bibinfo
  {author} {\bibfnamefont {J.}~\bibnamefont {van~den Brink}},\ and\ \bibinfo
  {author} {\bibfnamefont {J.}~\bibnamefont {Sinova}},\ }\bibfield  {title}
  {\bibinfo {title} {Structure, control, and dynamics of altermagnetic
  textures},\ }\href {https://doi.org/10.1038/s44306-024-00042-3} {\bibfield
  {journal} {\bibinfo  {journal} {npj Spintronics}\ }\textbf {\bibinfo {volume}
  {2}},\ \bibinfo {pages} {35} (\bibinfo {year} {2024})}\BibitemShut {NoStop}%
\bibitem [{\citenamefont {Kravchuk}\ \emph {et~al.}(2025)\citenamefont
  {Kravchuk}, \citenamefont {Yershov}, \citenamefont {Facio}, \citenamefont
  {Guo}, \citenamefont {Janson}, \citenamefont {Gomonay}, \citenamefont
  {Sinova},\ and\ \citenamefont {van~den Brink}}]{Kravchuk2025PRB}%
  \BibitemOpen
  \bibfield  {author} {\bibinfo {author} {\bibfnamefont {V.~P.}\ \bibnamefont
  {Kravchuk}}, \bibinfo {author} {\bibfnamefont {K.~V.}\ \bibnamefont
  {Yershov}}, \bibinfo {author} {\bibfnamefont {J.~I.}\ \bibnamefont {Facio}},
  \bibinfo {author} {\bibfnamefont {Y.}~\bibnamefont {Guo}}, \bibinfo {author}
  {\bibfnamefont {O.}~\bibnamefont {Janson}}, \bibinfo {author} {\bibfnamefont
  {O.}~\bibnamefont {Gomonay}}, \bibinfo {author} {\bibfnamefont
  {J.}~\bibnamefont {Sinova}},\ and\ \bibinfo {author} {\bibfnamefont
  {J.}~\bibnamefont {van~den Brink}},\ }\bibfield  {title} {\bibinfo {title}
  {Chiral magnetic excitations and domain textures of {$g$}-wave
  altermagnets},\ }\bibfield  {journal} {\bibinfo  {journal} {Physical Review
  B}\ }\textbf {\bibinfo {volume} {112}},\ \href
  {https://doi.org/10.1103/zn8d-ft9b} {10.1103/zn8d-ft9b} (\bibinfo {year}
  {2025}),\ \Eprint {https://arxiv.org/abs/2504.05241} {arXiv:2504.05241
  [cond-mat.mtrl-sci]} \BibitemShut {NoStop}%
\bibitem [{\citenamefont {\ifmmode~\check{S}\else \v{S}\fi{}mejkal}\ \emph
  {et~al.}(2022)\citenamefont {\ifmmode~\check{S}\else \v{S}\fi{}mejkal},
  \citenamefont {Hellenes}, \citenamefont {Gonz\'alez-Hern\'andez},
  \citenamefont {Sinova},\ and\ \citenamefont {Jungwirth}}]{mejkal2022}%
  \BibitemOpen
  \bibfield  {author} {\bibinfo {author} {\bibfnamefont {L.}~\bibnamefont
  {\ifmmode~\check{S}\else \v{S}\fi{}mejkal}}, \bibinfo {author} {\bibfnamefont
  {A.~B.}\ \bibnamefont {Hellenes}}, \bibinfo {author} {\bibfnamefont
  {R.}~\bibnamefont {Gonz\'alez-Hern\'andez}}, \bibinfo {author} {\bibfnamefont
  {J.}~\bibnamefont {Sinova}},\ and\ \bibinfo {author} {\bibfnamefont
  {T.}~\bibnamefont {Jungwirth}},\ }\bibfield  {title} {\bibinfo {title} {Giant
  and tunneling magnetoresistance in unconventional collinear antiferromagnets
  with nonrelativistic spin-momentum coupling},\ }\href
  {https://doi.org/10.1103/PhysRevX.12.011028} {\bibfield  {journal} {\bibinfo
  {journal} {Phys. Rev. X}\ }\textbf {\bibinfo {volume} {12}},\ \bibinfo
  {pages} {011028} (\bibinfo {year} {2022})}\BibitemShut {NoStop}%
\bibitem [{\citenamefont {Zhang}\ \emph {et~al.}(2024)\citenamefont {Zhang},
  \citenamefont {Hu},\ and\ \citenamefont {Neupert}}]{Zhang2024}%
  \BibitemOpen
  \bibfield  {author} {\bibinfo {author} {\bibfnamefont {S.-B.}\ \bibnamefont
  {Zhang}}, \bibinfo {author} {\bibfnamefont {L.-H.}\ \bibnamefont {Hu}},\ and\
  \bibinfo {author} {\bibfnamefont {T.}~\bibnamefont {Neupert}},\ }\bibfield
  {title} {\bibinfo {title} {Finite-momentum {Cooper} pairing in proximitized
  altermagnets},\ }\href {https://doi.org/10.1038/s41467-024-45951-3}
  {\bibfield  {journal} {\bibinfo  {journal} {Nat. Commun.}\ }\textbf {\bibinfo
  {volume} {15}},\ \bibinfo {pages} {1801} (\bibinfo {year}
  {2024})}\BibitemShut {NoStop}%
\bibitem [{\citenamefont {Weber}\ \emph {et~al.}()\citenamefont {Weber},
  \citenamefont {Leckron}, \citenamefont {Haag}, \citenamefont
  {Jaeschke-Ubiergo}, \citenamefont {\v{S}mejkal}, \citenamefont {Sinova},\
  and\ \citenamefont {Schneider}}]{arxiv_KRu4O8}%
  \BibitemOpen
  \bibfield  {author} {\bibinfo {author} {\bibfnamefont {M.}~\bibnamefont
  {Weber}}, \bibinfo {author} {\bibfnamefont {K.}~\bibnamefont {Leckron}},
  \bibinfo {author} {\bibfnamefont {L.}~\bibnamefont {Haag}}, \bibinfo {author}
  {\bibfnamefont {R.}~\bibnamefont {Jaeschke-Ubiergo}}, \bibinfo {author}
  {\bibfnamefont {L.}~\bibnamefont {\v{S}mejkal}}, \bibinfo {author}
  {\bibfnamefont {J.}~\bibnamefont {Sinova}},\ and\ \bibinfo {author}
  {\bibfnamefont {H.~C.}\ \bibnamefont {Schneider}},\ }\bibfield  {title}
  {\bibinfo {title} {Ultrafast electron dynamics in altermagnetic materials},\
  }\href {https://arxiv.org/abs/2411.08160} {\bibinfo  {journal}
  {arXiv:2411.08160}\ }\BibitemShut {NoStop}%
\bibitem [{\citenamefont {Tang}\ \emph {et~al.}(2025)\citenamefont {Tang},
  \citenamefont {Zhang},\ and\ \citenamefont {Cheng}}]{Tang2025}%
  \BibitemOpen
\bibfield  {journal} {  }\bibfield  {author} {\bibinfo {author} {\bibfnamefont
  {J.}~\bibnamefont {Tang}}, \bibinfo {author} {\bibfnamefont {H.}~\bibnamefont
  {Zhang}},\ and\ \bibinfo {author} {\bibfnamefont {R.}~\bibnamefont {Cheng}},\
  }\bibfield  {title} {\bibinfo {title} {{N\'eel} spin-orbit torque in
  antiferromagnetic quantum spin and anomalous {Hall} insulators},\ }\href
  {https://doi.org/10.1038/s41467-025-63171-1} {\bibfield  {journal} {\bibinfo
  {journal} {Nat. Commun.}\ }\textbf {\bibinfo {volume} {16}},\ \bibinfo
  {pages} {7790} (\bibinfo {year} {2025})}\BibitemShut {NoStop}%
\bibitem [{\citenamefont {Parkin}\ and\ \citenamefont
  {Yang}(2015)}]{Parkin2015MemoryOnTheRacetrack}%
  \BibitemOpen
  \bibfield  {author} {\bibinfo {author} {\bibfnamefont {S.~S.~P.}\
  \bibnamefont {Parkin}}\ and\ \bibinfo {author} {\bibfnamefont {S.-H.}\
  \bibnamefont {Yang}},\ }\bibfield  {title} {\bibinfo {title} {Memory on the
  racetrack},\ }\href {https://doi.org/10.1038/nnano.2015.41} {\bibfield
  {journal} {\bibinfo  {journal} {Nat. Nanotechnol.}\ }\textbf {\bibinfo
  {volume} {10}},\ \bibinfo {pages} {195} (\bibinfo {year} {2015})}\BibitemShut
  {NoStop}%
\bibitem [{\citenamefont {Tinkham}(2004)}]{Tinkham2004}%
  \BibitemOpen
  \bibfield  {author} {\bibinfo {author} {\bibfnamefont {M.}~\bibnamefont
  {Tinkham}},\ }\href@noop {} {\emph {\bibinfo {title} {Introduction to
  Superconductivity}}},\ \bibinfo {edition} {2nd}\ ed.\ (\bibinfo  {publisher}
  {Dover Publications},\ \bibinfo {address} {Mineola, NY},\ \bibinfo {year}
  {2004})\BibitemShut {NoStop}%
\bibitem [{\citenamefont {Henrich}\ \emph {et~al.}(2012)\citenamefont
  {Henrich}, \citenamefont {Reichensperger}, \citenamefont {Hofherr},
  \citenamefont {Meckbach}, \citenamefont {Ilin}, \citenamefont {Siegel},
  \citenamefont {Semenov}, \citenamefont {Zotova},\ and\ \citenamefont
  {Vodolazov}}]{Henrich2012PRB}%
  \BibitemOpen
  \bibfield  {author} {\bibinfo {author} {\bibfnamefont {D.}~\bibnamefont
  {Henrich}}, \bibinfo {author} {\bibfnamefont {P.}~\bibnamefont
  {Reichensperger}}, \bibinfo {author} {\bibfnamefont {M.}~\bibnamefont
  {Hofherr}}, \bibinfo {author} {\bibfnamefont {J.~M.}\ \bibnamefont
  {Meckbach}}, \bibinfo {author} {\bibfnamefont {K.}~\bibnamefont {Ilin}},
  \bibinfo {author} {\bibfnamefont {M.}~\bibnamefont {Siegel}}, \bibinfo
  {author} {\bibfnamefont {A.}~\bibnamefont {Semenov}}, \bibinfo {author}
  {\bibfnamefont {A.}~\bibnamefont {Zotova}},\ and\ \bibinfo {author}
  {\bibfnamefont {D.~Y.}\ \bibnamefont {Vodolazov}},\ }\bibfield  {title}
  {\bibinfo {title} {Geometry-induced reduction of the critical current in
  superconducting {NbN} nanowires},\ }\href
  {https://doi.org/10.1103/PhysRevB.86.144504} {\bibfield  {journal} {\bibinfo
  {journal} {Physical Review B}\ }\textbf {\bibinfo {volume} {86}},\ \bibinfo
  {pages} {144504} (\bibinfo {year} {2012})}\BibitemShut {NoStop}%
\bibitem [{\citenamefont {Clem}\ and\ \citenamefont
  {Berggren}(2011)}]{ClemBerggren2011PRB}%
  \BibitemOpen
  \bibfield  {author} {\bibinfo {author} {\bibfnamefont {J.~R.}\ \bibnamefont
  {Clem}}\ and\ \bibinfo {author} {\bibfnamefont {K.~K.}\ \bibnamefont
  {Berggren}},\ }\bibfield  {title} {\bibinfo {title} {Geometry-dependent
  critical currents in superconducting nanocircuits},\ }\href
  {https://doi.org/10.1103/PhysRevB.84.174510} {\bibfield  {journal} {\bibinfo
  {journal} {Physical Review B}\ }\textbf {\bibinfo {volume} {84}},\ \bibinfo
  {pages} {174510} (\bibinfo {year} {2011})}\BibitemShut {NoStop}%
\bibitem [{\citenamefont {Bobkov}\ \emph {et~al.}(2021)\citenamefont {Bobkov},
  \citenamefont {Bobkova}, \citenamefont {Bobkov},\ and\ \citenamefont
  {Kamra}}]{Bobkov2020}%
  \BibitemOpen
  \bibfield  {author} {\bibinfo {author} {\bibfnamefont {G.~A.}\ \bibnamefont
  {Bobkov}}, \bibinfo {author} {\bibfnamefont {I.~V.}\ \bibnamefont {Bobkova}},
  \bibinfo {author} {\bibfnamefont {A.~M.}\ \bibnamefont {Bobkov}},\ and\
  \bibinfo {author} {\bibfnamefont {A.}~\bibnamefont {Kamra}},\ }\bibfield
  {title} {\bibinfo {title} {Thermally induced spin torque and domain-wall
  motion in superconductor/antiferromagnetic-insulator bilayers},\ }\href
  {https://doi.org/10.1103/PhysRevB.103.094506} {\bibfield  {journal} {\bibinfo
   {journal} {Phys. Rev. B}\ }\textbf {\bibinfo {volume} {103}},\ \bibinfo
  {pages} {094506} (\bibinfo {year} {2021})}\BibitemShut {NoStop}%
\bibitem [{\citenamefont {Reichlova}\ \emph {et~al.}(2024)\citenamefont
  {Reichlova}, \citenamefont {Seeger}, \citenamefont {Gonzalez-Hernandez},
  \citenamefont {Kounta}, \citenamefont {Schlitz}, \citenamefont {Kriegner},
  \citenamefont {Ritzinger}, \citenamefont {Lammel}, \citenamefont {Leiviska},
  \citenamefont {Hellenes}, \citenamefont {Olejnik}, \citenamefont {Petricek},
  \citenamefont {Dolezal}, \citenamefont {Horak}, \citenamefont
  {Schmoranzerova}, \citenamefont {Badura}, \citenamefont {Bertaina},
  \citenamefont {Thomas}, \citenamefont {Baltz}, \citenamefont {Michez},
  \citenamefont {Sinova}, \citenamefont {Goennenwein}, \citenamefont
  {Jungwirth},\ and\ \citenamefont {\v{S}mejkal}}]{Reichlova2024Mn5Si3AHE}%
  \BibitemOpen
  \bibfield  {author} {\bibinfo {author} {\bibfnamefont {H.}~\bibnamefont
  {Reichlova}}, \bibinfo {author} {\bibfnamefont {R.~L.}\ \bibnamefont
  {Seeger}}, \bibinfo {author} {\bibfnamefont {R.}~\bibnamefont
  {Gonzalez-Hernandez}}, \bibinfo {author} {\bibfnamefont {I.}~\bibnamefont
  {Kounta}}, \bibinfo {author} {\bibfnamefont {R.}~\bibnamefont {Schlitz}},
  \bibinfo {author} {\bibfnamefont {D.}~\bibnamefont {Kriegner}}, \bibinfo
  {author} {\bibfnamefont {P.}~\bibnamefont {Ritzinger}}, \bibinfo {author}
  {\bibfnamefont {M.}~\bibnamefont {Lammel}}, \bibinfo {author} {\bibfnamefont
  {M.}~\bibnamefont {Leiviska}}, \bibinfo {author} {\bibfnamefont {A.~B.}\
  \bibnamefont {Hellenes}}, \bibinfo {author} {\bibfnamefont {K.}~\bibnamefont
  {Olejnik}}, \bibinfo {author} {\bibfnamefont {V.}~\bibnamefont {Petricek}},
  \bibinfo {author} {\bibfnamefont {P.}~\bibnamefont {Dolezal}}, \bibinfo
  {author} {\bibfnamefont {L.}~\bibnamefont {Horak}}, \bibinfo {author}
  {\bibfnamefont {E.}~\bibnamefont {Schmoranzerova}}, \bibinfo {author}
  {\bibfnamefont {A.}~\bibnamefont {Badura}}, \bibinfo {author} {\bibfnamefont
  {S.}~\bibnamefont {Bertaina}}, \bibinfo {author} {\bibfnamefont
  {A.}~\bibnamefont {Thomas}}, \bibinfo {author} {\bibfnamefont
  {V.}~\bibnamefont {Baltz}}, \bibinfo {author} {\bibfnamefont
  {L.}~\bibnamefont {Michez}}, \bibinfo {author} {\bibfnamefont
  {J.}~\bibnamefont {Sinova}}, \bibinfo {author} {\bibfnamefont {S.~T.~B.}\
  \bibnamefont {Goennenwein}}, \bibinfo {author} {\bibfnamefont
  {T.}~\bibnamefont {Jungwirth}},\ and\ \bibinfo {author} {\bibfnamefont
  {L.}~\bibnamefont {\v{S}mejkal}},\ }\bibfield  {title} {\bibinfo {title}
  {Observation of a spontaneous anomalous {Hall} response in the {Mn$_5$Si$_3$}
  $d$-wave altermagnet candidate},\ }\href
  {https://doi.org/10.1038/s41467-024-48493-w} {\bibfield  {journal} {\bibinfo
  {journal} {Nat. Commun.}\ }\textbf {\bibinfo {volume} {15}},\ \bibinfo
  {pages} {4961} (\bibinfo {year} {2024})}\BibitemShut {NoStop}%
\bibitem [{\citenamefont {Zhu}\ \emph {et~al.}(2025)\citenamefont {Zhu},
  \citenamefont {Li}, \citenamefont {Chen}, \citenamefont {Yu},\ and\
  \citenamefont {Liu}}]{Zhu2025MagneticGeometry}%
  \BibitemOpen
  \bibfield  {author} {\bibinfo {author} {\bibfnamefont {H.}~\bibnamefont
  {Zhu}}, \bibinfo {author} {\bibfnamefont {J.}~\bibnamefont {Li}}, \bibinfo
  {author} {\bibfnamefont {X.}~\bibnamefont {Chen}}, \bibinfo {author}
  {\bibfnamefont {Y.}~\bibnamefont {Yu}},\ and\ \bibinfo {author}
  {\bibfnamefont {Q.}~\bibnamefont {Liu}},\ }\bibfield  {title} {\bibinfo
  {title} {Magnetic geometry induced quantum geometry and nonlinear
  transports},\ }\href {https://doi.org/10.1038/s41467-025-60128-2} {\bibfield
  {journal} {\bibinfo  {journal} {Nat. Commun.}\ }\textbf {\bibinfo {volume}
  {16}},\ \bibinfo {pages} {4882} (\bibinfo {year} {2025})}\BibitemShut
  {NoStop}%
\bibitem [{\citenamefont {Reimers}\ \emph {et~al.}(2024)\citenamefont
  {Reimers}, \citenamefont {Odenbreit}, \citenamefont {\v{S}mejkal},
  \citenamefont {Strocov}, \citenamefont {Constantinou}, \citenamefont
  {Hellenes}, \citenamefont {Jaeschke~Ubiergo}, \citenamefont {Campos},
  \citenamefont {Bharadwaj}, \citenamefont {Chakraborty}, \citenamefont
  {Denneulin}, \citenamefont {Shi}, \citenamefont {Dunin-Borkowski},
  \citenamefont {Das}, \citenamefont {Kl\"{a}ui}, \citenamefont {Sinova},\ and\
  \citenamefont {Jourdan}}]{Reimers2024}%
  \BibitemOpen
  \bibfield  {author} {\bibinfo {author} {\bibfnamefont {S.}~\bibnamefont
  {Reimers}}, \bibinfo {author} {\bibfnamefont {L.}~\bibnamefont {Odenbreit}},
  \bibinfo {author} {\bibfnamefont {L.}~\bibnamefont {\v{S}mejkal}}, \bibinfo
  {author} {\bibfnamefont {V.~N.}\ \bibnamefont {Strocov}}, \bibinfo {author}
  {\bibfnamefont {P.}~\bibnamefont {Constantinou}}, \bibinfo {author}
  {\bibfnamefont {A.~B.}\ \bibnamefont {Hellenes}}, \bibinfo {author}
  {\bibfnamefont {R.}~\bibnamefont {Jaeschke~Ubiergo}}, \bibinfo {author}
  {\bibfnamefont {W.~H.}\ \bibnamefont {Campos}}, \bibinfo {author}
  {\bibfnamefont {V.~K.}\ \bibnamefont {Bharadwaj}}, \bibinfo {author}
  {\bibfnamefont {A.}~\bibnamefont {Chakraborty}}, \bibinfo {author}
  {\bibfnamefont {T.}~\bibnamefont {Denneulin}}, \bibinfo {author}
  {\bibfnamefont {W.}~\bibnamefont {Shi}}, \bibinfo {author} {\bibfnamefont
  {R.~E.}\ \bibnamefont {Dunin-Borkowski}}, \bibinfo {author} {\bibfnamefont
  {S.}~\bibnamefont {Das}}, \bibinfo {author} {\bibfnamefont {M.}~\bibnamefont
  {Kl\"{a}ui}}, \bibinfo {author} {\bibfnamefont {J.}~\bibnamefont {Sinova}},\
  and\ \bibinfo {author} {\bibfnamefont {M.}~\bibnamefont {Jourdan}},\
  }\bibfield  {title} {\bibinfo {title} {Direct observation of altermagnetic
  band splitting in {CrSb} thin films},\ }\href
  {https://doi.org/10.1038/s41467-024-46476-5} {\bibfield  {journal} {\bibinfo
  {journal} {Nat. Commun.}\ }\textbf {\bibinfo {volume} {15}},\ \bibinfo
  {pages} {2116} (\bibinfo {year} {2024})}\BibitemShut {NoStop}%
\bibitem [{\citenamefont {Ryu}\ \emph {et~al.}(2013)\citenamefont {Ryu},
  \citenamefont {Thomas}, \citenamefont {Yang},\ and\ \citenamefont
  {Parkin}}]{Ryu2013ChiralSpinTorque}%
  \BibitemOpen
  \bibfield  {author} {\bibinfo {author} {\bibfnamefont {K.-S.}\ \bibnamefont
  {Ryu}}, \bibinfo {author} {\bibfnamefont {L.}~\bibnamefont {Thomas}},
  \bibinfo {author} {\bibfnamefont {S.-H.}\ \bibnamefont {Yang}},\ and\
  \bibinfo {author} {\bibfnamefont {S.}~\bibnamefont {Parkin}},\ }\bibfield
  {title} {\bibinfo {title} {Chiral spin torque at magnetic domain walls},\
  }\href {https://doi.org/10.1038/nnano.2013.102} {\bibfield  {journal}
  {\bibinfo  {journal} {Nat. Nanotechnol.}\ }\textbf {\bibinfo {volume} {8}},\
  \bibinfo {pages} {527} (\bibinfo {year} {2013})}\BibitemShut {NoStop}%
\bibitem [{\citenamefont {Hess}\ \emph {et~al.}(2023)\citenamefont {Hess},
  \citenamefont {Legg}, \citenamefont {Loss},\ and\ \citenamefont
  {Klinovaja}}]{PhysRevB.108.174516}%
  \BibitemOpen
  \bibfield  {author} {\bibinfo {author} {\bibfnamefont {R.}~\bibnamefont
  {Hess}}, \bibinfo {author} {\bibfnamefont {H.~F.}\ \bibnamefont {Legg}},
  \bibinfo {author} {\bibfnamefont {D.}~\bibnamefont {Loss}},\ and\ \bibinfo
  {author} {\bibfnamefont {J.}~\bibnamefont {Klinovaja}},\ }\bibfield  {title}
  {\bibinfo {title} {{Josephson} transistor from the superconducting diode
  effect in domain wall and skyrmion magnetic racetracks},\ }\href
  {https://doi.org/10.1103/PhysRevB.108.174516} {\bibfield  {journal} {\bibinfo
   {journal} {Phys. Rev. B}\ }\textbf {\bibinfo {volume} {108}},\ \bibinfo
  {pages} {174516} (\bibinfo {year} {2023})}\BibitemShut {NoStop}%
\bibitem [{\citenamefont {Edwards}\ \emph {et~al.}()\citenamefont {Edwards},
  \citenamefont {Le}, \citenamefont {Smith}, \citenamefont {Usih},
  \citenamefont {Thomas}, \citenamefont {Richardson}, \citenamefont
  {Blumenschein}, \citenamefont {Hanbicki}, \citenamefont {Friedman},\ and\
  \citenamefont {Friedman}}]{Edwards2026}%
  \BibitemOpen
  \bibfield  {author} {\bibinfo {author} {\bibfnamefont {A.~J.}\ \bibnamefont
  {Edwards}}, \bibinfo {author} {\bibfnamefont {S.~T.}\ \bibnamefont {Le}},
  \bibinfo {author} {\bibfnamefont {N.~W.~G.}\ \bibnamefont {Smith}}, \bibinfo
  {author} {\bibfnamefont {E.~C.}\ \bibnamefont {Usih}}, \bibinfo {author}
  {\bibfnamefont {A.}~\bibnamefont {Thomas}}, \bibinfo {author} {\bibfnamefont
  {C.~J.~K.}\ \bibnamefont {Richardson}}, \bibinfo {author} {\bibfnamefont
  {N.~A.}\ \bibnamefont {Blumenschein}}, \bibinfo {author} {\bibfnamefont
  {A.~T.}\ \bibnamefont {Hanbicki}}, \bibinfo {author} {\bibfnamefont {A.~L.}\
  \bibnamefont {Friedman}},\ and\ \bibinfo {author} {\bibfnamefont {J.~S.}\
  \bibnamefont {Friedman}},\ }\bibfield  {title} {\bibinfo {title} {Magnetic
  field-mediated superconducting logic},\ }\href
  {https://arxiv.org/abs/2602.07146} {\bibinfo  {journal} {arXiv:2602.07146}\
  }\BibitemShut {NoStop}%
\end{thebibliography}%

\clearpage
\onecolumngrid
\setcounter{equation}{0}
\setcounter{figure}{0}
\setcounter{table}{0}
\setcounter{section}{0}
\setcounter{subsection}{0}
\setcounter{subsubsection}{0}
\renewcommand{\theequation}{S\arabic{equation}}
\renewcommand{\thefigure}{S\arabic{figure}}
\renewcommand{\thetable}{S\arabic{table}}
\renewcommand{\theHequation}{supp.\arabic{equation}}
\renewcommand{\theHfigure}{supp.\arabic{figure}}
\renewcommand{\theHtable}{supp.\arabic{table}}
\renewcommand{\theHsection}{supp.\arabic{section}}
\renewcommand{\theHsubsection}{\theHsection.\arabic{subsection}}
\renewcommand{\theHsubsubsection}{\theHsubsection.\arabic{subsubsection}}
\pdfbookmark[0]{Supplemental Material}{supplemental-material}
\begin{center}
{\large\bfseries
Supplemental Material: Supercurrent-Driven N\'eel Torque in
Superconductor/Altermagnet Hybrids\par}
\vspace{0.8em}
Hamed Vakili\textsuperscript{1},
Moaz Ali\textsuperscript{1},
Igor \v{Z}uti\'{c}\textsuperscript{2}, and
Alexey A. Kovalev\textsuperscript{1}\par
\vspace{0.4em}
{\small\itshape
\textsuperscript{1}Department of Physics and Astronomy and Nebraska Center
for Materials and Nanoscience, University of Nebraska, Lincoln,
Nebraska 68588, USA\par
\textsuperscript{2}University at Buffalo, State University of New York,
Buffalo, New York 14260-1500, USA\par}
\end{center}
\medskip

\begingroup
\makeatletter
\renewcommand{\@extra@b@citeb}{supp}
\renewcommand{\@extra@binfo}{supp}
\makeatother
\begin{bibunit}[apsrev4-2]

\makeatletter
\@startbibunitorrelax
\if@filesw
  \immediate\write\@bibunitaux{\string\citation{apsrev42Control}}
\fi
\makeatother

\section{Analytical derivation for the four-band model}
\label{app:4b_soc_single}

We begin with the two-dimensional lattice model of Eq.~(18) in
Ref.~\cite{Minimal_models},
\begin{equation}
H_{2\mathrm D}(\mathbf k)
=
-t(\cos k_x+\cos k_y)+t_2\cos k_x\cos k_y
+t_3\cos\frac{k_x}{2}\cos\frac{k_y}{2}\,\tau_x
+t_4(\cos k_x-\cos k_y)\tau_z-\widetilde\mu .
\label{eq:4b_lattice_eq18}
\end{equation}
Here, $\widetilde\mu$ is the chemical potential; $t$, $t_2$, and
$t_4$ parameterize intrasublattice hopping, whereas $t_3$
parameterizes intersublattice hopping. Expanding about
$\mathbf k=\mathbf 0$ gives
\begin{equation}
H_{2\mathrm D}(\mathbf k)
\simeq
\xi_k\tau_0+\epsilon_k\tau_x
+\frac{t_4}{2}d_{\mathbf k}\tau_z,
\label{eq:4b_continuum_eq18}
\end{equation}
where
\begin{equation}
\xi_k=\bar t k^2-\widetilde\mu-2t+t_2,
\qquad
\bar t\equiv\frac{t-t_2}{2},
\qquad
\epsilon_k\equiv t_3\left(1-\frac{k^2}{8}\right),
\qquad
d_{\mathbf k}=k_y^2-k_x^2,
\qquad
k^2=k_x^2+k_y^2.
\label{eq:4b_model_defs}
\end{equation}
The continuum theory is understood with an ultraviolet cutoff
$|\mathbf k|<k_c$.

We next add the staggered exchange field and consider one
symmetry-allowed spin--orbit-coupling (SOC) channel at a time,
\begin{equation}
h(\mathbf k)
=
\xi_k\tau_0\sigma_0
+\epsilon_k\tau_x\sigma_0
+\frac{t_4}{2}d_{\mathbf k}\tau_z\sigma_0
+H_{\mathrm{SOC}}(\mathbf k)
+J_{\mathrm{ex}}\tau_z
(\hat{\mathbf n}\!\cdot\!\boldsymbol\sigma),
\label{eq:4b_soc_model_normal}
\end{equation}
with
\begin{align}
H_{\mathrm{SOC}}^{(0)}(\mathbf k)
&=
\alpha_0(k_y\sigma_x-k_x\sigma_y)\tau_0,
\label{eq:4b_soc_alpha0}
\\
H_{\mathrm{SOC}}^{(x)}(\mathbf k)
&=
\alpha_x(k_y\sigma_x-k_x\sigma_y)\tau_x,
\label{eq:4b_soc_alphax}
\\
H_{\mathrm{SOC}}^{(z)}(\mathbf k)
&=
\beta_z(k_y\sigma_x+k_x\sigma_y)\tau_z.
\label{eq:4b_soc_betaz}
\end{align}
Here, $J_{\mathrm{ex}}$ is the exchange coupling, $\tau_i$ act in
sublattice space, and $\sigma_i$ act in spin space. The parameters
$\alpha_0$, $\alpha_x$, and $\beta_z$ denote the three SOC
amplitudes retained in the continuum theory.

We assume a uniform, sublattice-even spin-singlet gap
$\widehat\Delta=\Delta\tau_0 i\sigma_y$ and choose a gauge in which
$\Delta$ is real. For a Cooper-pair momentum $\mathbf q$, introduce
the time-reversal Nambu spinor
\begin{equation}
\Phi_{\mathbf k,\mathbf q}
=
\begin{pmatrix}
\psi_{\mathbf k+\mathbf q/2}\\[1mm]
\mathcal U_T
\big[\psi_{-\mathbf k+\mathbf q/2}^{\dagger}\big]^T
\end{pmatrix},
\qquad
\mathcal U_T=\tau_0\otimes i\sigma_y,
\qquad
\psi_{\mathbf k}
=
\begin{pmatrix}
c_{\mathbf k,A\uparrow}\\
c_{\mathbf k,A\downarrow}\\
c_{\mathbf k,B\uparrow}\\
c_{\mathbf k,B\downarrow}
\end{pmatrix}.
\label{eq:4b_rotated_basis}
\end{equation}
Writing $h_0(\mathbf k)$ for the time-reversal-even part of
Eq.~\eqref{eq:4b_soc_model_normal}, i.e., the normal-state
Hamiltonian without the exchange term, the BdG Hamiltonian in this
basis is
\begin{equation}
\mathcal H_{\mathbf k,\mathbf q}
=
\begin{pmatrix}
h_0(\mathbf k+\mathbf q/2)
+J_{\mathrm{ex}}\tau_z
(\hat{\mathbf n}\!\cdot\!\boldsymbol\sigma)
&
\Delta
\\[1mm]
\Delta
&
-h_0(\mathbf k-\mathbf q/2)
+J_{\mathrm{ex}}\tau_z
(\hat{\mathbf n}\!\cdot\!\boldsymbol\sigma)
\end{pmatrix}.
\label{eq:4b_BdG_q}
\end{equation}
Identity matrices in spin, sublattice, and Nambu spaces are left
implicit when unambiguous. The exchange field is proportional to
$\rho_0$ in this Nambu basis because it is odd under time reversal.

Setting $k_B=1$, the orientation-dependent part of the free energy
may be obtained from
\begin{equation}
\mathcal F(\mathbf q)
=
-\frac{T}{2}\sum_{\omega_n}\int_{\mathbf k}
\ln\det\!\left[i\omega_n-\mathcal H_{\mathbf k,\mathbf q}\right],
\qquad
\int_{\mathbf k}
\equiv
\int_{|\mathbf k|<k_c}\frac{d^2k}{(2\pi)^2},
\qquad
\omega_n=(2n+1)\pi T,
\qquad
\beta=T^{-1}.
\label{eq:4b_free_energy}
\end{equation}
To linear order in $\mathbf q$,
\begin{equation}
\mathcal H_{\mathbf k,\mathbf q}
=
\mathcal H_{0,\mathbf k}
+J_{\mathrm{ex}}\mathcal V_m
+q_i v_i(\mathbf k)
+\mathcal O(q^2),
\label{eq:4b_H0_split}
\end{equation}
where
\begin{equation}
\mathcal H_{0,\mathbf k}
=h_0(\mathbf k)\rho_z+\Delta\rho_x,
\qquad
\mathcal V_m
=\tau_z(\hat{\mathbf n}\!\cdot\!\boldsymbol\sigma)\rho_0,
\qquad
v_i(\mathbf k)
=\frac12\,\partial_{k_i}h_0(\mathbf k)\rho_0.
\label{eq:4b_Vm_vi_general}
\end{equation}
Let $\Pi_a(\mathbf k)$ denote the spin--sublattice projector onto a
normal-state branch $a$. Then
\begin{equation}
\mathcal H_{0,\mathbf k}
=
\sum_a\Pi_a(\mathbf k)
\left[\xi_a(\mathbf k)\rho_z+\Delta\rho_x\right],
\end{equation}
and
\begin{equation}
\mathcal G_0(\mathbf k,i\omega_n)
\equiv
\left[i\omega_n-\mathcal H_{0,\mathbf k}\right]^{-1}
=
-\sum_a
\frac{i\omega_n+\xi_a(\mathbf k)\rho_z+\Delta\rho_x}
{\omega_n^2+E_a^2(\mathbf k)}
\Pi_a(\mathbf k),
\qquad
E_a(\mathbf k)=\sqrt{\xi_a^2(\mathbf k)+\Delta^2}.
\end{equation}

Expanding the logarithm in $J_{\mathrm{ex}}$ and $\mathbf q$ gives
\begin{align}
\delta\mathcal F^{(1)}(\mathbf 0)
&=
\frac{J_{\mathrm{ex}}T}{2}
\sum_{\omega_n}\int_{\mathbf k}
\operatorname{Tr}\!\left[\mathcal G_0\mathcal V_m\right]
=0,
\label{eq:4b_F1_expand}
\\
\delta\mathcal F^{(2)}(\mathbf 0)
&=
\frac{J_{\mathrm{ex}}^2T}{4}
\sum_{\omega_n}\int_{\mathbf k}
\operatorname{Tr}\!\left[
\mathcal G_0\mathcal V_m
\mathcal G_0\mathcal V_m
\right],
\label{eq:4b_F0_expand}
\\
\delta\mathcal F_{\mathrm{lin}}(\mathbf q)
&=
\frac{J_{\mathrm{ex}}q_iT}{2}
\sum_{\omega_n}\int_{\mathbf k}
\operatorname{Tr}\!\left[
\mathcal G_0\mathcal V_m
\mathcal G_0 v_i
\right]
+\text{higher-order terms}.
\label{eq:4b_Flin_expand}
\end{align}
The term linear in $J_{\mathrm{ex}}$ vanishes by time-reversal
symmetry:
$\mathcal F(\mathbf 0,\hat{\mathbf n})
=\mathcal F(\mathbf 0,-\hat{\mathbf n})$.
Equivalently, the integrand of Eq.~\eqref{eq:4b_F1_expand} cancels
after momentum integration. It does not generally vanish from the
spin trace at a fixed momentum when SOC is present. Consequently,
\begin{equation}
\mathcal F(\mathbf 0)
=
\mathcal F^{(0)}(\mathbf 0)
+\delta\mathcal F^{(2)}(\mathbf 0)
+\mathcal O(J_{\mathrm{ex}}^4).
\end{equation}

The zero-superflow anisotropy can be constrained directly by the
$[001]$ surface antisymmetry point group
${}^{2}4\,{}^{1}m_x\,{}^{2}m_d$. The sublattice-preserving mirrors
${}^{1}m_x$ and ${}^{1}m_y$ eliminate the mixed quadratic terms,
whereas the sublattice-exchanging fourfold rotation
${}^{2}C_{4z}$ requires equal in-plane coefficients. Thus,
\begin{equation}
\delta\mathcal F(\mathbf 0,\hat{\mathbf n})
=
J_{\mathrm{ex}}^2
\left[A_{\parallel}(n_x^2+n_y^2)+A_{\perp}n_z^2\right]
=
J_{\mathrm{ex}}^2\left[A_0+A_z n_z^2\right],
\label{eq:4b_sym_F0_final}
\end{equation}
where $A_0=A_{\parallel}$ and
$A_z=A_{\perp}-A_{\parallel}$.

For the superflow response, write
\begin{equation}
\mathcal F(\mathbf q)
=
\mathcal F(\mathbf 0)+\Gamma_i q_i+\mathcal O(q^2),
\qquad
\Gamma_i
=\left.\frac{\partial\mathcal F}{\partial q_i}\right|_{\mathbf q=0}.
\label{eq:4b_Gamma_def}
\end{equation}
To first order in both $\mathbf q$ and $J_{\mathrm{ex}}$, the most
general in-plane term is
\begin{equation}
\delta\mathcal F_{\mathrm{lin}}(\mathbf q,\hat{\mathbf n})
=
J_{\mathrm{ex}}\Lambda_{ij}q_i n_j.
\label{eq:4b_sym_lin_tensor}
\end{equation}
Here, $\mathbf q$ is a polar vector, whereas $\hat{\mathbf n}$ is
a sublattice-odd axial vector. Under the two
sublattice-preserving mirrors,
\begin{align}
{}^{1}m_x:&\quad
(q_x,q_y)\mapsto(-q_x,q_y),
&
(n_x,n_y,n_z)\mapsto(n_x,-n_y,-n_z),
\\
{}^{1}m_y:&\quad
(q_x,q_y)\mapsto(q_x,-q_y),
&
(n_x,n_y,n_z)\mapsto(-n_x,n_y,-n_z).
\end{align}
These mirrors allow only $q_xn_y$ and $q_yn_x$. Under the
sublattice-exchanging fourfold rotation,
\begin{equation}
{}^{2}C_{4z}:\qquad
(q_x,q_y)\mapsto(-q_y,q_x),
\qquad
(n_x,n_y,n_z)\mapsto(n_y,-n_x,-n_z),
\end{equation}
the two terms are interchanged, which fixes their coefficients to
be equal. Therefore,
\begin{equation}
\delta\mathcal F_{\mathrm{lin}}(\mathbf q,\hat{\mathbf n})
=
J_{\mathrm{ex}}\gamma
(q_xn_y+q_yn_x).
\label{eq:4b_sym_lin_final}
\end{equation}
The microscopic calculations below determine the channel-dependent
coefficient $\gamma$. For the $\alpha_0$ and $\alpha_x$ channels,
$\gamma$ is odd in $t_4$ and therefore begins at order $t_4$.
For the $\beta_z$ channel, $\gamma$ is even in $t_4$ and can remain
nonzero in the limit $t_4\to0$.

For later use, define
\begin{equation}
S_1(E)
\equiv
T\sum_{\omega_n}\frac{1}{\omega_n^2+E^2}
=
\frac{1}{2E}\tanh\!\left(\frac{\beta E}{2}\right).
\label{eq:4b_S1}
\end{equation}
For a branch $a$, introduce
\begin{equation}
g_a(\mathbf k,i\omega_n)
=
\frac{i\omega_n+\xi_a(\mathbf k)\rho_z+\Delta\rho_x}
{\omega_n^2+E_a^2(\mathbf k)},
\label{eq:4b_ga}
\end{equation}
so that the corresponding projector-resolved contribution to
$\mathcal G_0$ is $-g_a\Pi_a$. We use the symmetric branch kernel
\begin{equation}
\widetilde S_{ab}(\mathbf k)
\equiv
T\sum_{\omega_n}
\operatorname{Tr}_{\rho}\!\left[
 g_a(\mathbf k,i\omega_n)g_b(\mathbf k,i\omega_n)
\right].
\label{eq:4b_Splus_tilde}
\end{equation}
A direct Matsubara summation gives
\begin{equation}
\widetilde S_{ab}(\mathbf k)
=
\left[(\xi_a+\xi_b)^2+4\Delta^2\right]
\frac{S_1(E_a)-S_1(E_b)}{E_b^2-E_a^2}
-\left[S_1(E_a)+S_1(E_b)\right],
\label{eq:4b_Sab}
\end{equation}
with the diagonal limit
\begin{equation}
\widetilde S_{aa}(\mathbf k)
=
-\frac{\beta}{2}
\operatorname{sech}^2\!\left(\frac{\beta E_a}{2}\right).
\label{eq:4b_Saa}
\end{equation}
For the small-$t_4$ expansions below, we also define
\begin{align}
\mathcal K^{(1)}_{abc}(\mathbf k)
&\equiv
T\sum_{\omega_n}
\operatorname{Tr}_{\rho}\!\left[
 g_a g_b\rho_z g_c
\right],
\label{eq:4b_K1_def}
\\
\mathcal K^{(2)}_{abc}(\mathbf k)
&\equiv
T\sum_{\omega_n}
\operatorname{Tr}_{\rho}\!\left[
 g_a\rho_z g_b g_c
\right],
\label{eq:4b_K2_def}
\end{align}
where all Green-function blocks in a given trace carry the same
arguments $(\mathbf k,i\omega_n)$.

The $\alpha_0$ and $\alpha_x$ channels share the helicity axis
\begin{equation}
\hat{\mathbf g}_{\mathbf k}
=\frac{(k_y,-k_x,0)}{k},
\qquad
P_\eta(\mathbf k)
=\frac12\left(
\sigma_0+\eta\hat{\mathbf g}_{\mathbf k}\!\cdot\!\boldsymbol\sigma
\right),
\qquad
\eta=\pm.
\label{eq:4b_alpha_family_proj}
\end{equation}
For $\mathbf n_{\parallel}=(n_x,n_y,0)$,
\begin{align}
\left|\langle\eta,\mathbf k|
\hat{\mathbf n}\!\cdot\!\boldsymbol\sigma
|\eta,\mathbf k\rangle\right|^2
&=
(\mathbf n_{\parallel}\!\cdot\!\hat{\mathbf g}_{\mathbf k})^2,
\label{eq:4b_alpha_spin_same}
\\
\left|\langle\eta,\mathbf k|
\hat{\mathbf n}\!\cdot\!\boldsymbol\sigma
|-\eta,\mathbf k\rangle\right|^2
&=
1-(\mathbf n_{\parallel}\!\cdot\!\hat{\mathbf g}_{\mathbf k})^2.
\label{eq:4b_alpha_spin_off}
\end{align}
For an angular weight with the $C_4$ symmetry of the present model,
\begin{equation}
\left\langle
(\mathbf n_{\parallel}\!\cdot\!\hat{\mathbf g}_{\mathbf k})^2
\right\rangle_{\varphi_k}
=\frac{1-n_z^2}{2},
\qquad
\left\langle
1-(\mathbf n_{\parallel}\!\cdot\!\hat{\mathbf g}_{\mathbf k})^2
\right\rangle_{\varphi_k}
=\frac{1+n_z^2}{2}.
\label{eq:4b_alpha_angleavg}
\end{equation}

\subsection{\(\alpha_0\)-type SOC}
\label{app:4b_soc_alpha0}

\subsubsection{Zeroth-order free energy at \(\mathbf q=\mathbf 0\)}
\label{app:4b_alpha0_t3_F0}

We take $H_{\rm SOC}^{(0)}(\mathbf k)
=
\alpha_0(k_y\sigma_x-k_x\sigma_y)\tau_0,$
write \(\mathbf k=(k\cos\varphi_k,k\sin\varphi_k)\), so that
\begin{equation}
d_{\mathbf k}=-k^2\cos 2\varphi_k,
\qquad
\hat{\mathbf g}_{\mathbf k}=(\sin\varphi_k,-\cos\varphi_k,0).
\label{eq:4b_a0_kpolar}
\end{equation}
Since the \(\alpha_0\) SOC term is proportional to \(\tau_0\), the spin helicity projectors remain
\(P_\eta(\mathbf k)\) from Eq.~\eqref{eq:4b_alpha_family_proj}. The exact sublattice projectors are
\begin{equation}
 B_{\mathbf k}^{(0)}\equiv\sqrt{\epsilon_{k}^2+\left(\frac{t_4}{2}\right)^2 d_{\mathbf k}^2},
\qquad
 c_{\mathbf k}^{(0)}\equiv\frac{\epsilon_{k}}{B_{\mathbf k}^{(0)}},
\qquad
 b_{\mathbf k}^{(0)}\equiv\frac{\frac{t_4}{2} d_{\mathbf k}}{B_{\mathbf k}^{(0)}},
\qquad
 R_s^{(0)}(\mathbf k)\equiv
 \frac12\Big[\tau_0+s\big(c_{\mathbf k}^{(0)}\tau_x+b_{\mathbf k}^{(0)}\tau_z\big)\Big],
\qquad
 s=\pm,
\label{eq:4b_a0_Qs}
\end{equation}
with branch projectors
\begin{equation}
\Pi_{s\eta}^{(0)}(\mathbf k)
=
\begin{pmatrix}
R_s^{(0)}(\mathbf k)P_\eta(\mathbf k) & 0\\
0 & R_s^{(0)}(\mathbf k)P_\eta(\mathbf k)
\end{pmatrix}.
\label{eq:4b_a0_Pi}
\end{equation}
The four quasiparticle branches are
\begin{equation}
\xi_{s\eta}^{(0)}(\mathbf k)=\xi_k+s\,B_{\mathbf k}^{(0)}+\eta\,\alpha_0 k,
\qquad
E_{s\eta}^{(0)}(\mathbf k)=\sqrt{\big(\xi_{s\eta}^{(0)}(\mathbf k)\big)^2+\Delta^2}.
\label{eq:4b_a0_disp}
\end{equation}
\(\tau_z\) no longer flips a pure sublattice branch. Instead, the orbital matrix element entering
\(\delta\mathcal F^{(2)}\) is
\begin{equation}
W_{ss'}^{(0)}(\mathbf k)
\equiv
\operatorname{Tr}_\tau\!\Big[R_s^{(0)}(\mathbf k)\tau_zR_{s'}^{(0)}(\mathbf k)\tau_z\Big]
=
\frac12\Big[1+ss'\big((b_{\mathbf k}^{(0)})^2-(c_{\mathbf k}^{(0)})^2\big)\Big]
=
(b_{\mathbf k}^{(0)})^2\delta_{ss'}+(c_{\mathbf k}^{(0)})^2\delta_{s,-s'}.
\label{eq:4b_a0_tauz_flip}
\end{equation}
Using the spin matrix elements of
Eqs.~\eqref{eq:4b_alpha_spin_same}--\eqref{eq:4b_alpha_spin_off}, the quadratic term can be written as
\begin{equation}
\delta\mathcal F^{(2)}_{(\alpha_0,t_4)}(\mathbf 0,\hat{\mathbf n})
=
J_{\mathrm{ex}}^2\!\int\!\frac{k\,dk\,d\varphi_k}{4(2\pi)^2}
\Big[
\Sigma_{\rm same}^{(0)}(\mathbf k)\,
(\mathbf n_{\parallel}\!\cdot\!\hat{\mathbf g}_{\mathbf k})^2
+
\Sigma_{\rm off}^{(0)}(\mathbf k)\,
\Big(1-(\mathbf n_{\parallel}\!\cdot\!\hat{\mathbf g}_{\mathbf k})^2\Big)
\Big],
\label{eq:4b_a0_F0_final}
\end{equation}
with
\begin{equation}
\Sigma_{\rm same}^{(0)}(\mathbf k)
=
\sum_{s,s'=\pm}\sum_{\eta=\pm}
W_{ss'}^{(0)}(\mathbf k)\,
\widetilde S_{(s,\eta),(s',\eta)}^{(0)}(\mathbf k),
\qquad
\Sigma_{\rm off}^{(0)}(\mathbf k)
=
\sum_{s,s'=\pm}\sum_{\eta=\pm}
W_{ss'}^{(0)}(\mathbf k)\,
\widetilde S_{(s,\eta),(s',-\eta)}^{(0)}(\mathbf k).
\label{eq:4b_a0_sigmas}
\end{equation}
Because \(B_{\mathbf k}^{(0)}\), \(b_{\mathbf k}^{(0)}\), and \(c_{\mathbf k}^{(0)}\) depend on
\(\varphi_k\) only through \(\cos^2 2\varphi_k\), the angular integral obeys
\begin{equation}
\int_0^{2\pi}\!d\varphi_k\,f\!\left(\cos^2 2\varphi_k\right)
(\mathbf n_{\parallel}\!\cdot\!\hat{\mathbf g}_{\mathbf k})^2
=
\frac{1-n_z^2}{2}\int_0^{2\pi}\!d\varphi_k\,f\!\left(\cos^2 2\varphi_k\right).
\label{eq:4b_a0_phiavg}
\end{equation}
Therefore, the N\'eel-vector dependence still separates as
\begin{equation}
\delta\mathcal F^{(2)}_{(\alpha_0,t_4)}(\mathbf 0,\hat{\mathbf n})
=
J_{\mathrm{ex}}^2\Big[A_0^{(\alpha_0,t_4)}+A_z^{(\alpha_0,t_4)}\,n_z^2\Big]
+\mathcal O(J_{\mathrm{ex}}^4),
\end{equation}
with the exact coefficients within the continuum model
\begin{align}
A_0^{(\alpha_0,t_4)}
&=
\frac{1}{32\pi^2}\int_0^{k_c}\!dk\,k\int_0^{2\pi}\!d\varphi_k\,
\Big[\Sigma_{\rm same}^{(0)}(\mathbf k)+\Sigma_{\rm off}^{(0)}(\mathbf k)\Big],
\label{eq:4b_a0_A0}
\\
A_z^{(\alpha_0,t_4)}
&=
-\frac{1}{32\pi^2}\int_0^{k_c}\!dk\,k\int_0^{2\pi}\!d\varphi_k\,
\Big[\Sigma_{\rm same}^{(0)}(\mathbf k)-\Sigma_{\rm off}^{(0)}(\mathbf k)\Big].
\label{eq:4b_a0_Az}
\end{align}

\subsubsection{Derivation of \(\Gamma_i\)}
\label{app:4b_alpha0_t3_Gamma}

To linear order in \(\mathbf q\), the Hamiltonian is
\begin{equation}
\mathcal H_{0,\mathbf k,\mathbf q}^{(0)}
=
\mathcal H_{0,\mathbf k}^{(0)}+q_i\,v_i^{(\alpha_0)}(\mathbf k)+\mathcal O(q^2),
\label{eq:4b_a0_Hq}
\end{equation}
with
\begin{align}
v_x^{(\alpha_0)}(\mathbf k)
&=
\Big[
\bar{t} k\cos\varphi_k\,\tau_0
-\frac{t_3}{8}k\cos\varphi_k\,\tau_x
-\frac{t_4}{2} k\cos\varphi_k\,\tau_z
-\frac{\alpha_0}{2}\sigma_y\tau_0
\Big]\rho_0,
\label{eq:4b_a0_vx}
\\
v_y^{(\alpha_0)}(\mathbf k)
&=
\Big[
\bar{t} k\sin\varphi_k\,\tau_0
-\frac{t_3}{8}k\sin\varphi_k\,\tau_x
+\frac{t_4}{2} k\sin\varphi_k\,\tau_z
+\frac{\alpha_0}{2}\sigma_x\tau_0
\Big]\rho_0.
\label{eq:4b_a0_vy}
\end{align}
Equation~\eqref{eq:4b_Flin_expand} then gives the explicit bubble
\begin{equation}
\Gamma_i^{(\alpha_0,t_4)}
=
\frac{J_{\mathrm{ex}}T}{2}\sum_{\omega_n}\int_{\mathbf k}
\operatorname{Tr}\!\left[
\mathcal G_0^{(0)}(\mathbf k,\omega_n)\,
\mathcal V_m\,
\mathcal G_0^{(0)}(\mathbf k,\omega_n)\,
v_i^{(\alpha_0)}(\mathbf k)
\right].
\label{eq:4b_a0_Gamma_bubble}
\end{equation}
After the spin--sublattice trace and the \(\varphi_k\) integral, the only allowed in-plane tensor structure is
\begin{equation}
\Gamma_x^{(\alpha_0,t_4)}=J_{\mathrm{ex}}\,\gamma_{\alpha_0}^{\rm ex}(t_4)\,n_y,
\qquad
\Gamma_y^{(\alpha_0,t_4)}=J_{\mathrm{ex}}\,\gamma_{\alpha_0}^{\rm ex}(t_4)\,n_x,
\label{eq:4b_a0_Gamma_tensor}
\end{equation}
and the exact coefficient may be defined by choosing \(\hat{\mathbf n}=\hat{\mathbf y}\),
\begin{equation}
\gamma_{\alpha_0}^{\rm ex}(t_4)
=
\frac{T}{2}\sum_{\omega_n}\int_{\mathbf k}
\operatorname{Tr}\!\left[
\mathcal G_0^{(0)}(\mathbf k,\omega_n)\,
\tau_z\sigma_y\,
\mathcal G_0^{(0)}(\mathbf k,\omega_n)\,
v_x^{(\alpha_0)}(\mathbf k)
\right].
\label{eq:4b_a0_gamma_split}
\end{equation}
Therefore
\begin{equation}
\delta\mathcal F_{{\rm lin},(\alpha_0,t_4)}(\mathbf q,\hat{\mathbf n})
=
J_{\mathrm{ex}}\,\gamma_{\alpha_0}^{\rm ex}(t_4)\,(q_xn_y+q_yn_x).
\label{eq:4b_a0_Flin_final}
\end{equation}

In the small-$t_4$ limit, and away from accidental band
crossings, the response coefficient is odd in $t_4$,
\begin{equation}
\gamma_{\alpha_0}^{\mathrm{ex}}(t_4)
=
\frac{t_4}{2}\,\gamma_{\alpha_0}^{(1)}
+\mathcal O(t_4^3).
\end{equation}
For this expansion, all kernels below are evaluated at $t_4=0$,
where the branches may be labeled by the eigenvalue $s=\pm$ of
$\tau_x$ and the helicity $\eta=\pm$,
\begin{equation}
\xi_{s\eta}^{(0,0)}(k)
=\xi_k+s\epsilon_k+\eta\alpha_0 k.
\end{equation}
Define
\begin{equation}
\mathcal X_{s,\eta}^{(\alpha_0)}(k)
\equiv
\mathcal K^{(1)}_{(s,\eta),(-s,\eta),(s,\eta)}{}^{(\alpha_0,0)}(k)
+
\mathcal K^{(2)}_{(s,\eta),(-s,\eta),(s,\eta)}{}^{(\alpha_0,0)}(k),
\end{equation}
and
\begin{align}
\mathcal W_{{\rm same},s}^{(\alpha_0)}(k)
&\equiv
\sum_{\eta=\pm}
\Big[
\mathcal K^{(1)}_{(s,\eta),(-s,\eta),(s,\eta)}{}^{(\alpha_0,0)}(k)
+
\mathcal K^{(2)}_{(s,\eta),(-s,\eta),(s,\eta)}{}^{(\alpha_0,0)}(k)
\Big],
\\
\mathcal W_{{\rm off},s}^{(\alpha_0)}(k)
&\equiv
\sum_{\eta=\pm}
\Big[
\mathcal K^{(1)}_{(s,\eta),(-s,-\eta),(s,-\eta)}{}^{(\alpha_0,0)}(k)
+
\mathcal K^{(2)}_{(s,\eta),(-s,\eta),(s,-\eta)}{}^{(\alpha_0,0)}(k)
\Big].
\end{align}
The linear coefficient separates into four radial contributions,
\begin{equation}
\gamma_{\alpha_0}^{(1)}
=
\gamma_{\rm contact}^{(\alpha_0)}
+\gamma_{\rm kin}^{(\alpha_0)}
+\gamma_{t_3}^{(\alpha_0)}
+\gamma_{\rm soc}^{(\alpha_0)},
\end{equation}
where
\begin{align}
\gamma_{\rm contact}^{(\alpha_0)}
&=
\frac{1}{8\pi}\int_0^{k_c}\!dk\,k^2
\sum_{s=\pm}
\left[
\widetilde S_{(s,+),(-s,+)}^{(0,0)}(k)
-
\widetilde S_{(s,-),(-s,-)}^{(0,0)}(k)
\right],
\\
\gamma_{\rm kin}^{(\alpha_0)}
&=
-\frac{\bar t}{16\pi}\int_0^{k_c}\!dk\,k^4
\sum_{s=\pm}
\left[
\mathcal X_{s,+}^{(\alpha_0)}(k)
-
\mathcal X_{s,-}^{(\alpha_0)}(k)
\right],
\\
\gamma_{t_3}^{(\alpha_0)}
&=
\frac{t_3}{128\pi}\int_0^{k_c}\!dk\,k^4
\sum_{s=\pm}s
\left[
\mathcal X_{s,+}^{(\alpha_0)}(k)
-
\mathcal X_{s,-}^{(\alpha_0)}(k)
\right],
\\
\gamma_{\rm soc}^{(\alpha_0)}
&=
\frac{\alpha_0}{32\pi}\int_0^{k_c}\!dk\,k^3
\sum_{s=\pm}
\left[
\mathcal W_{{\rm off},s}^{(\alpha_0)}(k)
-
\mathcal W_{{\rm same},s}^{(\alpha_0)}(k)
\right].
\end{align}

\subsection{\(\alpha_x\)-type SOC}
\label{app:4b_soc_alphax}

\subsubsection{Zeroth-order free energy at \(\mathbf q=\mathbf 0\)}
\label{app:4b_alphax_t3_F0}

For the \(\alpha_x\) SOC, using the same spin helicity projectors \(P_\eta(\mathbf k)\), the \(\eta=\pm\) sector sees the exact
sublattice pseudospin field
\begin{equation}
A_{\eta,k}^{(x)}\equiv \epsilon_{k}+\eta\,\alpha_x k,
\qquad
B_{\eta,\mathbf k}^{(x)}\equiv
\sqrt{\big(A_{\eta,k}^{(x)}\big)^2+\frac{t_4^2}{4} d_{\mathbf k}^2},
\qquad
c_{\eta,\mathbf k}^{(x)}\equiv\frac{A_{\eta,k}^{(x)}}{B_{\eta,\mathbf k}^{(x)}},
\qquad
b_{\eta,\mathbf k}^{(x)}\equiv\frac{t_4 d_{\mathbf k}}{2B_{\eta,\mathbf k}^{(x)}},
\label{eq:4b_ax_Qs}
\end{equation}
and sublattice projectors
\begin{equation}
\!\!
R_{s\eta}^{(x)}(\mathbf k)\equiv
\frac12\Big[\tau_0+s\big(c_{\eta,\mathbf k}^{(x)}\tau_x+b_{\eta,\mathbf k}^{(x)}\tau_z\big)\Big],
\qquad
\Pi_{s\eta}^{(x)}(\mathbf k)
=
\begin{pmatrix}
R_{s\eta}^{(x)}(\mathbf k)P_\eta(\mathbf k) & 0\\
0 & R_{s\eta}^{(x)}(\mathbf k)P_\eta(\mathbf k)
\end{pmatrix},
\label{eq:4b_ax_disp}
\end{equation}
so that the four branches are
\begin{equation}
\xi_{s\eta}^{(x)}(\mathbf k)=\xi_k+s\,B_{\eta,\mathbf k}^{(x)},
\qquad
E_{s\eta}^{(x)}(\mathbf k)=\sqrt{\big(\xi_{s\eta}^{(x)}(\mathbf k)\big)^2+\Delta^2}.
\end{equation}
Because the sublattice projector now depends on \(\eta\), the exchange vertex carries the weight
\begin{equation}
W_{ss'}^{(x)}(\mathbf k;\eta,\eta')
\equiv
\operatorname{Tr}_\tau\!\Big[
R_{s\eta}^{(x)}(\mathbf k)\tau_z
R_{s'\eta'}^{(x)}(\mathbf k)\tau_z
\Big]
=
\frac12\Big[
1+ss'\big(
b_{\eta,\mathbf k}^{(x)}b_{\eta',\mathbf k}^{(x)}
-c_{\eta,\mathbf k}^{(x)}c_{\eta',\mathbf k}^{(x)}
\big)
\Big].
\end{equation}
The quadratic free energy is therefore
\begin{equation}
\delta\mathcal F^{(2)}_{(\alpha_x,t_4)}(\mathbf 0,\hat{\mathbf n})
=
J_{\mathrm{ex}}^2\!\int\!\frac{k\,dk\,d\varphi_k}{4(2\pi)^2}
\Big[
\Sigma_{\rm same}^{(x)}(\mathbf k)\,
(\mathbf n_{\parallel}\!\cdot\!\hat{\mathbf g}_{\mathbf k})^2
+
\Sigma_{\rm off}^{(x)}(\mathbf k)\,
\Big(1-(\mathbf n_{\parallel}\!\cdot\!\hat{\mathbf g}_{\mathbf k})^2\Big)
\Big],
\end{equation}
with
\begin{equation}
\Sigma_{\rm same}^{(x)}(\mathbf k)
=
\sum_{s,s'=\pm}\sum_{\eta=\pm}
W_{ss'}^{(x)}(\mathbf k;\eta,\eta)\,
\widetilde S_{(s,\eta),(s',\eta)}^{(x)}(\mathbf k),
\qquad
\Sigma_{\rm off}^{(x)}(\mathbf k)
=
\sum_{s,s'=\pm}\sum_{\eta=\pm}
W_{ss'}^{(x)}(\mathbf k;\eta,-\eta)\,
\widetilde S_{(s,\eta),(s',-\eta)}^{(x)}(\mathbf k),
\label{eq:4b_ax_sigmas}
\end{equation}
and the same angular-symmetry argument as in Eq.~\eqref{eq:4b_a0_phiavg} yields
\begin{equation}
\delta\mathcal F^{(2)}_{(\alpha_x,t_4)}(\mathbf 0,\hat{\mathbf n})
=
J_{\mathrm{ex}}^2\Big[A_0^{(\alpha_x,t_4)}+A_z^{(\alpha_x,t_4)}\,n_z^2\Big]
+\mathcal O(J_{\mathrm{ex}}^4),
\label{eq:4b_ax_F0_final}
\end{equation}
with the exact coefficients within the continuum model
\begin{align}
A_0^{(\alpha_x,t_4)}
&=
\frac{1}{32\pi^2}\int_0^{k_c}\!dk\,k\int_0^{2\pi}\!d\varphi_k\,
\Big[\Sigma_{\rm same}^{(x)}(\mathbf k)+\Sigma_{\rm off}^{(x)}(\mathbf k)\Big],
\label{eq:4b_ax_A0}
\\
A_z^{(\alpha_x,t_4)}
&=
-\frac{1}{32\pi^2}\int_0^{k_c}\!dk\,k\int_0^{2\pi}\!d\varphi_k\,
\Big[\Sigma_{\rm same}^{(x)}(\mathbf k)-\Sigma_{\rm off}^{(x)}(\mathbf k)\Big].
\label{eq:4b_ax_Az}
\end{align}

\subsubsection{Derivation of \(\Gamma_i\)}
\label{app:4b_alphax_t3_Gamma}

To linear order in \(\mathbf q\), we write
\begin{equation}
\mathcal H_{0,\mathbf k,\mathbf q}^{(x)}
=
\mathcal H_{0,\mathbf k}^{(x)}+q_i\,v_i^{(\alpha_x)}(\mathbf k)+\mathcal O(q^2),
\end{equation}
with
\begin{align}
v_x^{(\alpha_x)}(\mathbf k)
&=
\Big[
\bar{t} k\cos\varphi_k\,\tau_0
-\frac{t_3}{8}k\cos\varphi_k\,\tau_x
-\frac{t_4}{2} k\cos\varphi_k\,\tau_z
-\frac{\alpha_x}{2}\sigma_y\tau_x
\Big]\rho_0,
\label{eq:4b_ax_vx}
\\
v_y^{(\alpha_x)}(\mathbf k)
&=
\Big[
\bar{t} k\sin\varphi_k\,\tau_0
-\frac{t_3}{8}k\sin\varphi_k\,\tau_x
+\frac{t_4}{2} k\sin\varphi_k\,\tau_z
+\frac{\alpha_x}{2}\sigma_x\tau_x
\Big]\rho_0.
\label{eq:4b_ax_vy}
\end{align}
Equation~\eqref{eq:4b_Flin_expand} gives
\begin{equation}
\Gamma_i^{(\alpha_x,t_4)}
=
\frac{J_{\mathrm{ex}}T}{2}\sum_{\omega_n}\int_{\mathbf k}
\operatorname{Tr}\!\left[
\mathcal G_0^{(x)}(\mathbf k,\omega_n)\,
\mathcal V_m\,
\mathcal G_0^{(x)}(\mathbf k,\omega_n)\,
v_i^{(\alpha_x)}(\mathbf k)
\right].
\end{equation}
After evaluating the spin--sublattice trace and performing the angular integration, one again obtains the unique in-plane structure
\begin{equation}
\Gamma_x^{(\alpha_x,t_4)}=J_{\mathrm{ex}}\,\gamma_{\alpha_x}^{\rm ex}(t_4)\,n_y,
\qquad
\Gamma_y^{(\alpha_x,t_4)}=J_{\mathrm{ex}}\,\gamma_{\alpha_x}^{\rm ex}(t_4)\,n_x,
\label{eq:4b_ax_Gamma_tensor}
\end{equation}
where the exact coefficient may again be defined by choosing \(\hat{\mathbf n}=\hat{\mathbf y}\),
\begin{equation}
\gamma_{\alpha_x}^{\rm ex}(t_4)
=
\frac{T}{2}\sum_{\omega_n}\int_{\mathbf k}
\operatorname{Tr}\!\left[
\mathcal G_0^{(x)}(\mathbf k,\omega_n)\,
\tau_z\sigma_y\,
\mathcal G_0^{(x)}(\mathbf k,\omega_n)\,
v_x^{(\alpha_x)}(\mathbf k)
\right].
\label{eq:4b_ax_gamma_split}
\end{equation}
Hence
\begin{equation}
\delta\mathcal F_{{\rm lin},(\alpha_x,t_4)}(\mathbf q,\hat{\mathbf n})
=
J_{\mathrm{ex}}\,\gamma_{\alpha_x}^{\rm ex}(t_4)\,(q_xn_y+q_yn_x).
\label{eq:4b_ax_Flin_final}
\end{equation}

In the small-$t_4$ limit,
\begin{equation}
\gamma_{\alpha_x}^{\mathrm{ex}}(t_4)
=
\frac{t_4}{2}\,\gamma_{\alpha_x}^{(1)}
+\mathcal O(t_4^3).
\end{equation}
At $t_4=0$, it is convenient to label the branches by the
$\tau_x$ eigenvalue $s$ and helicity $\eta$,
\begin{equation}
\xi_{s\eta}^{(x,0)}(k)
=\xi_k+s\big(\epsilon_k+\eta\alpha_x k\big).
\end{equation}
Define
\begin{equation}
\mathcal X_{s,\eta}^{(\alpha_x)}(k)
\equiv
\mathcal K^{(1)}_{(s,\eta),(-s,\eta),(s,\eta)}{}^{(\alpha_x,0)}(k)
+
\mathcal K^{(2)}_{(s,\eta),(-s,\eta),(s,\eta)}{}^{(\alpha_x,0)}(k),
\end{equation}
and
\begin{align}
\mathcal W_{{\rm same},s}^{(\alpha_x)}(k)
&\equiv
\sum_{\eta=\pm}
\Big[
\mathcal K^{(1)}_{(s,\eta),(-s,\eta),(s,\eta)}{}^{(\alpha_x,0)}(k)
+
\mathcal K^{(2)}_{(s,\eta),(-s,\eta),(s,\eta)}{}^{(\alpha_x,0)}(k)
\Big],
\\
\mathcal W_{{\rm off},s}^{(\alpha_x)}(k)
&\equiv
\sum_{\eta=\pm}
\Big[
\mathcal K^{(1)}_{(s,\eta),(-s,-\eta),(s,-\eta)}{}^{(\alpha_x,0)}(k)
+
\mathcal K^{(2)}_{(s,\eta),(-s,\eta),(s,-\eta)}{}^{(\alpha_x,0)}(k)
\Big].
\end{align}
Then
\begin{equation}
\gamma_{\alpha_x}^{(1)}
=
\gamma_{\rm contact}^{(\alpha_x)}
+\gamma_{\rm kin}^{(\alpha_x)}
+\gamma_{t_3}^{(\alpha_x)}
+\gamma_{\rm soc}^{(\alpha_x)},
\end{equation}
with
\begin{align}
\gamma_{\rm contact}^{(\alpha_x)}
&=
\frac{1}{8\pi}\int_0^{k_c}\!dk\,k^2
\sum_{s=\pm}
\left[
\widetilde S_{(s,+),(-s,+)}^{(x,0)}(k)
-
\widetilde S_{(s,-),(-s,-)}^{(x,0)}(k)
\right],
\\
\gamma_{\rm kin}^{(\alpha_x)}
&=
-\frac{\bar t}{16\pi}\int_0^{k_c}\!dk\,k^4
\sum_{s=\pm}
\left[
\mathcal X_{s,+}^{(\alpha_x)}(k)
-
\mathcal X_{s,-}^{(\alpha_x)}(k)
\right],
\\
\gamma_{t_3}^{(\alpha_x)}
&=
\frac{t_3}{128\pi}\int_0^{k_c}\!dk\,k^4
\sum_{s=\pm}s
\left[
\mathcal X_{s,+}^{(\alpha_x)}(k)
-
\mathcal X_{s,-}^{(\alpha_x)}(k)
\right],
\\
\gamma_{\rm soc}^{(\alpha_x)}
&=
\frac{\alpha_x}{32\pi}\int_0^{k_c}\!dk\,k^3
\sum_{s=\pm}s
\left[
\mathcal W_{{\rm off},s}^{(\alpha_x)}(k)
-
\mathcal W_{{\rm same},s}^{(\alpha_x)}(k)
\right].
\end{align}
At $t_3=0$, the $t_4=0$ spectrum is invariant under
$(s,\eta)\mapsto(-s,-\eta)$, and the four contributions above
cancel after summing over the branches. Consequently,
$\gamma_{\alpha_x}^{\mathrm{ex}}(t_4)=0$ at $t_3=0$ within the
present model.

\subsection{\(\beta_z\)-type SOC}
\label{app:4b_soc_betaz}

For the \(\beta_z\) SOC, using \(\mathbf k=(k\cos\varphi_k,k\sin\varphi_k)\), define the spin-helicity axis
\begin{equation}
\hat{\mathbf u}_{\mathbf k}=\frac{(k_y,k_x,0)}{k}
=(\sin\varphi_k,\cos\varphi_k,0),
\qquad
U_\eta^{(z)}(\mathbf k)=
\frac12\Big(\sigma_0+\eta\,\hat{\mathbf u}_{\mathbf k}\!\cdot\!\boldsymbol\sigma\Big),
\qquad
\eta=\pm.
\label{eq:4b_bz_Bk}
\end{equation}
In the \(\eta\) sector the sublattice pseudospin field is
\begin{equation}
\!\!
A_{\eta,\mathbf k}^{(z)}\equiv \frac{t_4}{2} d_{\mathbf k}+\eta\,\beta_z k,
\qquad
B_{\eta,\mathbf k}^{(z)}\equiv\sqrt{\epsilon_{k}^2+\big(A_{\eta,\mathbf k}^{(z)}\big)^2},
\qquad
c_{\eta,\mathbf k}^{(z)}\equiv\frac{\epsilon_{k}}{B_{\eta,\mathbf k}^{(z)}},
\qquad
b_{\eta,\mathbf k}^{(z)}\equiv\frac{A_{\eta,\mathbf k}^{(z)}}{B_{\eta,\mathbf k}^{(z)}}.
\label{eq:4b_bz_delta}
\end{equation}
The exact sublattice projectors are
\begin{equation}
R_{s\eta}^{(z)}(\mathbf k)\equiv
\frac12\Big[\tau_0+s\big(c_{\eta,\mathbf k}^{(z)}\tau_x+b_{\eta,\mathbf k}^{(z)}\tau_z\big)\Big],
\qquad
\Pi_{s\eta}^{(z)}(\mathbf k)
=
\begin{pmatrix}
R_{s\eta}^{(z)}(\mathbf k)U_\eta^{(z)}(\mathbf k) & 0\\
0 & R_{s\eta}^{(z)}(\mathbf k)U_\eta^{(z)}(\mathbf k)
\end{pmatrix},
\qquad
s=\pm,
\label{eq:4b_bz_proj}
\end{equation}
with four branches
\begin{equation}
\xi_{s\eta}^{(z)}(\mathbf k)=\xi_k+s\,B_{\eta,\mathbf k}^{(z)},
\qquad
E_{s\eta}^{(z)}(\mathbf k)=\sqrt{\big(\xi_{s\eta}^{(z)}(\mathbf k)\big)^2+\Delta^2}.
\label{eq:4b_bz_disp}
\end{equation}
The orbital part of the exchange vertex contributes the weight
\begin{equation}
W_{ss'}^{(z)}(\mathbf k;\eta,\eta')
\equiv
\operatorname{Tr}_\tau\!\Big[
R_{s\eta}^{(z)}(\mathbf k)\tau_z
R_{s'\eta'}^{(z)}(\mathbf k)\tau_z
\Big]
=
\frac12\Big[
1+ss'\big(
b_{\eta,\mathbf k}^{(z)}b_{\eta',\mathbf k}^{(z)}
-c_{\eta,\mathbf k}^{(z)}c_{\eta',\mathbf k}^{(z)}
\big)
\Big].
\label{eq:4b_bz_Sdef}
\end{equation}
The quadratic term takes the form
\begin{equation}
\delta\mathcal F^{(2)}_{(\beta_z,t_4)}(\mathbf 0,\hat{\mathbf n})
=
J_{\mathrm{ex}}^2\!\int\!\frac{k\,dk\,d\varphi_k}{4(2\pi)^2}
\Big[
\Sigma_{\rm same}^{(z)}(\mathbf k)\,
(\mathbf n_{\parallel}\!\cdot\!\hat{\mathbf u}_{\mathbf k})^2
+
\Sigma_{\rm off}^{(z)}(\mathbf k)\,
\Big(1-(\mathbf n_{\parallel}\!\cdot\!\hat{\mathbf u}_{\mathbf k})^2\Big)
\Big]
+\mathcal O(J_{\mathrm{ex}}^4),
\label{eq:4b_bz_F0_final}
\end{equation}
with
\begin{align}
\Sigma_{\rm same}^{(z)}(\mathbf k)
&=
\sum_{s,s'=\pm}\sum_{\eta=\pm}
W_{ss'}^{(z)}(\mathbf k;\eta,\eta)\,
\widetilde S_{(s,\eta),(s',\eta)}^{(z)}(\mathbf k),
\label{eq:4b_bz_A0}
\\
\Sigma_{\rm off}^{(z)}(\mathbf k)
&=
\sum_{s,s'=\pm}\sum_{\eta=\pm}
W_{ss'}^{(z)}(\mathbf k;\eta,-\eta)\,
\widetilde S_{(s,\eta),(s',-\eta)}^{(z)}(\mathbf k).
\label{eq:4b_bz_Az}
\end{align}
After summing over \(\eta=\pm\), the \(\varphi_k\) dependence is \(C_4\)-symmetric, so the
N\'eel-vector dependence again separates as
\begin{equation}
\delta\mathcal F^{(2)}_{(\beta_z,t_4)}(\mathbf 0,\hat{\mathbf n})
=
J_{\mathrm{ex}}^2\Big[A_0^{(\beta_z,t_4)}+A_z^{(\beta_z,t_4)}\,n_z^2\Big]
+\mathcal O(J_{\mathrm{ex}}^4),
\end{equation}
with the exact coefficients within the continuum model
\begin{align}
A_0^{(\beta_z,t_4)}
&=
\frac{1}{32\pi^2}\int_0^{k_c}\!dk\,k\int_0^{2\pi}\!d\varphi_k\,
\Big[\Sigma_{\rm same}^{(z)}(\mathbf k)+\Sigma_{\rm off}^{(z)}(\mathbf k)\Big],
\\
A_z^{(\beta_z,t_4)}
&=
-\frac{1}{32\pi^2}\int_0^{k_c}\!dk\,k\int_0^{2\pi}\!d\varphi_k\,
\Big[\Sigma_{\rm same}^{(z)}(\mathbf k)-\Sigma_{\rm off}^{(z)}(\mathbf k)\Big].
\end{align}

\subsubsection{Derivation of \(\Gamma_i\)}
\label{app:4b_betaz_t3_Gamma}

For the \(\beta_z\) channel, the linear response can remain nonzero already at \(O(J_{\mathrm{ex}}q)\), even in the limit \(t_4\to0\).
The \(q\)-expansion reads
\begin{equation}
\mathcal H_{0,\mathbf k,\mathbf q}^{(z)}
=
\mathcal H_{0,\mathbf k}^{(z)}+q_i\,v_i^{(\beta_z)}(\mathbf k)+\mathcal O(q^2),
\end{equation}
with
\begin{align}
v_x^{(\beta_z)}(\mathbf k)
&=
\Big[
\bar{t} k\cos\varphi_k\,\tau_0
-\frac{t_3}{8}k\cos\varphi_k\,\tau_x
-\frac{t_4}{2} k\cos\varphi_k\,\tau_z
+\frac{\beta_z}{2}\sigma_y\tau_z
\Big]\rho_0,
\label{eq:4b_bz_gamma_kin}
\\
v_y^{(\beta_z)}(\mathbf k)
&=
\Big[
\bar{t} k\sin\varphi_k\,\tau_0
-\frac{t_3}{8}k\sin\varphi_k\,\tau_x
+\frac{t_4}{2} k\sin\varphi_k\,\tau_z
+\frac{\beta_z}{2}\sigma_x\tau_z
\Big]\rho_0.
\label{eq:4b_bz_gamma_t3}
\end{align}
Equation~\eqref{eq:4b_Flin_expand} gives
\begin{equation}
\Gamma_i^{(\beta_z,t_4)}
=
\frac{J_{\mathrm{ex}}T}{2}\sum_{\omega_n}\int_{\mathbf k}
\operatorname{Tr}\!\left[
\mathcal G_0^{(z)}(\mathbf k,\omega_n)\,
\mathcal V_m\,
\mathcal G_0^{(z)}(\mathbf k,\omega_n)\,
v_i^{(\beta_z)}(\mathbf k)
\right].
\label{eq:4b_bz_gamma_soc}
\end{equation}
After evaluating the spin--sublattice trace and performing the angular average, the linear coefficient has the in-plane form
\begin{equation}
\Gamma_x^{(\beta_z,t_4)}=J_{\mathrm{ex}}\,\gamma_{\beta_z}^{\rm ex}(t_4)\,n_y,
\qquad
\Gamma_y^{(\beta_z,t_4)}=J_{\mathrm{ex}}\,\gamma_{\beta_z}^{\rm ex}(t_4)\,n_x,
\label{eq:4b_bz_Gamma_tensor}
\end{equation}
where the exact coefficient may be extracted by choosing \(\hat{\mathbf n}=\hat{\mathbf y}\),
\begin{equation}
\gamma_{\beta_z}^{\rm ex}(t_4)
=
\frac{T}{2}\sum_{\omega_n}\int_{\mathbf k}
\operatorname{Tr}\!\left[
\mathcal G_0^{(z)}(\mathbf k,\omega_n)\,
\tau_z\sigma_y\,
\mathcal G_0^{(z)}(\mathbf k,\omega_n)\,
v_x^{(\beta_z)}(\mathbf k)
\right].
\label{eq:4b_bz_gamma_split}
\end{equation}
Therefore
\begin{equation}
\delta\mathcal F_{{\rm lin},(\beta_z,t_4)}(\mathbf q,\hat{\mathbf n})
=
J_{\mathrm{ex}}\,\gamma_{\beta_z}^{\rm ex}(t_4)\,(q_xn_y+q_yn_x).
\label{eq:4b_bz_Flin_final}
\end{equation}

In the small-$t_4$ limit, the coefficient is even in $t_4$,
\begin{equation}
\gamma_{\beta_z}^{\mathrm{ex}}(t_4)
=
\gamma_{\beta_z}^{(0)}
+\mathcal O(t_4^2),
\qquad
\gamma_{\beta_z}^{(0)}
=
\gamma_{\rm kin}^{(\beta_z)}
+\gamma_{t_3}^{(\beta_z)}
+\gamma_{\rm soc}^{(\beta_z)}.
\end{equation}
At $t_4=0$, define
\begin{equation}
B_k^{(z,0)}
=\sqrt{\epsilon_k^2+\beta_z^2k^2},
\qquad
b_k^{(z,0)}
=\frac{\beta_z k}{B_k^{(z,0)}},
\qquad
c_k^{(z,0)}
=\frac{\epsilon_k}{B_k^{(z,0)}},
\end{equation}
\begin{equation}
\xi_s^{(z,0)}(k)=\xi_k+sB_k^{(z,0)},
\qquad
E_s^{(z,0)}(k)
=\sqrt{\big(\xi_s^{(z,0)}(k)\big)^2+\Delta^2},
\end{equation}
and let
$\widetilde S_{ss'}^{(z,0)}(k)$ denote the kernel in
Eq.~\eqref{eq:4b_Splus_tilde} evaluated for these two branches.
The three radial contributions are
\begin{align}
\gamma_{\rm kin}^{(\beta_z)}
&=
\frac{\bar t}{4\pi}\int_0^{k_c}\!dk\,k^2
b_k^{(z,0)}
\left[
\widetilde S_{++}^{(z,0)}(k)
-
\widetilde S_{--}^{(z,0)}(k)
\right],
\\
\gamma_{t_3}^{(\beta_z)}
&=
-\frac{t_3}{32\pi}\int_0^{k_c}\!dk\,k^2
b_k^{(z,0)}c_k^{(z,0)}
\left[
\widetilde S_{++}^{(z,0)}(k)
+
\widetilde S_{--}^{(z,0)}(k)
-2\widetilde S_{+-}^{(z,0)}(k)
\right],
\\
\gamma_{\rm soc}^{(\beta_z)}
&=
\frac{\beta_z}{8\pi}\int_0^{k_c}\!dk\,k
\Bigg[
\big(b_k^{(z,0)}\big)^2
\left(
\widetilde S_{++}^{(z,0)}(k)
+
\widetilde S_{--}^{(z,0)}(k)
\right)
\\
&\hspace{42mm}
+2\left(1+\big(c_k^{(z,0)}\big)^2\right)
\widetilde S_{+-}^{(z,0)}(k)
\Bigg].
\end{align}

\section{Symmetry Arguments}
\label{socsymm}

\subsection{$d$-wave model with a $(001)$ interface}

We now determine the symmetry-allowed SOC terms for the extended
four-band model. The $(001)$ surface, whose normal is along
$[001]$, has the surface antisymmetry point group
${}^{2}4\,{}^{1}m_x\,{}^{2}m_d$~Ref.~\cite{Deterministic_AM}. This group is generated by the
sublattice-exchanging fourfold rotation ${}^{2}C_{4z}$ and the
sublattice-preserving mirror ${}^{1}m_x$, corresponding to
reflection in the $yz$ plane. In the basis
\begin{equation}
\Psi_{\mathbf{k}}
=
\left(
c_{A\uparrow},
c_{A\downarrow},
c_{B\uparrow},
c_{B\downarrow}
\right)^{T},
\end{equation}
we choose the representations
\begin{align}
{}^{2}C_{4z}:&\quad
(k_x,k_y)\mapsto(-k_y,k_x),
\qquad
U_{4}=\tau_x e^{-i\pi\sigma_z/4},
\\
{}^{1}m_x:&\quad
(k_x,k_y)\mapsto(-k_x,k_y),
\qquad
U_{m}=i\tau_0\sigma_x.
\end{align}
They give
\begin{align}
U_4(\sigma_x,\sigma_y,\sigma_z)U_4^\dagger
&=
(\sigma_y,-\sigma_x,\sigma_z),
\\
U_4(\tau_x,\tau_y,\tau_z)U_4^\dagger
&=
(\tau_x,-\tau_y,-\tau_z),
\\
U_m(\sigma_x,\sigma_y,\sigma_z)U_m^\dagger
&=
(\sigma_x,-\sigma_y,-\sigma_z),
\end{align}
while the sublattice Pauli matrices are unchanged under
${}^{1}m_x$.

The intersublattice hopping in Eq.~(7),
\begin{equation}
H_{t_3}(\mathbf{k})
=
t_3
\cos\frac{k_x}{2}
\cos\frac{k_y}{2}\,
\tau_x,
\end{equation}
is invariant under both generators:
\begin{equation}
U_4H_{t_3}(\mathbf{k})U_4^\dagger
=
H_{t_3}(-k_y,k_x),
\qquad
U_mH_{t_3}(\mathbf{k})U_m^\dagger
=
H_{t_3}(-k_x,k_y).
\end{equation}
Thus, the $t_3$ hopping does not reduce the surface antisymmetry
group and does not alter the symmetry classification of the SOC.

We next consider the most general Hermitian,
N\'eel-vector-independent spin-dependent term linear in momentum,
\begin{equation}
H_{\mathrm{SOC}}^{\mathrm{lin}}(\mathbf{k})
=
\sum_{a=0,x,y,z}
\sum_{\mu=x,y}
\sum_{\nu=x,y,z}
c_{\mu\nu}^{(a)}
k_\mu\tau_a\sigma_\nu.
\end{equation}
We require this SOC sector to be time-reversal even,
$\Theta H_{\mathrm{SOC}}(\mathbf{k})\Theta^{-1}
=H_{\mathrm{SOC}}(-\mathbf{k})$, with
$\Theta=i\sigma_yK$. For the linear-in-$\mathbf{k}$ ansatz above,
this condition eliminates all $\tau_y$ structures. The mirror
${}^{1}m_x$ further leaves only
\begin{equation}
k_y\sigma_x,\qquad
k_x\sigma_y,\qquad
k_x\sigma_z.
\end{equation}

The fourfold operation eliminates the remaining $\sigma_z$
channel. Introducing
\begin{equation}
\mathcal R(\mathbf{k})
=
k_y\sigma_x-k_x\sigma_y,
\qquad
\mathcal D(\mathbf{k})
=
k_y\sigma_x+k_x\sigma_y,
\end{equation}
we find
\begin{equation}
U_4\mathcal R(\mathbf{k})U_4^\dagger
=
\mathcal R(C_{4z}\mathbf{k}),
\qquad
U_4\mathcal D(\mathbf{k})U_4^\dagger
=
-\mathcal D(C_{4z}\mathbf{k}).
\end{equation}
Because $\tau_0$ and $\tau_x$ are even under ${}^{2}C_{4z}$,
whereas $\tau_z$ is odd, the symmetry-allowed products are
\begin{equation}
\mathcal R(\mathbf{k})\tau_0,
\qquad
\mathcal R(\mathbf{k})\tau_x,
\qquad
\mathcal D(\mathbf{k})\tau_z.
\end{equation}
Consequently, the most general time-reversal-even,
N\'eel-vector-independent SOC Hamiltonian to linear order in
momentum is
\begin{equation}
H_{\mathrm{SOC}}^{\mathrm{lin}}(\mathbf{k})
=
(\alpha_0+\alpha_x\tau_x)
(k_y\sigma_x-k_x\sigma_y)
+
\beta_z\tau_z
(k_y\sigma_x+k_x\sigma_y).
\end{equation}
The lattice regularization $k_\mu\mapsto\sin k_\mu$ then gives
the SOC Hamiltonian used in Eq.~(8), with
$\beta_z$ replaced by $\lambda\beta_z$ when the structural control
parameter $\lambda$ is retained explicitly.

\subsection{$g$-wave model with a [100] side surface}
\label{app:gwave_socsymm}

We now apply the surface-antisymmetry analysis to the hexagonal
$g$-wave model relevant to CrSb and MnTe
\cite{Minimal_models,MnTe_strain}. The bulk antisymmetry point group
of this $B_{1g}$ altermagnetic state is
\begin{equation}
G_g
=
{}^{2}6/{}^{2}m\,{}^{2}m_y\,{}^{1}m_x
=
\left\langle
{}^{2}C_{6z},
{}^{1}C_{2x},
{}^{1}I
\right\rangle ,
\label{eq:gwave_bulk_asg}
\end{equation}
where the superscripts ${}^{1}$ and ${}^{2}$ denote operations that
preserve and exchange the two magnetic sublattices, respectively.
In particular,
${}^{1}C_{3z}=({}^{2}C_{6z})^{2}$.

For an oriented surface with normal along $[100]$, only operations
whose spatial parts leave $\hat{\mathbf{x}}$ invariant belong to the
surface antisymmetry point group (SAPG)
\cite{Deterministic_AM}. Therefore,
\begin{equation}
\begin{aligned}
G_s([100])
&=
\left\{
g\in G_g:
R_g\hat{\mathbf{x}}=\hat{\mathbf{x}}
\right\}
\\
&=
\left\{
{}^{1}E,\,
{}^{1}C_{2x},\,
{}^{2}m_y,\,
{}^{2}m_z
\right\}
=
\left\langle
{}^{2}m_y,\,
{}^{2}m_z
\right\rangle .
\end{aligned}
\label{eq:gwave_surface_asg}
\end{equation}
Here, $m_y$ denotes reflection in the $xz$ plane and $m_z$
denotes reflection in the $xy$ plane. Their product is the
sublattice-preserving twofold rotation
${}^{2}m_y\,{}^{2}m_z={}^{1}C_{2x}$.
The sublattice-preserving mirror ${}^{1}m_x$ of the bulk group is
not retained because it reverses the oriented surface normal.

In the basis
\begin{equation}
\Psi_{\mathbf{k}}
=
\left(
c_{A\uparrow},
c_{A\downarrow},
c_{B\uparrow},
c_{B\downarrow}
\right)^{T},
\label{eq:gwave_symm_basis}
\end{equation}
the two surface generators may be represented as
\begin{equation}
\begin{aligned}
{}^{2}m_y:\quad
&(k_y,k_z)\mapsto(-k_y,k_z),
\\
&(\sigma_x,\sigma_y,\sigma_z)
\mapsto
(-\sigma_x,\sigma_y,-\sigma_z),
\\
&(\tau_x,\tau_y,\tau_z)
\mapsto
(\tau_x,-\tau_y,-\tau_z),
\\
&U_{{}^{2}m_y}
=
i\tau_x\sigma_y ,
\end{aligned}
\label{eq:gwave_symm_my}
\end{equation}
and
\begin{equation}
\begin{aligned}
{}^{2}m_z:\quad
&(k_y,k_z)\mapsto(k_y,-k_z),
\\
&(\sigma_x,\sigma_y,\sigma_z)
\mapsto
(-\sigma_x,-\sigma_y,\sigma_z),
\\
&(\tau_x,\tau_y,\tau_z)
\mapsto
(\tau_x,-\tau_y,-\tau_z),
\\
&U_{{}^{2}m_z}
=
i\tau_x\sigma_z .
\end{aligned}
\label{eq:gwave_symm_mz}
\end{equation}

We next classify the N\'eel-vector-independent interfacial SOC terms
that are linear in the tangential momentum
$\mathbf{k}_{\parallel}=(k_y,k_z)$. We require this SOC sector to be
even under ordinary time reversal,
\begin{equation}
\Theta
H_{\mathrm{SOC}}(\mathbf{k}_{\parallel})
\Theta^{-1}
=
H_{\mathrm{SOC}}(-\mathbf{k}_{\parallel}),
\qquad
\Theta=i\sigma_yK .
\label{eq:gwave_SOC_TR}
\end{equation}
For terms linear in both momentum and spin,
$k_\mu\sigma_\nu$ is time-reversal even, whereas $\tau_y$ is odd
under complex conjugation. Time-reversal symmetry therefore excludes
all $\tau_y$ structures. The most general remaining Hermitian term is
\begin{equation}
H_{\mathrm{SOC},[100]}^{\mathrm{gen}}
(\mathbf{k}_{\parallel})
=
\sum_{a\in\{0,x,z\}}
\sum_{\mu\in\{y,z\}}
\sum_{\nu\in\{x,y,z\}}
c_{\mu\nu}^{(a)}
k_\mu\tau_a\sigma_\nu.
\label{eq:gwave_symm_generalSOC}
\end{equation}

Imposing the ${}^{2}m_y$ symmetry first leaves the structures
\begin{equation}
\begin{aligned}
k_y:\quad&
\tau_0\sigma_x,\quad
\tau_x\sigma_x,\quad
\tau_0\sigma_z,\quad
\tau_x\sigma_z,\quad
\tau_z\sigma_y,
\\
k_z:\quad&
\tau_0\sigma_y,\quad
\tau_x\sigma_y,\quad
\tau_z\sigma_x,\quad
\tau_z\sigma_z.
\end{aligned}
\label{eq:gwave_symm_my_allowed}
\end{equation}
The second generator, ${}^{2}m_z$, eliminates the two
$\sigma_x$ channels. The common invariant subspace of the two
surface generators is therefore
\begin{equation}
\begin{gathered}
k_y\tau_0\sigma_z,\qquad
k_y\tau_x\sigma_z,\qquad
k_z\tau_0\sigma_y,\qquad
k_z\tau_x\sigma_y,
\\
k_y\tau_z\sigma_y,\qquad
k_z\tau_z\sigma_z.
\end{gathered}
\label{eq:gwave_symm_surface_allowed}
\end{equation}

Consequently, the most general time-reversal-even,
N\'eel-vector-independent SOC Hamiltonian that is linear in
$\mathbf{k}_{\parallel}$ and compatible with the ideal $[100]$
SAPG is
\begin{equation}
\begin{aligned}
H_{\mathrm{SOC},[100]}^{(g)}
(\mathbf{k}_{\parallel})
={}&
\left(
\alpha_{0y}\tau_0+\alpha_{xy}\tau_x
\right)
k_y\sigma_z
+
\left(
\alpha_{0z}\tau_0+\alpha_{xz}\tau_x
\right)
k_z\sigma_y
\\
&+
\beta_{zy}\,k_y\tau_z\sigma_y
+
\beta_{zz}\,k_z\tau_z\sigma_z .
\end{aligned}
\label{eq:gwave_symm_SOCcompact}
\end{equation}
The terms proportional to $\tau_z$ are sublattice odd and provide
direct microscopic channels for a staggered N\'eel field linear in
the superflow momentum. At the level of symmetry, these couplings
are compatible with a sublattice-diagonal pairing potential and do
not require an intersublattice pairing amplitude. Symmetry permits
such responses but does not, by itself, require the corresponding
microscopic coefficients to be nonzero.

We next apply Neumann's principle to determine the allowed
superflow-induced torque response. With tangential condensate
momentum
\begin{equation}
\mathbf{q}_{\parallel}=(q_y,q_z),
\end{equation}
the most general free-energy contribution that is linear in both
$\mathbf{q}_{\parallel}$ and the N\'eel vector $\mathbf{n}$ can be
written as
\begin{equation}
\delta\mathcal{F}_{\mathrm{lin}}^{(g,[100])}
=
J_{\mathrm{ex}}
\sum_{i=x,y,z}
\sum_{\mu=y,z}
n_i\Lambda_{i\mu}q_\mu .
\label{eq:g_general_tensor}
\end{equation}
Explicitly,
\begin{equation}
\begin{aligned}
\delta\mathcal{F}_{\mathrm{lin}}^{(g,[100])}
=
J_{\mathrm{ex}}\Big(
&\Lambda_{xy}\,n_xq_y
+\Lambda_{xz}\,n_xq_z
+\Lambda_{yy}\,n_yq_y
+\Lambda_{yz}\,n_yq_z
\\
&+\Lambda_{zy}\,n_zq_y
+\Lambda_{zz}\,n_zq_z
\Big).
\end{aligned}
\label{eq:g_general_all}
\end{equation}

Because $\mathbf{n}$ is a sublattice-odd axial vector, its
transformation under an antisymmetry operation $g$ is
\begin{equation}
\mathbf{n}
\mapsto
\eta_g(\det R_g)R_g\mathbf{n},
\label{eq:gwave_n_transform_general}
\end{equation}
where $\eta_g=-1$ for a sublattice-exchanging operation. For both
${}^{2}m_y$ and ${}^{2}m_z$, one has
$\eta_g=-1$ and $\det R_g=-1$. Therefore,
\begin{equation}
{}^{2}m_y:\qquad
(q_y,q_z)\mapsto(-q_y,q_z),
\qquad
(n_x,n_y,n_z)\mapsto(n_x,-n_y,n_z).
\label{eq:gwave_response_my}
\end{equation}
Invariance under ${}^{2}m_y$ leaves only
\begin{equation}
\Lambda_{yy},\qquad
\Lambda_{xz},\qquad
\Lambda_{zz}.
\end{equation}
The second generator acts as
\begin{equation}
{}^{2}m_z:\qquad
(q_y,q_z)\mapsto(q_y,-q_z),
\qquad
(n_x,n_y,n_z)\mapsto(n_x,n_y,-n_z).
\label{eq:gwave_response_mz}
\end{equation}
Applying this constraint to the remaining terms requires
$\Lambda_{xz}=0$. Hence, the symmetry-allowed response tensor is
\begin{equation}
\hat{\Lambda}^{(g,[100])}
=
\begin{pmatrix}
0 & 0 \\
\Lambda_{yy} & 0 \\
0 & \Lambda_{zz}
\end{pmatrix},
\label{eq:gwave_response_tensor}
\end{equation}
where the rows correspond to $(n_x,n_y,n_z)$ and the columns to
$(q_y,q_z)$. The resulting linear free-energy coupling is
\begin{equation}
\boxed{
\delta\mathcal{F}_{\mathrm{lin}}^{(g,[100])}
(\mathbf{q}_{\parallel},\mathbf{n})
=
J_{\mathrm{ex}}
\left(
\Lambda_{yy}\,q_y n_y
+
\Lambda_{zz}\,q_z n_z
\right).
}
\label{eq:gwave_symm_torque_final}
\end{equation}
The corresponding staggered N\'eel field is
\begin{equation}
\mathbf{h}^{(-)}
=
-\frac{\partial
\delta\mathcal{F}_{\mathrm{lin}}^{(g,[100])}}
{\partial\mathbf{n}}
=
-J_{\mathrm{ex}}
\left(
0,\,
\Lambda_{yy}q_y,\,
\Lambda_{zz}q_z
\right),
\label{eq:gwave_symm_neel_field}
\end{equation}
and the field-like torque is
$\boldsymbol{\tau}_{\mathbf{n}}
=\mathbf{n}\times\mathbf{h}^{(-)}$.

Thus, the $d$-wave $[001]$ surface supports a symmetric
off-diagonal response in the $(x,y)$ plane,
Eq.~\eqref{eq:4b_sym_lin_final}, whereas the $g$-wave $[100]$
side surface supports a diagonal response in the tangential
$(y,z)$ basis, with independent coefficients
$\Lambda_{yy}$ and $\Lambda_{zz}$.

\section{Heterostructure model for the N\'eel field estimate}
\label{app:stack}

\begin{figure}[t]
\centering
\includegraphics[width=0.85\linewidth]{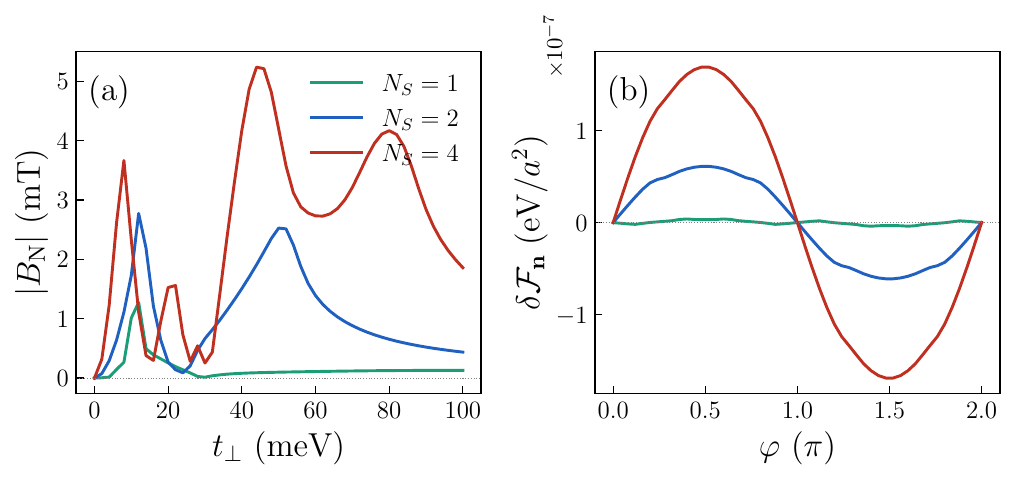}
\caption{%
(a) Magnitude of the supercurrent-induced N\'eel-field difference in the
S/AM stack,
$B_{\rm N}=a^2|\delta\mathcal F_{\hat{\mathbf y}}^{\rm odd}|/m_s$,
evaluated for $\mathbf n=\hat{\mathbf y}$ and
$\mathbf q\parallel\hat{\mathbf x}$, as a function of the interface
hopping $t_\perp$. Results are shown for $N_S=1$, $2$, and $4$
superconducting layers, with adjacent superconducting layers coupled by
$t_S=t$.
(b) In-plane angular dependence of the odd-in-$\mathbf q$ free-energy
density $\delta\mathcal F_{\mathbf n}^{\rm odd}$, defined in
Eq.~\eqref{eq:stack_Fodd}, at $t_\perp=40~\mathrm{meV}$ for
$N_S=1$, $2$, and $4$, with
$\mathbf n=(\cos\varphi,\sin\varphi,0)$. For $N_S=2$ and $4$, the BdG
result closely follows the $\sin\varphi$ dependence implied by
Eq.~\eqref{eq:4b_sym_lin_final}, with higher harmonics contributing at
the few-percent level. For $N_S=1$, the coupling is negligible at this
value of $t_\perp$.
The parameters are $t=0.1~\mathrm{eV}$, $t_4=0.15~\mathrm{eV}$,
$J_{\rm ex}=0.18~\mathrm{eV}$, $\beta_z=10~\mathrm{meV}$,
$\Delta_0=2~\mathrm{meV}$, $\mu=0.03~\mathrm{eV}$,
$a=0.99~\mathrm{nm}$, $q=10^{-3}\,\mathring{\mathrm A}^{-1}$,
$T=0.1T_c$, and $m_s=0.7\,\mu_{\rm B}$.%
}
\label{fig:stack_tp}
\end{figure}

The magnitude estimate presented in the main text is obtained from an
explicit multilayer model of the S/AM heterostructure. The system consists
of $N_S$ superconducting layers with nearest-neighbor interlayer hopping
$t_S$, coupled through an interface hopping $t_\perp$ to a single
altermagnetic layer. This construction follows the tunneling description
of proximity-coupled S/AM heterostructures in
Ref.~\cite{HeinsdorfFranz2026}. The normal-state Hamiltonian is
\begin{align}
H_{\rm stack}^{\rm N}(\mathbf k)
={}&
\sum_{l=1}^{N_S+1}
\Psi_{l,\mathbf k}^{\dagger}
H_l^{\rm N}(\mathbf k)
\Psi_{l,\mathbf k}
\nonumber\\
&+
t_S\sum_{l=1}^{N_S-1}
\left(
\Psi_{l+1,\mathbf k}^{\dagger}
\tau_0\sigma_0
\Psi_{l,\mathbf k}
+\mathrm{H.c.}
\right)
\nonumber\\
&+
t_\perp
\left(
\Psi_{N_S+1,\mathbf k}^{\dagger}
\tau_0\sigma_0
\Psi_{N_S,\mathbf k}
+\mathrm{H.c.}
\right),
\label{eq:H_stack_supp}
\end{align}
where
\begin{equation}
\Psi_{l,\mathbf k}
=
\left(
a_{l\mathbf k\uparrow},
a_{l\mathbf k\downarrow},
b_{l\mathbf k\uparrow},
b_{l\mathbf k\downarrow}
\right)^T
\end{equation}
is the sublattice--spin spinor in layer $l$.

The superconducting layers, $l\leq N_S$, are described by the spin- and
sublattice-independent dispersion
\begin{equation}
H_l^{\rm N}(\mathbf k)
=
\epsilon_{0,\mathbf k}\tau_0\sigma_0,
\qquad
\epsilon_{0,\mathbf k}
=
-t(\cos k_x+\cos k_y)-\mu.
\label{eq:H_S_layer_supp}
\end{equation}
The altermagnetic layer, $l=N_S+1$, is described by the reduced form of
Eqs.~(7) and (8) of the main text,
\begin{align}
H_{N_S+1}^{\rm N}(\mathbf k)
={}&
\epsilon_{0,\mathbf k}\tau_0\sigma_0
+t_4(\cos k_x-\cos k_y)\tau_z\sigma_0
\nonumber\\
&+
J_{\rm ex}\tau_z
(\mathbf n\cdot\boldsymbol\sigma)
+
\beta_z\tau_z
\left(
\sin k_y\,\sigma_x+\sin k_x\,\sigma_y
\right).
\label{eq:H_AM_layer_supp}
\end{align}
For simplicity, we set
$t_2=t_3=\alpha_0=\alpha_x=0$.
All layers are assigned the same in-plane dispersion and chemical
potential.

In Eq.~\eqref{eq:H_AM_layer_supp}, the inversion-breaking SOC is placed
on the altermagnetic side of the interface. We have also verified that
placing an $\alpha_0$-type Rashba SOC term on the adjacent
superconducting layer instead yields qualitatively similar results. The
spin-singlet $s$-wave pairing potential is nonzero only in the
superconducting layers,
\begin{equation}
\widehat\Delta_l(T)
=
\begin{cases}
\Delta(T)\,\tau_0 i\sigma_y, & l\leq N_S,\\[2mm]
0, & l=N_S+1,
\end{cases}
\label{eq:stack_pairing}
\end{equation}
where $\Delta(T)$ follows the same BCS temperature dependence as in the
main text. Thus, the pairing potential vanishes in the altermagnetic
layer, although anomalous superconducting correlations are induced there
through the interface hopping $t_\perp$.

We set $t_S=t=0.1~\mathrm{eV}$, so that the superconducting film has
comparable in-plane and interlayer hopping scales. The interface hopping
$t_\perp$ controls the S/AM hybridization and serves as a microscopic
proxy for the interface transparency.

A uniform condensate momentum $\hbar\mathbf q$, with
$\mathbf q=q\hat{\mathbf x}$, is imposed in the superconducting film. In
the gauge in which the pairing potential is spatially uniform, the
finite-$\mathbf q$ BdG Hamiltonian is
\begin{equation}
\mathcal H_{\rm BdG}^{\mathbf k,\mathbf q}
=
\begin{pmatrix}
H_{\rm stack}^{\rm N}(\mathbf k+\mathbf q/2)
&
\widehat\Delta(T)
\\[1mm]
\widehat\Delta^{\dagger}(T)
&
-H_{\rm stack}^{\rm N*}(-\mathbf k+\mathbf q/2)
\end{pmatrix},
\label{eq:H_stack_BdG}
\end{equation}
where
\begin{equation}
\widehat\Delta(T)
=
\operatorname{diag}
\left(
\widehat\Delta_1,\ldots,\widehat\Delta_{N_S},0
\right).
\end{equation}
The Hamiltonian in Eq.~\eqref{eq:H_stack_BdG} has dimension
$8(N_S+1)$. The free-energy density per unit interfacial area is
evaluated using Eq.~(2) of the main text, with $\Delta(T)$ following the
same BCS temperature dependence as in the single-layer calculation.

To isolate the supercurrent-induced N\'eel-vector-dependent contribution,
we define
\begin{equation}
\delta\mathcal F_{\mathbf n}(\mathbf q)
\equiv
\mathcal F(\mathbf q,\mathbf n)
-
\mathcal F(\mathbf q,\hat{\mathbf x}),
\label{eq:stack_deltaF}
\end{equation}
and its part odd under reversal of the condensate momentum,
\begin{equation}
\delta\mathcal F_{\mathbf n}^{\rm odd}(\mathbf q)
\equiv
\frac{
\delta\mathcal F_{\mathbf n}(\mathbf q)
-
\delta\mathcal F_{\mathbf n}(-\mathbf q)
}{2}.
\label{eq:stack_Fodd}
\end{equation}
For $\mathbf q=q\hat{\mathbf x}$ and
$\mathbf n=\hat{\mathbf y}$, the leading symmetry-allowed term in
Eq.~\eqref{eq:4b_sym_lin_final} gives
\begin{equation}
\delta\mathcal F_{\hat{\mathbf y}}^{\rm odd}
=
J_{\rm ex}\gamma q.
\label{eq:stack_Fodd_y}
\end{equation}

To make the field convention explicit, we define the N\'eel-field
difference as
\begin{equation}
\mathbf B_{\rm N}
\equiv
\mathbf B_{{\rm N}1}-\mathbf B_{{\rm N}2}.
\label{eq:stack_BN_definition}
\end{equation}
If the two sublattice moments are
$\mathbf M_{1,2}=\pm m_s\mathbf n$, then
$|\mathbf M_1-\mathbf M_2|=2m_s$. The effective field defined in the
main text therefore satisfies
\begin{equation}
\mathbf h
=
|\mathbf M_1-\mathbf M_2|
\frac{
\mathbf B_{{\rm N}1}-\mathbf B_{{\rm N}2}
}{2}
=
m_s\mathbf B_{\rm N}.
\label{eq:stack_h_BN}
\end{equation}
Since $\mathcal F$ is a free-energy density per unit area, the
corresponding N\'eel-field difference is
\begin{equation}
B_{\rm N}
=
\frac{
a^2
\left|
\delta\mathcal F_{\hat{\mathbf y}}^{\rm odd}
\right|
}{m_s}.
\label{eq:stack_Bneel}
\end{equation}
For equal and opposite fields on the two sublattices, the magnitude of
the field acting on either individual sublattice is
$B_{\rm sub}=B_{\rm N}/2$.

Figure~\ref{fig:stack_tp} shows the resulting N\'eel-field difference as
a function of the interface hopping $t_\perp$ for $N_S=1$, $2$, and
$4$. Its dependence on both $N_S$ and $t_\perp$ originates from the
hybridization of the superconducting and altermagnetic bands.
Hybridization redistributes the low-energy quasiparticle weight between
the layers and thereby changes the effective exchange sampled by the
proximitized quasiparticles, analogously to the thickness-averaged
exchange field of thin S/F bilayers~\cite{Bergeret2005}.

At the parameter point used for the estimate in the main text,
$t_\perp=10~\mathrm{meV}$ and
$q=10^{-3}\,\mathring{\mathrm A}^{-1}$, the calculation gives
\begin{equation}
\left|
\delta\mathcal F_{\hat{\mathbf y}}^{\rm odd}
\right|
\simeq
10^{-7}~\mathrm{eV}/a^2.
\label{eq:stack_F_estimate}
\end{equation}
Using $m_s=0.7\,\mu_{\rm B}$ and
$\mu_{\rm B}=5.788\times10^{-5}~\mathrm{eV/T}$ in
Eq.~\eqref{eq:stack_Bneel}, we obtain
\begin{equation}
B_{\rm N}
\simeq
2.2~\mathrm{mT}.
\label{eq:stack_B_estimate}
\end{equation}
Equivalently, the field magnitude on either individual sublattice is
$B_{\rm sub}\simeq1.1~\mathrm{mT}$. Thus, the calculation predicts a
staggered field on the millitesla scale.

The underlying signed response coefficient $\gamma$ depends
nonmonotonically on $t_\perp$ and can pass through zero as the interface
hybridization is increased. Because Fig.~\ref{fig:stack_tp}(a) displays
$|B_{\rm N}|$, a reversal of the signed response appears as a minimum of
the plotted magnitude rather than as a negative field. This zero crossing
reflects competing quasiparticle contributions and does not indicate a
change in the symmetry of the torque.

\begingroup
\makeatletter
\let\@auxout\@bibunitaux
\let\thebibliography\NAT@thebibliography
\let\endthebibliography\endNAT@thebibliography
\makeatother
\renewcommand{\bibsection}{\section*{References}}
\putbib[\jobname Notes,SAM_revised]
\endgroup
\end{bibunit}
\endgroup

\end{document}